\pdfoutput=1
\documentclass[reprint,aps,prl,superscriptaddress,amsmath,amssymb,noeprint,floatfix]{revtex4-2}

\usepackage[T1]{fontenc}
\usepackage{graphicx}
\usepackage{bm}
\usepackage{mathtools}
\usepackage{microtype}
\usepackage{mathrsfs}
\usepackage{xcolor}
\makeatletter
\newcommand{\FloatBarrier}{\ifx\@deferlist\@empty\else\clearpage\fi}
\makeatother
\newcommand{\SupplementFigure}[1]{%
 \IfFileExists{#1}{\includegraphics[width=0.82\textwidth,height=0.43\textheight,keepaspectratio]{#1}}{%
  \par\noindent\textit{Figure asset pending.}\par
  \noindent\texttt{\detokenize{#1}}\par}}
\usepackage[colorlinks=true,linkcolor=blue,citecolor=blue,urlcolor=blue]{hyperref}

\newcommand{\ii}{\mathrm{i}}
\newcommand{\dd}{\mathrm{d}}
\newcommand{\ee}{\mathrm{e}}
\newcommand{\Tr}{\operatorname{Tr}}
\renewcommand{\Re}{\operatorname{Re}}
\renewcommand{\Im}{\operatorname{Im}}

\begin{document}

\title{Nonlinear Quantum Oscillations from Triangular Landau Orbits}

\author{H. Geng}
\email{genghao@nuaa.edu.cn}
\affiliation{College of Physics, Key Laboratory of Aerospace Information Perception and Physics, MIIT, Nanjing University of Aeronautics and Astronautics, Nanjing 211106, China}

\author{G. Y. Qi}
\affiliation{College of Electrical Engineering, Zhejiang University of Water Resources and Electric Power, Hangzhou 310018, China}

\author{K. X. Jia}
\email{jia_kaixiang@fudan.edu.cn}
\affiliation{Interdisciplinary Center for Theoretical Physics and Information Sciences (ICTPIS) and State Key Laboratory of Surface Physics, Fudan University, Shanghai 200433, China}

\author{L. Sheng}
\affiliation{National Laboratory of Solid State Microstructures, School of Physics, and Collaborative Innovation Center of Advanced Microstructures, Nanjing University, Nanjing 210093, China}

\author{D. Y. Xing}
\affiliation{National Laboratory of Solid State Microstructures, School of Physics, and Collaborative Innovation Center of Advanced Microstructures, Nanjing University, Nanjing 210093, China}

\date{\today}

\begin{abstract}
Linear Shubnikov--de Haas oscillations map Fermi surfaces but leave crystal polarity unresolved. Here we formulate a microscopic quantum theory of nonlinear quantum oscillations (NQOs) under a perpendicular magnetic field $B_z$ and establish Landau-level parity selection rules classifying which crystals allow NQOs. Applying this framework to threefold-symmetric crystals, we find that breaking inversion symmetry deforms Landau states into real-space triangles. Extending guiding-center diffusion to second order, we show that these triangular orbits rectify scattering steps into two orthogonal NQO channels: the asymmetric real-space orbit shape drives a $B_z$-even oscillation, whereas chiral cyclotron motion drives a $B_z$-odd oscillation. From this theory, we derive a closed-form nonlinear Lifshitz--Kosevich formula. It shows that NQOs acquire an intrinsic $\pi/2$ phase shift in $1/B_z$ relative to linear oscillations and decay more slowly in amplitude as $B_z$ decreases, consistent with recent measurements in $\alpha\text{-Sn}(111)$. Finally, we extend this mechanism to systems without spin-orbit coupling and predict gate-switchable, purely orbital NQOs in Bernal bilayer graphene. Together, these results expand quantum oscillations beyond Fermi-surface mapping into a direct electrical probe of Landau-orbit geometry and crystal polarity.
\end{abstract}

\maketitle
\addtocontents{toc}{\protect\iffalse}
\renewcommand{\topfraction}{0.99}
\renewcommand{\bottomfraction}{0.99}
\renewcommand{\textfraction}{0.01}
\renewcommand{\floatpagefraction}{0.88}
\renewcommand{\dbltopfraction}{0.95}
\renewcommand{\dblfloatpagefraction}{0.85}
\setlength{\textfloatsep}{7pt plus 1.5pt minus 1.5pt}
\setlength{\dbltextfloatsep}{7pt plus 1.5pt minus 1.5pt}
\setlength{\intextsep}{6pt plus 1.5pt minus 1.5pt}
\setlength{\abovecaptionskip}{4pt}
\setlength{\belowcaptionskip}{0pt}

\textit{Introduction.}---Linear Shubnikov--de Haas (SdH) and de Haas--van Alphen oscillations map extremal Fermi-surface areas and cyclotron masses in a perpendicular magnetic field~\cite{Onsager1952,LifshitzKosevich1956,Shoenberg1984}, yet linear conductivity is inversion-even~\cite{Birss1964} and cannot distinguish opposite inversion partners~\cite{TokuraNagaosa2018,NagaosaYanase2024}. Resolving crystal polarity requires a second-order longitudinal current, $j_\parallel^{(2)}=\sigma_\parallel^{(2)}E_\parallel^2$, as in electrical magnetochiral anisotropy~\cite{Rikken2001,Rikken2005}. Most studies of second-order nonlinear transport address the weak-field semiclassical regime, invoking asymmetric dispersions and spin textures~\cite{Ideue2017,He2018NatPhys,Zhang2022SciAdv}, asymmetric scattering~\cite{IshizukaNagaosa2020NatCommun,IsobeXuFu2020,HeIsobe2021}, or quantum geometry~\cite{GaoYangNiu2014,Wang2023Nature,KaplanHolderYan2024PRL,LiuLuXie2025NSR}. Much less is understood in the strong-field Landau-quantized regime under a purely perpendicular field $\bm{B}=B_z\bm{e}_z$, where Landau quantization replaces the continuous in-plane dispersion with discrete levels, obscuring the asymmetry of the zero-field bands.

In an isotropic two-dimensional electron gas (2DEG), circular Landau orbits preserve $180^\circ$ rotation $C_{2z}$, so opposite electric fields $\pm E_\parallel$ produce identical responses and $\sigma_\parallel^{(2)}$ vanishes. Obtaining a second-order current under a purely perpendicular field therefore requires breaking $C_{2z}$. Recent experiments reported pronounced perpendicular-field nonlinear quantum oscillations (NQOs) in trigonal $\alpha\text{-Sn}(111)$~\cite{Kalappattil2026} and, transversely via Brown--Zak states, in graphene moir\'e superlattices~\cite{Zhong2026}. Existing studies treat phenomenological Boltzmann transport~\cite{Kalappattil2026}, optical second-harmonic generation~\cite{Chang2025}, or general Landau-level Keldysh formulas requiring an in-plane magnetic-field component~\cite{Nakazawa2026}, leaving open which crystal symmetries permit NQOs in a purely perpendicular magnetic field, how symmetry breaking manifests in real-space Landau orbits and generates a second-order longitudinal current through guiding-center scattering, and whether the resulting oscillations admit a closed-form nonlinear Lifshitz--Kosevich (NLLK) description.

\begin{figure*}[!t]
\centering
\includegraphics[width=\textwidth]{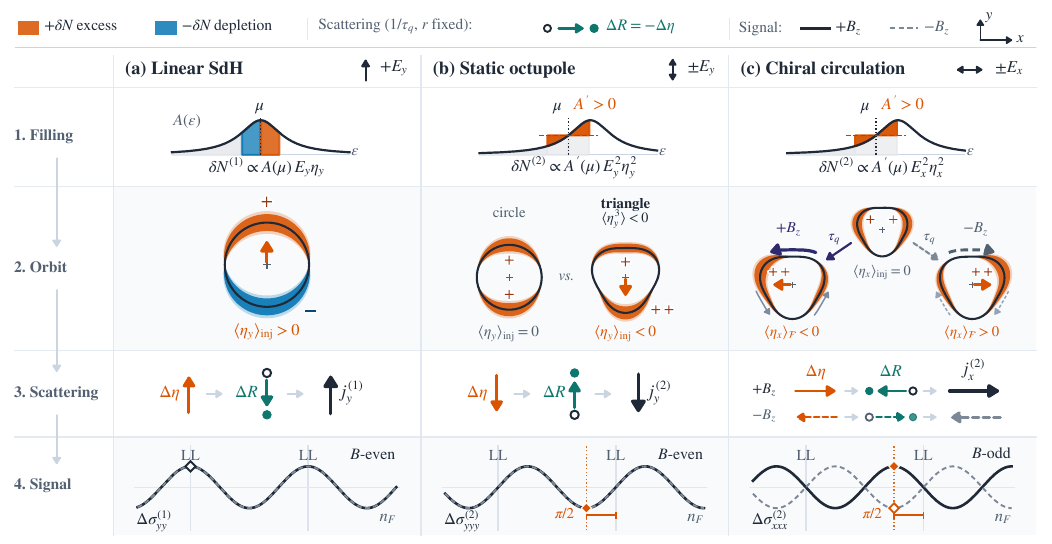}
\caption{\label{fig:schematic}Guiding-center scattering mechanism unifying linear SdH oscillations and NQOs in a $C_{3v}$ crystal under $\bm{B}=B_z\bm{e}_z$ ($e>0$, $\bm{r}=\bm{R}+\bm{\eta}$). Columns trace level filling $A(\varepsilon)$, orbital occupation $\delta N$, local scattering at rate $1/\tau_q$ jumping $\Delta\bm{R}=-\Delta\bm{\eta}$ [$\Delta\bm{\eta}\equiv\langle\bm{\eta}\rangle_{\mathrm{inj}}-\langle\bm{\eta}\rangle_{\mathcal{F}}$, Eq.~\eqref{eq:master_exact}], and oscillatory conductivity vs.\ $n_F$. (a)~Linear SdH ($\bm{E}=+E_y\bm{e}_y$, circle): odd filling at a peak gives $\langle\eta_y\rangle_{\mathrm{inj}}>0$, stepping $\Delta R_y<0$ and $j_y^{(1)}>0$ ($\propto\cos 2\pi n_F$). (b)~$B_z$-even static-octupole channel ($\bm{E}=\pm E_y\bm{e}_y$): on a sloped flank ($A'>0$), quadratic filling weights the sharp $-y$ corner over the flat $+y$ side ($\langle\eta_y^3\rangle<0$), rectifying $j_y^{(2)}<0$ shifted by $\pi/2$ ($\propto +\sin 2\pi n_F$). (c)~$B_z$-odd chiral-circulation channel ($\bm{E}=\pm E_x\bm{e}_x$): symmetric filling piles onto $+y$ shoulders ($\langle\eta_x\rangle_{\mathrm{inj}}=0$), while $\pm B_z$ cyclotron circulation produces opposite steady-state shifts $\langle\eta_x\rangle_{\mathcal{F}}\lessgtr 0$ and reverses $j_x^{(2)}$ ($\propto -s_B\sin 2\pi n_F$).}
\end{figure*}

Here we formulate a microscopic quantum theory of perpendicular-field NQOs. We derive the second-order longitudinal conductivity and a generalized Landau-level parity rule classifying allowed crystal symmetries (Table~\ref{tab:symmetry}). We then show how symmetry breaking is encoded in the real-space geometry of Landau orbits and how this geometry enters the second-order longitudinal conductivity in $C_{3v}$ crystals, where inversion breaking deforms Landau states into real-space triangles. Extending guiding-center diffusion~\cite{Kubo1959,AndoUemura1974} to second order reveals two orthogonal channels at electric-field angle $\phi$ (Fig.~\ref{fig:schematic}): a $B_z$-even channel from the asymmetric triangular orbit shape ($\sin 3\phi$) and a $B_z$-odd channel from chiral cyclotron circulation ($\cos 3\phi$), yielding a closed-form NLLK formula. Applied to warped Dirac surfaces, this theory reproduces the threefold angular sign structure and captures the slower relative $1/B_z$ damping ratio in $\alpha\text{-Sn}(111)$~\cite{Kalappattil2026}. Finally, in gated Bernal bilayer graphene (BLG), where a vertical bias lowers $D_{3d}$ to $C_{3v}$~\cite{McCann2006PRB,Zhang2009Nature,McCannKoshino2013RPP}, we predict purely orbital NQOs without spin-orbit coupling that vanish at zero bias and invert with gate polarity.

\textit{Nonlinear conductivity in quantized Landau levels.}---Consider a 2D electron system described by $H_B(\bm{\Pi})$ under $\bm{B}=B_z\bm{e}_z$, with kinetic momentum $\bm{\Pi}=-\ii\hbar\bm{\nabla}+e\bm{\mathcal{A}}$ obeying $[\Pi_x,\Pi_y]=-\ii s_B\hbar^2/\ell_B^2$ ($s_B\equiv\operatorname{sgn}B_z$, $\ell_B=\sqrt{\hbar/(e|B_z|)}$). Splitting $\bm{r}=\bm{R}+\bm{\eta}$ into the guiding center $\bm{R}$ (areal degeneracy $1/2\pi\ell_B^2$) and cyclotron coordinate $\bm{\eta}=-s_B(\ell_B^2/\hbar)\bm{e}_z\times\bm{\Pi}$~\cite{Kubo1959,Goerbig2011RMP}, a uniform electric field $\bm{E}=E_\parallel\bm{e}_\phi$ along $\bm{e}_\phi=(\cos\phi,\sin\phi)$ splits $e\bm{E}\cdot\bm{r}=eE_\parallel R_\parallel+eE_\parallel\eta_\parallel$. Since $\bm{R}$ commutes with $H_B(\bm{\Pi})$ and $\bm{\eta}$, $eE_\parallel R_\parallel$ is an energy offset per sector and Landau levels couple solely via $eE_\parallel\eta_\parallel$. Without scattering, $R_\parallel$ is conserved while $[R_x,R_y]=\ii s_B\ell_B^2$ gives purely transverse drift and zero longitudinal current.

Dissipative longitudinal transport requires scattering that breaks $R_\parallel$ conservation. Within a local, wide-band fermionic-bath model with scalar self-energy $\Sigma^{r,a}=\mp\ii\Gamma$ [$\Gamma=\hbar/(2\tau_q)$]~\cite{Han2013PRB,HanLi2013PRB,SM}, the bath distribution shifts with the local scalar potential, eliminating $R_\parallel$ identically and yielding the exact length-gauge steady state $\hat{\rho}_{\mathcal{F}}=(\Gamma/\pi)\int\dd E\,G_{\mathcal{F}}^r(E)f_0(E-\mathcal{F}\eta_\parallel)G_{\mathcal{F}}^a(E)$, where $\mathcal{F}\equiv eE_\parallel$, $G_{\mathcal{F}}^{r,a}(E)=(E-H_B-\mathcal{F}\eta_\parallel\pm\ii\Gamma)^{-1}$, and $f_0$ is the Fermi--Dirac distribution at $(\mu,T)$. It satisfies the stationary quantum kinetic equation
\begin{equation}\label{eq:kinetic_eq}
\big[H_B + \mathcal{F}\eta_\parallel,\,\hat{\rho}_{\mathcal{F}}\big] - 2\ii\Gamma\big(\hat{\rho}_{\mathcal{F}} - \hat{\rho}_{\mathrm{inj}}\big) = 0,
\end{equation}
with Hermitian injection source operator $\hat{\rho}_{\mathrm{inj}}=\frac{\ii}{2\pi}\int\dd E\,[G_{\mathcal{F}}^r(E)f_0(E-\mathcal{F}\eta_\parallel)-f_0(E-\mathcal{F}\eta_\parallel)G_{\mathcal{F}}^a(E)]$ (not necessarily positive semidefinite) encoding the local chemical potential $\mu+\mathcal{F}\eta_\parallel$. Integrating Eq.~\eqref{eq:kinetic_eq} gives the survival-time average $\hat{\rho}_{\mathcal{F}}=\int_0^\infty(\dd t/\tau_q)\,\ee^{-t/\tau_q}\,U(t)\hat{\rho}_{\mathrm{inj}}U^\dagger(t)$ under $U(t)=\ee^{-\ii(H_B+\mathcal{F}\eta_\parallel)t/\hbar}$. Between successive scattering events of mean lifetime $\tau_q$, $U(t)$ coherently rotates the injected density $\hat{\rho}_{\mathrm{inj}}$ along the chiral cyclotron orbit (with frequency $\omega_c=eB_z/m_c$ and direction set by $s_B$), averaging it over many turns when $|\omega_c|\tau_q\gg 1$ or cutting the rotation short when $|\omega_c|\tau_q\lesssim 1$. Tracing Eq.~\eqref{eq:kinetic_eq} against $\eta_\parallel$ via $v_\parallel=(\ii/\hbar)[H_B,\eta_\parallel]$ converts $j_\parallel=-[e/(2\pi\ell_B^2)]\Tr(v_\parallel\hat{\rho}_{\mathcal{F}})$ into
\begin{equation}\label{eq:master_exact}
\begin{aligned}
j_\parallel &= \frac{e}{2\pi\ell_B^2\tau_q}\Tr\!\big[\eta_\parallel\big(\hat{\rho}_{\mathrm{inj}}-\hat{\rho}_{\mathcal{F}}\big)\big] \\
            &= \frac{e}{2\pi\ell_B^2\tau_q}\big(\langle\eta_\parallel\rangle_{\mathrm{inj}} - \langle\eta_\parallel\rangle_{\mathcal{F}}\big).
\end{aligned}
\end{equation}
Here $\langle\eta_\parallel\rangle_{\mathrm{inj}}=\int\dd E\,\Tr[\eta_\parallel\hat{A}_{\mathcal{F}}(E)f_0(E-\mathcal{F}\eta_\parallel)]$ [$\hat{A}_{\mathcal{F}}=(\Gamma/\pi)G_{\mathcal{F}}^rG_{\mathcal{F}}^a$] and $\langle\eta_\parallel\rangle_{\mathcal{F}}=\Tr[\eta_\parallel\hat{\rho}_{\mathcal{F}}]$ are the real-space centroids of the locally injected and cyclotron-evolved steady-state distributions per guiding-center state. In Eq.~\eqref{eq:master_exact}, each local scattering event at fixed position $r_\parallel=R_\parallel+\eta_\parallel$ resets an electron from $\hat{\rho}_{\mathcal{F}}$ back to $\hat{\rho}_{\mathrm{inj}}$, shifting its guiding center by $\Delta R_\parallel=-\Delta\eta_\parallel\equiv-(\langle\eta_\parallel\rangle_{\mathrm{inj}}-\langle\eta_\parallel\rangle_{\mathcal{F}})$ (Fig.~\ref{fig:schematic}). For linear SdH on a circular orbit at a Landau peak [Fig.~\ref{fig:schematic}(a)], first-order local filling $\delta f^{(1)}\propto eE_y\eta_y(-f_0')$ shifts $\langle\eta_y\rangle_{\mathrm{inj}}>0$, while rapid cyclotron rotation averages out $\langle\eta_y\rangle_{\mathcal{F}}\to 0$, giving $\Delta R_y<0$. Formally, linearizing Eq.~\eqref{eq:master_exact} recovers the constant-broadening Kubo--Greenwood formula and, for sharp levels ($\Gamma\ll\hbar|\omega_c|$), Einstein guiding-center diffusion $\sigma_{\parallel\parallel,\alpha}^{(1)}\simeq e^2\int\dd E\,[-f_0']D_\alpha D_\alpha^*$ with $D_\alpha=A_\alpha/(2\pi\ell_B^2)$ and $D_\alpha^*=(\langle\eta_\parallel^2\rangle_\alpha-\langle\eta_\parallel\rangle_\alpha^2)/\tau_q$ (Appendix~A)~\cite{Kubo1957,Greenwood1958,Kubo1959,AndoUemura1974}. This diffusion constant measures only the orbit variance and stays unchanged under spatial inversion ($\eta_\parallel\to-\eta_\parallel$), so linear transport cannot distinguish opposite crystal polarities.

Expanding Eq.~\eqref{eq:master_exact} to second order in $E_\parallel$ cancels equilibrium $f_0(E)$ identically and splits $\sigma_\parallel^{(2)}=j_\parallel^{(2)}/E_\parallel^2$ into two terms [$G_0^{r,a}=(E-H_B\pm\ii\Gamma)^{-1}$, $\hat{A}_0=(\Gamma/\pi)G_0^rG_0^a$]~\cite{SM}:
\begin{equation}\label{eq:sigma2_decomp}
\begin{aligned}
\sigma_\parallel^{(2)} &= \left.\frac{\partial\sigma_{\parallel\parallel}^{(1)}[H_B+eE_\parallel\eta_\parallel]}{\partial E_\parallel}\right|_{E_\parallel=0} \\
&\quad +\frac{e^3}{4\pi\ell_B^2\tau_q}\!\int\!\dd E\,f_0''(E) \\
&\quad \times\!\left(\Tr[\eta_\parallel^3\hat{A}_0]-\frac{\Gamma}{\pi}\Tr[\eta_\parallel G_0^r\eta_\parallel^2 G_0^a]\right).
\end{aligned}
\end{equation}
Term~(I) captures orbital deformation of $\sigma_{\parallel\parallel}^{(1)}$ and, when $\langle\eta_\parallel\rangle_\alpha=0$ by symmetry, yields mainly the smooth background. Term~(II) stems from quadratic local filling $\delta f^{(2)}=\frac{1}{2}(eE_\parallel\eta_\parallel)^2 f_0''(E)$ on undeformed orbits and drives sharp NQOs. Its injection (II-a, first trace) and cyclotron (II-b, second trace) parts are cubic in $\eta_\parallel$ and flip under inversion ($\eta_\parallel\to-\eta_\parallel$), encoding the static-octupole ($\langle\eta_\parallel^3\rangle_\alpha$, $B_z$-even) and chiral-circulation ($B_z$-odd) channels. Integrating $f_0''(E)$ by parts replaces the SdH DOS peak $D_\alpha(E)$ with its flank slope $D_\alpha'(E)$, rectifying directional guiding-center drift $\pi/2$ out of phase with SdH [Figs.~\ref{fig:schematic}(b) and~\ref{fig:schematic}(c)].

\textit{Parity selection rule, triangular orbits, and dual channels.}---In terms of $a_B=\ell_B(\Pi_x-\ii s_B\Pi_y)/(\sqrt{2}\hbar)$ ($[a_B,a_B^\dagger]=1$, $\hat{n}=a_B^\dagger a_B$), $\eta_\parallel(\phi)=(\ii\ell_B/\sqrt{2})(\ee^{\ii s_B\phi}a_B-\ee^{-\ii s_B\phi}a_B^\dagger)$ couples only $\Delta n=\pm 1$. The oscillator parity operator $(-1)^{\hat{n}}$ reverses both $\bm{\Pi}$ and $\eta_\parallel$, acting as in-plane $C_{2z}$. Because every term in Eq.~\eqref{eq:sigma2_decomp} is cubic in $\eta_\parallel$, any $H_B$ commuting with $(-1)^{\hat{n}}$ multiplies every trace by $(-1)^3=-1$ and forces $\sigma_\parallel^{(2)}=0$. In a single-band spinless model ($\hbar\bm{k}\to\bm{\Pi}$), breaking parity requires odd-in-$\bm{k}$ warping such as $\lambda(k_x^3-3k_x k_y^2)\to\tilde{\lambda}(a_B^{\dagger 3}+a_B^3)$ ($\tilde{\lambda}=\sqrt{2}\lambda/\ell_B^3$). The cubic term $a_B^{\dagger 3}+a_B^3$ mixes $|n\rangle$ with $|n\pm 3\rangle$, which have opposite oscillator parity. In real space, interference between these components can generate a contribution to the orbital probability density that is odd under inversion about the guiding center. The same $\Delta n=\pm 3$ mixing allows three $\Delta n=\pm 1$ dipole transitions to form a closed loop, permitting a finite second-order longitudinal conductivity.

In $\Theta$-invariant systems, time reversal forbids spinless odd-in-$\bm{k}$ terms, so warping must couple to spin or valley. Yet this alone is insufficient: adding $\lambda(k_x^3-3k_x k_y^2)\sigma_z$ to a parabolic band~\cite{Fu2009,Ezawa2025PRB} warps both spins oppositely, preserving spin-swap parity $\sigma_x(-1)^{\hat{n}}$ so their signals cancel. More generally, any \textit{generalized Landau-level parity} $\tilde{\mathcal{P}}\equiv U(-1)^{\hat{n}}$ ($U^2=I$, $[U,\eta_\parallel]=0$) commuting with $H_B$ enforces $\sigma_\parallel^{(2)}=0$. Observing finite NQOs therefore requires breaking every such $\tilde{\mathcal{P}}$.

This rule classifies allowed crystal symmetries (Table~\ref{tab:symmetry}). Both $C_{2z}$ (2DEG, $C_{4v}$) and 3D inversion $\mathcal{P}$ (unbiased BLG) realize $\tilde{\mathcal{P}}$ at fixed $B_z$ and force $\sigma_\parallel^{(2)}=0$, so a crystal must lack both. Projecting $j_a^{(2)}=\sum_{bc}\sigma_{abc}^{(2)}E_b E_c$ onto the electric-field direction $\bm{e}_\phi=(\cos\phi,\sin\phi)$ yields only $1\phi$ and $3\phi$ harmonics. Polar $C_s$ crystals allow both, whereas threefold symmetry $C_{3z}$ isolates pure $3\phi$ oscillations in $C_{3v}$ $\alpha$-Sn and gated BLG~\cite{SM}.

In $C_{3v}$ crystals, a vertical mirror $M_x$ ($x\to-x$) flips $B_z$ and survives as magnetic mirror $M_x\Theta$ ($m'$ in Table~\ref{tab:symmetry}), reversing $j_x^{(2)}=\sigma_{xxx}^{(2)}E_x^2$ with $B_z$ while leaving $j_y^{(2)}=\sigma_{yyy}^{(2)}E_y^2$ even~\cite{Birss1964,WatanabeYanase2020PRR,NagaosaYanase2024}:
\begin{equation}\label{eq:sigma2_3phi}
\sigma_\parallel^{(2)}(B_z,\phi) = \chi_e(B_z)\sin 3\phi + \chi_o(B_z)\cos 3\phi,
\end{equation}
with $B_z$-even $\chi_e=-\sigma_{yyy}^{(2)}$ ($\bm{E}\parallel y$) and $B_z$-odd $\chi_o=\sigma_{xxx}^{(2)}$ ($\bm{E}\parallel x$). Chiral crystals without $M\Theta$ (class $3$) leave both $3\phi$ phases unconstrained.

\begin{table}[!tbp]
\caption{\label{tab:symmetry}Fixed-$B_z$ magnetic point groups, generalized Landau parity $\tilde{\mathcal{P}}=U(-1)^{\hat{n}}$, and vertical magnetic mirrors $M\Theta$. Here $U=I$ (2DEG), $\sigma_z$ (Rashba), and $\nu_x\Sigma_x$ (unbiased BLG: valley swap $\nu_x$, layer-sublattice swap $\Sigma_x$); gated BLG ($U_0\neq 0$) enters $3m'$. Finite $\sigma_\parallel^{(2)}$ requires breaking all $\tilde{\mathcal{P}}$ (broken$^*$ via $\Delta n=\pm 3$). The Berry-curvature dipole~\cite{SodemannFu2015} vanishes in $3m',3$ and is transverse in $m'$.}
\begin{ruledtabular}
\begin{tabular}{lllll}
Class & Example & $\tilde{\mathcal{P}}$ & $M\Theta$ & $\sigma_\parallel^{(2)}$ \\
\hline
$\infty m',4m'm'$ & 2DEG, $C_{4v}$ & $U(-1)^{\hat{n}}$ & Yes & $0$ \\
$\bar{3}m'$ & BLG ($U_0{=}0$) & $\nu_x\Sigma_x(-1)^{\hat{n}}$ & Yes & $0$ \\
$3m'$ & $\alpha$-Sn, BLG & Broken$^*$ & Yes & Locked $3\phi$ \\
$m'$ & Polar ($C_s$) & Model dep. & Yes & Locked $1\phi,3\phi$ \\
$3$ & Chiral ($C_3$) & Model dep. & No & Free $3\phi$ \\
\end{tabular}
\end{ruledtabular}
\end{table}

Applying $M_x\Theta$ directly to Eq.~\eqref{eq:sigma2_decomp} isolates the two channels (Appendix~A). Under $M_x\Theta$, $H_B$ stays invariant, $G_0^r$ and $G_0^a$ exchange, and $(\eta_x,\eta_y)\mapsto(-\eta_x,+\eta_y)$. For $\bm{E}\parallel y$, the static octupole $\langle\eta_y^3\rangle_\alpha\neq 0$ survives. For $\bm{E}\parallel x$, however, orbital deformation and $\langle\eta_x^3\rangle_\alpha$ vanish identically, and swapping $G_0^r\leftrightarrow G_0^a$ reduces the remaining trace to a single Fermi-surface expectation value. Integrating by parts yields~\cite{SM}
\begin{equation}\label{eq:dual_channels}
\begin{aligned}
\Delta\sigma_{yyy}^{(2)} &\simeq \frac{e^3}{4\pi\ell_B^2\tau_q}\!\int\!\dd E\,[-f_0']\,\partial_E\!\sum_\alpha A_\alpha\langle\eta_y^3\rangle_\alpha, \\
\sigma_{xxx}^{(2)} &= \frac{e^3}{4\pi\ell_B^2\tau_q}\!\int\!\dd E\,[-f_0']\,\partial_E\Tr\!\big[\eta_x^2\,\delta\hat{\rho}_{\mathcal{F}}^{(1)}(E)\big],
\end{aligned}
\end{equation}
where $\Delta\sigma_{yyy}^{(2)}$ retains the leading sharp-level, large-$n_F$ octupole term while $\sigma_{xxx}^{(2)}$ is exact for all $\Gamma$ and $B_z$, with linear non-equilibrium kernel $\delta\hat{\rho}_{\mathcal{F}}^{(1)}(E)=(\Gamma/\pi)G_0^r\eta_x G_0^a$.

Both channels come directly from the two-centroid difference $\langle\eta_\parallel\rangle_{\mathrm{inj}}-\langle\eta_\parallel\rangle_{\mathcal{F}}$ on a single deformed orbit. Here $\Delta n=\pm 3$ warping~\cite{Dresselhaus1974PRB,McCannFalko2006,RepinBurmistrov2015,Ju2020} beats circular states $\propto\ee^{\ii s_B\nu\theta_{\bm{r}}}$ with $\nu\pm 3$ neighbors into a $\sin 3\theta_{\bm{r}}$ triangle in $|\Psi_n(\bm{r})|^2$ encircled by a $B_z$-odd chiral current~\cite{SM}. Semiclassically, for a single $\Theta$-invariant pocket, $\bm{\eta}=-s_B(\ell_B^2/\hbar)\bm{e}_z\times\bm{\Pi}$ maps the field-skewed Fermi contour into the same $B_z$-even triangle ($s_B^2=+1$) with a wide flat $+y$ side and a sharp, farther $-y$ corner. On a rising Landau-level slope ($A'>0$), quadratic local filling $\delta f^{(2)}\propto\eta_\parallel^2 f_0''(E)$ adds excess density wherever $\eta_\parallel^2$ is largest, and the two field directions separate the two centroids in Eq.~\eqref{eq:master_exact}. Along $\bm{E}\parallel y$ [Fig.~\ref{fig:schematic}(b)], this excess concentrates at the farther, sharp $-y$ corner rather than the closer flat $+y$ side, pulling the injection centroid downward, $\langle\eta_y\rangle_{\mathrm{inj}}^{(2)}<0$ (proportional to $\langle\eta_y^3\rangle_\alpha<0$), with the same sign for $\pm E_y$ and $\pm B_z$. At weak damping ($\tilde{\Gamma}\ll 1$), multi-turn cyclotron rotation averages the steady-state centroid $\langle\eta_y\rangle_{\mathcal{F}}^{(2)}$ toward zero, leaving $\langle\eta_y\rangle_{\mathrm{inj}}^{(2)}$ to drive a $B_z$-even current. Along $\bm{E}\parallel x$ [Fig.~\ref{fig:schematic}(c)], by contrast, mirror symmetry $M_x$ places equal excess density on the left and right $+y$ shoulders, so the injection centroid vanishes, $\langle\eta_x\rangle_{\mathrm{inj}}^{(2)}=0$. As cyclotron motion rotates these two shoulders in the direction set by $s_B$, one moves across the flat $+y$ side while the other turns toward the sharp $-y$ corner. Over a finite lifetime $\tau_q$, this asymmetric motion breaks the left--right balance and produces a steady-state centroid $\langle\eta_x\rangle_{\mathcal{F}}^{(2)}\neq 0$ that reverses with $B_z$ (Appendix~B, Fig.~\ref{fig:endmatter_rho}). Equivalently under $M_x\Theta$ [Eq.~\eqref{eq:dual_channels}], transverse Lorentz deflection steers $\pm E_x$ carriers toward the wide $+y$ side vs.\ narrow $-y$ corner, coupling lateral width $\eta_x^2$ to $\partial_E$.

\textit{Nonlinear Lifshitz--Kosevich formula and Hall-bar rotation.}---In the semiclassical regime ($n_F\gg 1$) for a $3m'$ pocket, expanding Eq.~\eqref{eq:sigma2_decomp} in the warped eigenbasis and applying Poisson resummation to leading order in warping (SM Sec.~S3) yields the NLLK formula $\Delta\sigma_\parallel^{(2)}=\Delta\chi_e\sin 3\phi+\Delta\chi_o\cos 3\phi$~\cite{SM}:
\begin{equation}\label{eq:nllk}
\begin{aligned}
\Delta\sigma_\parallel^{(2)} &= \sum_{p=1}^\infty R_T(p\lambda_T) R_D(p\tilde{\Gamma})\,\mathcal{K}_p(B_z)\sin(2\pi p n_F) \\
&\quad \times \Big[s_B\tilde{\Gamma}\cos 3\phi + \big(1+2\tilde{\Gamma}^2\big)\sin 3\phi\Big],\\
\mathcal{K}_p(B_z) &= \frac{6p\,e^3\hbar^2 V_1^2 V_2\,\tilde{\Gamma}}{\ell_B^2(\hbar\omega_c)^4(1+\tilde{\Gamma}^2)(1+4\tilde{\Gamma}^2)}.
\end{aligned}
\end{equation}
Here $n_F=F/|B_z|-\gamma_B$ ($F=\hbar k_F^2/(2e)$, phase offset $\gamma_B$~\cite{Onsager1952,LifshitzKosevich1956,Shoenberg1984}), $\tilde{\Gamma}=\Gamma/(\hbar|\omega_c|)=1/(2|\omega_c|\tau_q)$, $R_D(p\tilde{\Gamma})=\ee^{-2\pi p\tilde{\Gamma}}$, $R_T(p\lambda_T)=p\lambda_T/\sinh(p\lambda_T)$ with $\lambda_T=2\pi^2 k_B T/(\hbar|\omega_c|)$, and $V_{1,2}$ are band-specific velocity amplitudes at $n_F$ [SM Sec.~S3; Eq.~\eqref{eq:V12}].

Equation~\eqref{eq:nllk} predicts three signatures. First, for a single $\Theta$-invariant pocket, NQOs share the SdH frequency $F$ but acquire a $\pi/2$ phase shift in $1/B_z$ [$\sin(2\pi p n_F)$ vs.\ $\cos(2\pi p n_F)$, Fig.~\ref{fig:schematic}; multivalley interference is treated in Appendix~C]. Second, band factors cancel in the single-pocket channel ratio, $\Delta\chi_o/\Delta\chi_e=s_B\tilde{\Gamma}/(1+2\tilde{\Gamma}^2)\simeq s_B/(2|\omega_c|\tau_q)$ at $\tilde{\Gamma}\ll 1$, so the $B_z$-even channel dominates sharp oscillations and directly reads out $\tau_q$ within the scalar-bath model. Third, at $\tilde{\Gamma}\ll 1$, algebraic $1/|B_z|$ growth of $\mathcal{K}_p(B_z)$ (scaling as $|B_z|^{-3}$ at fixed $\mu,\Gamma$ for linear Dirac cones) offsets $R_D$, making NQOs decay more slowly than linear SdH.

\begin{figure}[!tbp]
\centering
\includegraphics[width=0.90\columnwidth]{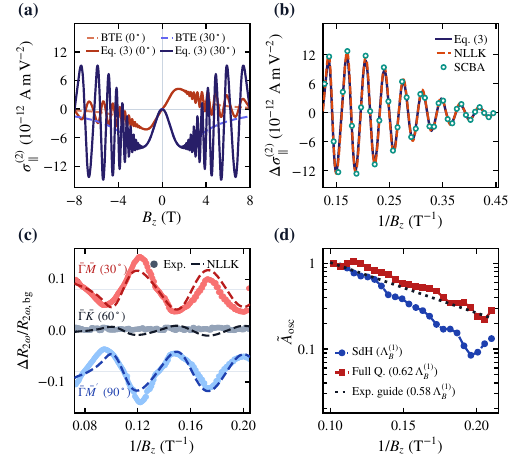}
\caption{\label{fig:dirac_exp}NQOs on a warped Dirac surface and comparison with $\alpha\text{-Sn}(111)$~\cite{SM}. (a)~Total $\sigma_\parallel^{(2)}(B_z)$ for Eq.~\eqref{eq:HB} at $\phi=0^\circ$ (red) and $30^\circ$ (indigo; $T=5\;\mathrm{K}$, $g_z=0$): Eq.~\eqref{eq:sigma2_decomp} (solid) vs.\ matched-$\tau_{\mathrm{tr}}$ Boltzmann limit (BTE, dashed)~\cite{SundaramNiu1999,XiaoChangNiu2010}. (b)~Oscillatory $\Delta\sigma_\parallel^{(2)}$ vs.\ $1/B_z$ at $\phi=30^\circ$: Eq.~\eqref{eq:sigma2_decomp} minus BTE (solid), Eq.~\eqref{eq:nllk} (dashed), and projected scalar SCBA (circles). (c)~Normalized $\Delta R_{2\omega}/R_{2\omega,\mathrm{bg}}$ ($R_{2\omega}\propto\rho_\parallel^{(2)}$) in $\alpha\text{-Sn}(111)$ ($T=5\;\mathrm{K}$) along $\bar{\Gamma}\bar{M}$ ($+0.08$), $\bar{\Gamma}\bar{K}$, and $\bar{\Gamma}\bar{M}'$ ($-0.08$): data~\cite{Kalappattil2026} (circles) and joint fits via Eqs.~\eqref{eq:nllk}--\eqref{eq:V12} (dashed, $F=19.1\;\mathrm{T}$). (d)~Normalized $1/B_z$ envelope decay ($B_0=9.8\;\mathrm{T}$): full-quantum NQO ($\Lambda_B^{(2)}$, squares), SdH ($\Lambda_B^{(1)}$, circles), and experimental ratio guide ($0.58\,\Lambda_B^{(1)}$, dotted)~\cite{Kalappattil2026}.}
\end{figure}

In constant-current Hall bars ($\bm{J}=J\bm{e}_\phi$), the open-circuit Hall field tilts $\bm{E}^{(1)}$ by the Hall angle $\theta_H(B_z)=\arctan(h/r)$ ($r\equiv\rho_{xx}^{(1)}$, $h\equiv\rho_{xy}^{(1)}=-\rho_{yx}^{(1)}$), shifting $\rho_\parallel^{(2)}\equiv\bm{e}_\phi\cdot\bm{E}^{(2)}/J^2$ by $\theta_H$~\cite{SM}:
\begin{equation}\label{eq:rho2}
\begin{aligned}
\rho_\parallel^{(2)}(B_z,\phi) &= -(r^2+h^2)^{3/2}\big[\chi_o(B_z)\cos(3\phi - \theta_H) \\
&\quad + \chi_e(B_z)\sin(3\phi - \theta_H)\big],
\end{aligned}
\end{equation}
rotating the six angular nodes by $\Delta\phi_{\mathrm{node}}=\theta_H/3$ while preserving both $B_z$ parities.

\textit{Warped Dirac surface states and $\alpha\text{-Sn}(111)$ experiment.}---We first consider a $C_{3v}$ topological Dirac surface state ($3m'$ under $B_z$), realized in $\alpha\text{-Sn}(111)$ and $\mathrm{Bi}_2\mathrm{Se}_3(111)$~\cite{Fu2009}:
\begin{equation}\label{eq:HB}
\begin{aligned}
H_B &= \frac{\bm{\Pi}^2}{2m^*}\sigma_0 + v_F(\Pi_x\sigma_y - \Pi_y\sigma_x) \\
    &\quad + \left[\frac{\lambda_w}{2\hbar^3}(\Pi_+^3 + \Pi_-^3) + \Delta_Z\right]\sigma_z,
\end{aligned}
\end{equation}
with $\Pi_\pm=\Pi_x\pm\ii\Pi_y$, curvature mass $m^*$, Dirac velocity $v_F$, hexagonal warping $\lambda_w$, and Zeeman energy $\Delta_Z=g_z\mu_B B_z/2$. Setting $\lambda_w=0$ restores $\tilde{\mathcal{P}}=\sigma_z(-1)^{\hat{n}}$, whereas setting $v_F=\Delta_Z=0$ leaves $\sigma_x(-1)^{\hat{n}}$. Combining nonzero $v_F$ and $\lambda_w$ breaks every $\tilde{\mathcal{P}}$ even at $\Delta_Z=0$.

In the linear-Dirac limit ($1/m^*=0$, $n_F\gg 1$; finite $1/m^*$ in SM~\cite{SM}), Eq.~\eqref{eq:HB} gives
\begin{equation}\label{eq:V12}
V_1 = \frac{v_F}{2}, \qquad
V_2 = -\frac{\lambda_w}{\hbar\ell_B^2}\,(1-\zeta),
\end{equation}
where $\zeta=\Delta_Z/(\hbar\omega_c)=g_z m_c/(4m_e)$, with free-electron mass $m_e$ and cyclotron mass $m_c=\hbar k_F/v_F$. The factor $(1-\zeta)$ captures orbital--Zeeman competition. In moderate-$g_z$ surfaces such as $\mathrm{Bi}_2\mathrm{Se}_3(111)$, the orbital term dominates ($\zeta<1$), whereas crossing $\zeta=1$ ($g_z>4m_e/m_c$) flips the NQO phase by $180^\circ$.

Evaluating Eq.~\eqref{eq:sigma2_decomp} reproduces the smooth Boltzmann background at weak fields and sharp NQOs at strong fields [Fig.~\ref{fig:dirac_exp}(a)], where Eq.~\eqref{eq:nllk} closely matches both the full quantum trace and projected scalar SCBA calculations with frozen vertices [Fig.~\ref{fig:dirac_exp}(b)]~\cite{AndoUemura1974,SM}. Turning to $\alpha\text{-Sn}(111)$~\cite{Kalappattil2026}, we fix $(m_c,\tau_q)$ from linear SdH, adopt a representative compensated Hall-bar model with an empirical smooth background, and fit three angle-independent constants ($F=19.1\;\mathrm{T}$, $\gamma_B=0.083$, and an overall scale). Under standard $C_{3v}$ axes ($\bar{\Gamma}\bar{M}$ at $30^\circ$, $\bar{\Gamma}\bar{K}$ at $60^\circ$, $\bar{\Gamma}\bar{M}'$ at $90^\circ$), the $B_z$-even $\sin(3\phi-\theta_H)$ channel captures the antiphase $B_z>0$ oscillations and strongly suppressed $\bar{\Gamma}\bar{K}$ response [Fig.~\ref{fig:dirac_exp}(c)], extrapolated to $n_F\approx 1\text{--}4$ (where the fitted $2\omega$ frequency lies below $1\omega$ SdH $\approx 28\;\mathrm{T}$~\cite{Kalappattil2026}). Full quantum evaluation at $F=28.6\;\mathrm{T}$ [Fig.~\ref{fig:dirac_exp}(d)] yields a decay-rate ratio $\Lambda_B^{(2)}/\Lambda_B^{(1)}=0.62$ ($0.55\text{--}0.65$ across plausible bulk shunts and detrending protocols) versus the measured $0.58$ (with absolute slopes differing by an overall broadening scale~\cite{Kalappattil2026,SM}).

Since Ref.~\cite{Kalappattil2026} recorded $\Delta R_{2\omega}$ only for $B_z>0$ and at a lower $2\omega$ frequency, two measurements provide decisive tests: reversing $B_z\to-B_z$ distinguishes $B_z$-even dominance [Eq.~\eqref{eq:nllk}] from $B_z$-odd mechanisms beyond scalar disorder (e.g., spin-current conversion or spin-dependent scattering~\cite{Zhang2022SciAdv,Kalappattil2026}), while comparing $1\omega$ and $2\omega$ traces at a shared frequency directly tests the $\pi/2$ phase shift.

\begin{figure}[!tbp]
\centering
\includegraphics[width=0.90\columnwidth]{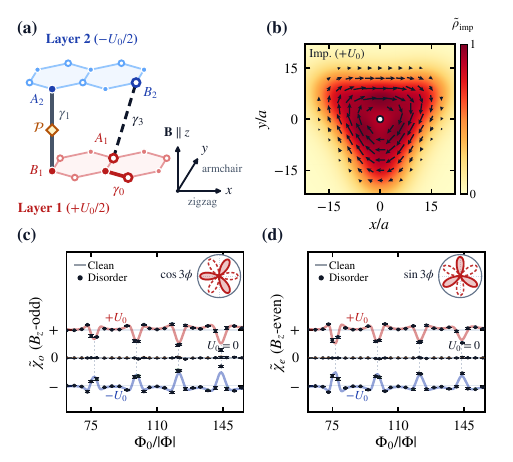}
\caption{\label{fig:blg}Gate-switchable orbital NQOs in Bernal BLG~\cite{SM}. (a)~AB-stacked lattice [Eq.~\eqref{eq:H_blg_tb}] under $\bm{B}=B_z\bm{e}_z$ ($x\parallel\text{zigzag}$, $y\parallel\text{armchair}$) showing $\pm U_0/2$, $(\gamma_0,\gamma_1,\gamma_3)$, and inversion center $\mathcal{P}$. (b)~Impurity-pinned LDOS $\tilde{\rho}_{\mathrm{imp}}$ and bond current $\bm{j}$ on a $61{,}268$-site disk ($U_0=+60\;\mathrm{meV}$, $\Phi/\Phi_0=1/180$, $E=99.19\;\mathrm{meV}$). (c),(d)~Rescaled $\tilde{\chi}_o$ ($\phi=0^\circ$) and $\tilde{\chi}_e$ ($\phi=30^\circ$) vs.\ $q=\Phi_0/|\Phi|$ at $\mu=0.24\;\mathrm{eV}$ for $+U_0$ (red), $0$ (gray), and $-U_0$ (blue, offset; $U_0=80\;\mathrm{meV}$): clean curves and 20-realization disorder averages (dots); insets: $3\phi$ profiles at $+U_0, q=100$ (scaled hoppings preserve exact inversion at high effective flux).}
\end{figure}

\textit{Gate-switchable orbital oscillations in bilayer gra\-phene.}---To demonstrate a purely orbital mechanism without spin-orbit coupling, we consider Bernal BLG~\cite{McCannFalko2006,McCann2006PRB,McCannKoshino2013RPP}, where a vertical bias $U_0$ electrically breaks inversion [Fig.~\ref{fig:blg}(a)]. Neglecting Zeeman splitting, its spin-independent Peierls tight-binding Hamiltonian with disorder $V_{\mathrm{imp}}(\bm{r}_i)$ reads~\cite{McCannKoshino2013RPP}
\begin{equation}\label{eq:H_blg_tb}
\begin{aligned}
H &= \sum_{i}\!\Big[\tfrac{1}{2}\xi_i U_0 + V_{\mathrm{imp}}(\bm{r}_i)\Big] c_i^\dagger c_i \\
  &\quad + \sum_{(i,j)}\!\big(t_{ij}\mathcal{U}_{ij} c_i^\dagger c_j + \mathrm{H.c.}\big),
\end{aligned}
\end{equation}
where $\xi_i=\pm 1$ labels layers~1 ($A_1,B_1$) and~2 ($A_2,B_2$), undirected intralayer ($-\gamma_0$), dimer ($+\gamma_1$), and skew ($-\gamma_3$) bonds enter once, and $\mathcal{U}_{ij}=\exp[-(\ii e/\hbar)\int_{\bm{r}_j}^{\bm{r}_i}\bm{\mathcal{A}}\cdot\dd\bm{r}]$ threads flux $\Phi$ per unit cell ($\Phi_0=2\pi\hbar/e$).

In the clean symmetric gauge $\bm{\mathcal{A}}=\frac{1}{2}\bm{B}\times\bm{r}$ about the interlayer inversion center $\mathcal{P}$, $\bm{r}_i\to-\bm{r}_i$ with layer-sublattice swap $\Sigma_x$ ($A_1\leftrightarrow B_2,B_1\leftrightarrow A_2$) leaves every $\mathcal{U}_{ij}$ invariant and maps $\mathcal{P}H(B_z,U_0)\mathcal{P}^{-1}=H(B_z,-U_0)$. At $U_0=0$ ($\bar{3}m'$), $\tilde{\mathcal{P}}_{\mathrm{BLG}}=\nu_x\Sigma_x(-1)^{\hat{n}}$ cancels opposite $\gamma_3$-warped triangular orbits ($\Delta n=\pm 3$)~\cite{McCannFalko2006,McCannKoshino2013RPP}. Applying $U_0\neq 0$ breaks $\mathcal{P}$ ($3m'$) and splits the $\bm{K}_\pm$ valleys via opposite orbital moments and dimer coupling~\cite{McCann2006PRB}. On an open disk, a smooth impurity then pins one valley-resolved triangular state with chiral bond currents [Fig.~\ref{fig:blg}(b)]~\cite{SM}. Because $\mathcal{P}$ reverses position and velocity at fixed $B_z$, linear and nonlinear conductivities obey opposite gate parities,
\begin{gather*}
\sigma_{\parallel\parallel}^{(1)}(B_z,-U_0)=+\sigma_{\parallel\parallel}^{(1)}(B_z,+U_0),\\
\chi_{o,e}(B_z,-U_0)=-\chi_{o,e}(B_z,+U_0).
\end{gather*}
Disorder-averaged finite-torus simulations in the projected $3\phi$ channels via the velocity-gauge form of Eq.~\eqref{eq:sigma2_decomp}~\cite{Parker2019,MichishitaPeters2021,Nakazawa2026} illustrate this electrical switch [Figs.~\ref{fig:blg}(c) and~\ref{fig:blg}(d)]. In these simulations, $\chi_{o,e}$ vanishes within disorder noise at $U_0=0$ and inverts with $U_0$, consistent with unprojected realistic-field continuum calculations ($v_F=10^6\;\mathrm{m/s}$, $B_z\approx 5\text{--}7\;\mathrm{T}$; Appendix~C, SM Fig.~S6)~\cite{SM}.

\textit{Conclusions.}---We established a microscopic quantum kinetic theory and generalized Landau-level parity rule ($\tilde{\mathcal{P}}=U(-1)^{\hat{n}}$) for perpendicular-field NQOs. Extending guiding-center scattering to second order reveals how $\Delta n=\pm 3$ triangular orbits drive orthogonal $B_z$-even ($\sin 3\phi$) and $B_z$-odd ($\cos 3\phi$) channels. The closed-form NLLK formula reproduces the threefold angular sign structure and captures the slower relative field damping in $\alpha\text{-Sn}(111)$ while pointing to decisive $\pm B_z$ and $\pi/2$-phase tests, and predicts gate-switchable orbital NQOs in Bernal BLG. Beyond $3m'$ (Table~\ref{tab:symmetry}), polar $m'$ crystals allow $1\phi,3\phi$ currents alongside transverse Berry-curvature-dipole responses, while chiral class $3$ leaves $3\phi$ phases free. Higher-order magnetotransport thus offers an electrical probe of orbital and magnetic multipoles in rhombohedral multilayers~\cite{KoshinoMcCann2009}, trigonal semiconductors~\cite{Ideue2017}, and unconventional magnets~\cite{Smejkal2022PRX1,FangCanoGhorashi2024,LiZhouWanChen2025}.

\begin{acknowledgments}
\textit{Acknowledgments.}---This work was supported by the National Natural Science Foundation of China under Grant No.~12304068 (H.G.), the Startup Fund of Nanjing University of Aeronautics and Astronautics under Grant No.~YAH24076 (H.G.), and the Fundamental Research Funds for the Central Universities under Grant No.~NS2026035 (H.G.). Large language models were used to assist with manuscript editing, code development, and algebraic cross-checking; the authors independently verified all derivations, numerical results, and text, and take full responsibility for the content of this work.
\end{acknowledgments}

\textit{Data availability.}---The simulation codes, figure scripts, and processed datasets supporting this Letter are available from the corresponding authors upon reasonable request and will be openly available in a public repository upon publication~\cite{SM}.

\bibliography{references_build_only}

\appendix
\onecolumngrid
\vspace{8pt}
\noindent\rule{\textwidth}{0.4pt}
\vspace{4pt}
\begin{center}
\textbf{\large End Matter}
\end{center}
\vspace{2pt}
\twocolumngrid
\setcounter{equation}{0}
\renewcommand{\theequation}{A\arabic{equation}}
\renewcommand{\theHequation}{A\arabic{equation}}

\textit{Appendix A: Linear Kubo--Greenwood limit, survival-time solution, and $M_x\Theta$ channel duality.}---Within the gauge-consistent local wide-band fermionic-bath model ($\Sigma^{r,a}=\mp\ii\Gamma$, $\Gamma=\hbar/(2\tau_q)$)~\cite{Han2013PRB,HanLi2013PRB}, transforming from the time-dependent velocity gauge to the static length gauge $H_{\mathrm{LG}}=H_B(\bm{\Pi})+\mathcal{F}\eta_\parallel$ shifts the bath distribution to $f_0(E-\mathcal{F}\eta_\parallel)$ and yields $\hat{\rho}_{\mathcal{F}}=(\Gamma/\pi)\int\dd E\,G_{\mathcal{F}}^r(E)f_0(E-\mathcal{F}\eta_\parallel)G_{\mathcal{F}}^a(E)$ inside every fixed-$R_\parallel$ sector. Using $H_{\mathrm{LG}}G_{\mathcal{F}}^r=(E+\ii\Gamma)G_{\mathcal{F}}^r-1$ and $G_{\mathcal{F}}^a H_{\mathrm{LG}}=(E-\ii\Gamma)G_{\mathcal{F}}^a-1$ directly gives the stationary kinetic equation~\eqref{eq:kinetic_eq}. The corresponding time-dependent relaxation equation $(\ii\hbar\partial_t+\ii\hbar/\tau_q)\hat{\rho}(t)-[H_{\mathrm{LG}},\hat{\rho}(t)]=(\ii\hbar/\tau_q)\hat{\rho}_{\mathrm{inj}}$ integrates in the interaction picture $U(t)=\ee^{-\ii H_{\mathrm{LG}}t/\hbar}$ to the survival-time average $\hat{\rho}_{\mathcal{F}}=\int_0^\infty(\dd t/\tau_q)\,\ee^{-t/\tau_q}\,U(t)\hat{\rho}_{\mathrm{inj}}U^\dagger(t)$.

Expanding the guiding-center master identity~\eqref{eq:master_exact} to linear order in $E_\parallel$ via $f_0(E-\mathcal{F}\eta_\parallel)=f_0(E)+[-f_0'(E)]\mathcal{F}\eta_\parallel+\cdots$ eliminates the equilibrium $f_0(E)$ part identically and yields
\begin{equation}\label{eq:app_sigma1}
\begin{aligned}
\sigma_{\parallel\parallel}^{(1)} &= \frac{e^2}{2\pi\ell_B^2\tau_q}\!\int\!\dd E\,[-f_0'(E)] \\
&\quad \times\!\left(\Tr[\eta_\parallel^2\hat{A}_0]-\frac{\Gamma}{\pi}\Tr[\eta_\parallel G_0^r\eta_\parallel G_0^a]\right).
\end{aligned}
\end{equation}

\begin{figure*}[!t]
\centering
\includegraphics[width=0.76\textwidth]{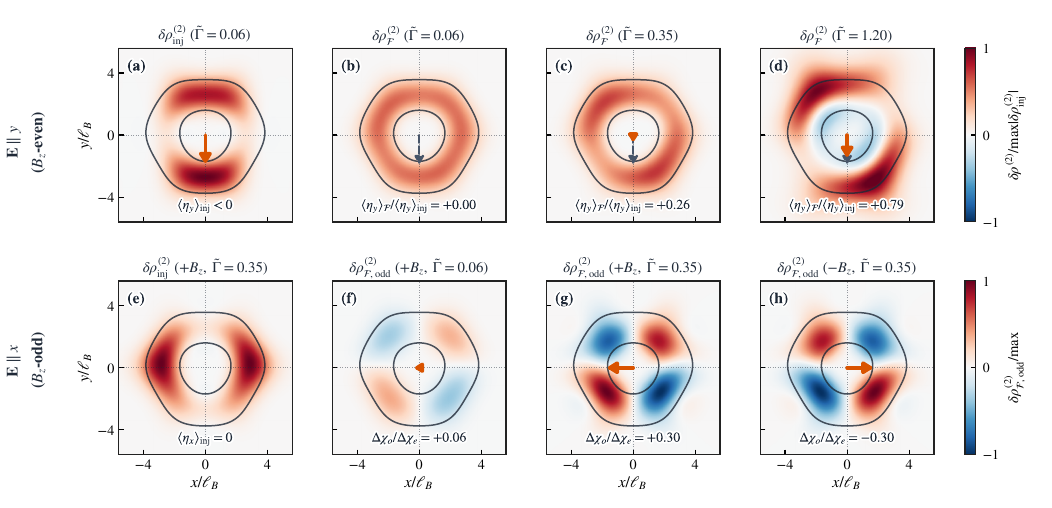}
\caption{\label{fig:endmatter_rho}Real-space illustration of the local-filling contribution in Eq.~\eqref{eq:sigma2_decomp} for the warped Dirac Landau state Eq.~\eqref{eq:HB} ($n=4$, $\lambda_w=+0.042$, $\Delta_Z=0$, $1/m^*=0$, $e=\hbar=v_F=|B_z|=1$; paired-doublet cutoff $N_{\mathrm{osc}}=22$) on the rising flank ($A_4'>0$, $\mu=E_4-\min[0.58\Gamma,0.25\hbar\omega_c]$; solid contour marks $0.35\max|\Psi_4|^2$). (a)--(d)~$\bm{E}\parallel y$ ($B_z$-even, $+B_z$): symmetrized local-filling representation $\delta\rho_{\mathrm{inj,sym}}^{(2)}\propto\partial_E(\frac{1}{2}\{\eta_y^2,\hat{A}_0\})$ at $\tilde{\Gamma}=0.06$ (a) vs.\ steady-state kernel $\delta\rho_{\mathcal{F}}^{(2)}\propto\partial_E[(\Gamma/\pi)G_0^r\eta_y^2 G_0^a]$ across $\tilde{\Gamma}=0.06,0.35,1.20$ (b)--(d); dashed/orange arrows compare $\langle\eta_y\rangle_{\mathrm{inj}}<0$ with $\langle\eta_y\rangle_{\mathcal{F}}$. Clean cyclotron rotation averages out $\langle\eta_y\rangle_{\mathcal{F}}$ ($\langle\eta_y\rangle_{\mathcal{F}}/\langle\eta_y\rangle_{\mathrm{inj}}=+0.00$), whereas strong broadening pulls $\delta\rho_{\mathcal{F}}^{(2)}$ back toward $\delta\rho_{\mathrm{inj,sym}}^{(2)}$ ($+0.79$). (e)--(h)~$\bm{E}\parallel x$ ($B_z$-odd): the symmetrized local-filling map is mirror-symmetric with vanishing longitudinal centroid ($\langle\eta_x\rangle_{\mathrm{inj}}=0$) (e), while the odd-$x$ steady-state shift $\delta\rho_{\mathcal{F},\mathrm{odd}}^{(2)}$ (minus $\lambda_w=0$ at fixed $E$) (f)--(h) is enhanced at intermediate damping ($\tilde{\Gamma}=0.35$) and inverts identically under $B_z\to-B_z$ ($\Delta\chi_o/\Delta\chi_e=+0.30\to-0.30$).}
\end{figure*}

Using $[G_0^r,\eta_\parallel]=-\ii\hbar G_0^r v_\parallel G_0^r$ and $\hat{A}_0=(\Gamma/\pi)G_0^r G_0^a$, Eq.~\eqref{eq:app_sigma1} equals the constant-broadening Kubo--Greenwood formula $\sigma_{\parallel\parallel}^{(1)}=[e^2\hbar/(2\ell_B^2)]\int\dd E\,[-f_0'(E)]\Tr[v_\parallel\hat{A}_0 v_\parallel\hat{A}_0]$~\cite{Kubo1957,Greenwood1958}. In the eigenbasis $H_B|\alpha\rangle=E_\alpha|\alpha\rangle$ with Lorentzian spectral density $A_\alpha(E)=(\Gamma/\pi)/[(E-E_\alpha)^2+\Gamma^2]$, the resonant contribution of each isolated sharp level ($\Gamma\ll\hbar|\omega_c|$) reduces to the Einstein guiding-center diffusion law $\sigma_{\parallel\parallel,\alpha}^{(1)}\simeq e^2\int\dd E\,[-f_0'(E)]D_\alpha(E)D_\alpha^*$~\cite{Kubo1959,AndoUemura1974}, with $D_\alpha(E)=A_\alpha(E)/(2\pi\ell_B^2)$ and $D_\alpha^*=\langle(\Delta R_\parallel)^2\rangle_\alpha/(2\tau_q)=(\langle\eta_\parallel^2\rangle_\alpha-\langle\eta_\parallel\rangle_\alpha^2)/\tau_q$.

For the $B_z$-odd channel ($\bm{E}\parallel x$, $\phi=0^\circ$) in Eq.~\eqref{eq:sigma2_decomp}, antiunitary magnetic mirror symmetry $\mathcal{M}_x=M_x\Theta$ satisfies $\mathcal{M}_x H_B\mathcal{M}_x^{-1}=H_B$, $\mathcal{M}_x G_0^r(E)\mathcal{M}_x^{-1}=G_0^a(E)$, and $\mathcal{M}_x\eta_x\mathcal{M}_x^{-1}=-\eta_x$. Consequently, $\sigma_{xx}^{(1)}[H_B+eE_x\eta_x]$ is even in $E_x$ (forcing the orbital-deformation derivative to vanish) and $\Tr[\eta_x^3\hat{A}_0(E)]=0$. In the surviving second trace, applying $\mathcal{M}_x$ and trace cyclicity reverses retarded and advanced Green's functions while flipping three factors of $\eta_x$:
\begin{equation}\label{eq:app_mx_flip}
\Tr\!\big[\eta_x G_0^r(E)\eta_x^2 G_0^a(E)\big] = -\Tr\!\big[\eta_x^2 G_0^r(E)\eta_x G_0^a(E)\big].
\end{equation}
Integrating $f_0''(E)$ once by parts and identifying $\delta\hat{\rho}_{\mathcal{F}}^{(1)}(E)=(\Gamma/\pi)G_0^r(E)\eta_x G_0^a(E)$---the linear-in-$eE_x$ non-equilibrium kernel of $\hat{\rho}_{\mathcal{F}}$---immediately yields the exact all-$\Gamma$, all-$B_z$ identity for $\sigma_{xxx}^{(2)}$ in Eq.~\eqref{eq:dual_channels}, demonstrating the exact duality between chiral-circulation shift ($\langle\eta_x\rangle_{\mathcal{F}}\neq 0$) and transverse Lorentz deflection.

\textit{Appendix B: Real-space local-filling illustration.}---To illustrate Figs.~\ref{fig:schematic}(b) and~\ref{fig:schematic}(c) from the local-filling term of Eq.~\eqref{eq:sigma2_decomp}, Fig.~\ref{fig:endmatter_rho} displays the real-space densities of the symmetrized local-filling kernel $\delta\hat{\rho}_{\mathrm{inj,sym}}^{(2)}(E)=\partial_E(\tfrac{1}{2}\{\eta_\parallel^2,\hat{A}_0(E)\})$ and steady-state kernel $\delta\hat{\rho}_{\mathcal{F}}^{(2)}(E)=\partial_E[(\Gamma/\pi)G_0^r(E)\eta_\parallel^2 G_0^a(E)]$ (in the $m=0$ guiding-center sector) on the rising flank ($A_4'>0$) of the $n=4$ warped Dirac Landau state [Eq.~\eqref{eq:HB}]. Note that $\delta\hat{\rho}_{\mathrm{inj,sym}}^{(2)}$ differs from the exact kinetic-equation source $\partial_E[\tfrac{\ii}{2\pi}(G_0^r\eta_\parallel^2-\eta_\parallel^2 G_0^a)]$ by the zero-dipole commutator $\partial_E[\tfrac{\ii}{4\pi}[G_0^r+G_0^a,\eta_\parallel^2]]$. Because $\Tr[\eta_\parallel[G_0^r+G_0^a,\eta_\parallel^2]]=0$, both yield identical longitudinal centroids $\langle\eta_\parallel\rangle_{\mathrm{inj}}$ and currents. Along $\bm{E}\parallel y$ (top row), opposite spins carry $\pm\lambda_w\sigma_z$ in Eq.~\eqref{eq:HB}, so the equilibrium contour $|\Psi_4(\bm{r})|^2$ combines a $C_{2z}$-even hexagonal $\cos 6\theta_{\bm{r}}$ modulation [$\mathcal{O}(\lambda_w^2)$, adding constructively across spin] with the $C_{2z}$-odd triangular $\sin 3\theta_{\bm{r}}$ distortion left by the $\nu=3,4$ spinor mismatch. The hexagonal harmonic contributes strictly zero to $\langle\eta_\parallel^3\rangle$ and $\sigma_\parallel^{(2)}$, whereas the sharp $-y$ triangular corner weights $\delta\rho_{\mathrm{inj,sym}}^{(2)}$ downward ($\langle\eta_y\rangle_{\mathrm{inj}}<0$, $\langle\eta_y^3\rangle_4=-0.44\,\ell_B^3$). Clean cyclotron motion ($\tilde{\Gamma}=0.06$) averages out $\langle\eta_y\rangle_{\mathcal{F}}/\langle\eta_y\rangle_{\mathrm{inj}}=+0.001$, whereas stronger broadening pulls it to $+0.26$ ($\tilde{\Gamma}=0.35$) and $+0.79$ ($\tilde{\Gamma}=1.20$). By contrast, single-valley $\gamma_3$ warping in four-band BLG [Eq.~\eqref{eq:app_blg4}] breaks $C_6$ directly at first order without opposite-spin cancellation, producing a purely triangular ring with the same relaxation behavior and $\pm B_z$ reversal (SM Fig.~S8)~\cite{SM}.

Along $\bm{E}\parallel x$ (bottom row), $\delta\rho_{\mathrm{inj,sym}}^{(2)}$ is mirror-symmetric with vanishing longitudinal centroid ($\langle\eta_x\rangle_{\mathrm{inj}}=0$ to $<10^{-13}$) and excess weight on the $+y$ shoulders. Subtracting the $\lambda_w=0$ background at the same $E$ isolates the odd-$x$ steady-state shift $\delta\rho_{\mathcal{F},\mathrm{odd}}^{(2)}$ [Figs.~\ref{fig:endmatter_rho}(f)--\ref{fig:endmatter_rho}(h)], which is small in the clean limit (local-filling ratio $\Delta\chi_o/\Delta\chi_e=+0.06$ at $\tilde{\Gamma}=0.06$), grows at intermediate damping ($+0.30$ at $\tilde{\Gamma}=0.35$, vs.\ NLLK ratio $\tilde{\Gamma}/(1+2\tilde{\Gamma}^2)=0.28$), and reverses identically under $B_z\to-B_z$ ($-0.30$, $\langle\eta_x\rangle_{\mathcal{F}}>0$), illustrating $\pm B_z$ branching in Fig.~\ref{fig:schematic}(c). Table~\ref{tab:gamma_regimes} summarizes how Terms~I, II-a, and II-b trade weight across $\tilde{\Gamma}$, with Term~I entering oscillations at $\mathcal{O}(1/\pi n_F)$ in phase with SdH (SM Sec.~S3)~\cite{SM}.

\setcounter{equation}{0}
\renewcommand{\theequation}{C\arabic{equation}}
\renewcommand{\theHequation}{C\arabic{equation}}

\textit{Appendix C: Effective continuum model and two-valley NLLK response of Bernal BLG.}---Expanding Eq.~\eqref{eq:H_blg_tb} ($V_{\mathrm{imp}}=0$) around $\bm{K}_{\tau_v}=(\tau_v\frac{4\pi}{3a},0)$ ($\tau_v=\pm 1$) in the basis $(A_1,B_2,A_2,B_1)^{\mathsf{T}}$ with $\pi_{\tau_v}\equiv\tau_v\Pi_x+\ii\Pi_y$ and $v_{F,3}=\sqrt{3}a\gamma_{0,3}/(2\hbar)$ yields the four-band continuum Hamiltonian~\cite{McCannFalko2006,McCannKoshino2013RPP}
\begin{table}[!t]
\caption{\label{tab:gamma_regimes}Regime breakdown of Eq.~\eqref{eq:sigma2_decomp} across $\tilde{\Gamma}=\Gamma/(\hbar|\omega_c|)$ into deformation (I), injection (II-a), and cyclotron (II-b) terms~\cite{SM}. II-b opposes II-a by $R_{\mathrm{wash}}=\tilde{\Gamma}^2(3+4\tilde{\Gamma}^2)/[(1+\tilde{\Gamma}^2)(1+4\tilde{\Gamma}^2)]$, $\chi_e^{(0)}$ is the $p=0$ part, and $\mathrm{I}=\mathrm{II\text{-}a}\equiv 0$ for $\bm{E}\parallel x$.}
\begin{ruledtabular}
\begin{tabular}{@{}llll@{}}
$\tilde{\Gamma}$ & $\bm{E}\parallel y$ (osc.) & $\bm{E}\parallel y$ ($p{=}0$) & $\bm{E}\parallel x$ (odd) \\
\hline
$\ll 1$ & II-a ($\approx\!100\%$) & I ($+2$), II-a ($-1$) & II-b ($100\%$) \\
(clean) & $R_{\mathrm{wash}}\!\simeq\!3\tilde{\Gamma}^2$ & of $\chi_e^{(0)}$ ($\propto\!\tilde{\Gamma}$) & $|\frac{\Delta\chi_o}{\Delta\chi_e}|\!\simeq\!\tilde{\Gamma}$ \\[2pt]
$0.3\text{--}1$ & II-a $+$ II-b & I $+$ II active & II-b ($100\%$) \\
(mid) & $R_{\mathrm{wash}}\!=\!0.20\text{--}0.70$ & II-a, II-b oppose & peak $0.35$ ($\frac{1}{\sqrt{2}}$) \\[2pt]
$\gg 1$ & $\ee^{-2\pi p\tilde{\Gamma}}\!\to\!0$ & I, II cancel $\tilde{\Gamma}^{-1}$, & II-b ($100\%$) \\
(BTE) & $R_{\mathrm{wash}}\!\to\!1\!-\!\frac{1}{2\tilde{\Gamma}^2}$ & yield $B_z^2\tau_q^3$ & BTE $\propto\!B_z\tau_q^2$ \\
\end{tabular}
\end{ruledtabular}
\end{table}
\begin{equation}\label{eq:app_blg4}
H_{\mathrm{BLG}}^{(\tau_v)} =
{\setlength{\arraycolsep}{3.2pt}\begin{pmatrix}
U_0/2 & v_3\pi_{\tau_v}^\dagger & 0 & v_F\pi_{\tau_v} \\
v_3\pi_{\tau_v} & -U_0/2 & v_F\pi_{\tau_v}^\dagger & 0 \\
0 & v_F\pi_{\tau_v} & -U_0/2 & \gamma_1 \\
v_F\pi_{\tau_v}^\dagger & 0 & \gamma_1 & U_0/2
\end{pmatrix}}.
\end{equation}
Integrating out the dimer sites $(A_2,B_1)$ at $|E|\ll\gamma_1$ reduces Eq.~\eqref{eq:app_blg4} to the two-band effective model on $(A_1,B_2)^{\mathsf{T}}$,
\begin{equation}\label{eq:app_blg2}
H_{\mathrm{eff}}^{(\tau_v)} =
\begin{pmatrix}
U_0/2 & v_3\pi_{\tau_v}^\dagger - \frac{v_F^2}{\gamma_1}\pi_{\tau_v}^2 \\
v_3\pi_{\tau_v} - \frac{v_F^2}{\gamma_1}(\pi_{\tau_v}^\dagger)^2 & -U_0/2
\end{pmatrix}.
\end{equation}
For $s_B=+1$ and $\tau_v=+1$, $\pi_+=\sqrt{2}\hbar a_B^\dagger/\ell_B$, so $(a_B^\dagger)^2$ pairs $|n\rangle$ on $A_1$ with $|n-2\rangle$ on $B_2$ ($\Delta\nu=2$). Restoring $v_3 a_B$ adds $\Delta\nu=\pm 1$, yielding $2+1=3$ and hence strict $\Delta n=\pm 3$ triangular hybridization without spin-orbit coupling.

Under inversion $\mathcal{P}=\nu_x\Sigma_x(-1)^{\hat{n}}$ [$\Sigma_x=I_2\otimes\sigma_x$, $\pi_{\tau_v}(-\bm{\Pi})=\pi_{-\tau_v}^\dagger(\bm{\Pi})$], Eq.~\eqref{eq:app_blg4} obeys $\Sigma_x(-1)^{\hat{n}} H_{\mathrm{BLG}}^{(\tau_v)}(U_0)[\Sigma_x(-1)^{\hat{n}}]^{-1}=H_{\mathrm{BLG}}^{(-\tau_v)}(-U_0)$, while $M_x$ maps $(K_+,+B_z)\leftrightarrow(K_-,-B_z)$. Whereas bare Eq.~\eqref{eq:app_blg2} retains $n\ge 2$ valley degeneracy at $v_3=0$, four-band dimer coupling in Eq.~\eqref{eq:app_blg4} produces both a $B_z$-odd valley phase splitting $\Delta\gamma_B\equiv\gamma_{B,+}-\gamma_{B,-}$ [$\Delta\gamma_B(-B_z)=-\Delta\gamma_B(B_z)$] and valley amplitude imbalance. In the leading common-spacing approximation ($g_\tau\equiv\Gamma_\tau/(\hbar|\omega_{c,\tau}|)\simeq\tilde{\Gamma}$) with $s_B$ factored from $\chi_o$, the common valley weights obey $W_{+p}(-B_z,U_0)=W_{-p}(+B_z,U_0)$ and $W_{\tau p}(B_z,-U_0)=-W_{-\tau,p}(B_z,+U_0)$, replacing $\mathcal{K}_p\sin(2\pi p n_F)$ in Eq.~\eqref{eq:nllk} by
\begin{equation}\label{eq:app_blg_nllk}
\begin{aligned}
&\sum_{\tau=\pm}\!W_{\tau p}\sin(2\pi p n_{F,\tau}) \\
&\quad = (W_{+p}+W_{-p})\cos(\pi p\,\Delta\gamma_B)\sin(2\pi p\,\bar{n}_F) \\
&\qquad -(W_{+p}-W_{-p})\sin(\pi p\,\Delta\gamma_B)\cos(2\pi p\,\bar{n}_F),
\end{aligned}
\end{equation}
with $\bar{n}_F=\frac{1}{2}(n_{F,+}+n_{F,-})$. Both terms are strictly $B_z$-even and $U_0$-odd (vanishing at $U_0=0$), mixing $\sin(2\pi p\bar{n}_F)$ (amplitude imbalance) and $\cos(2\pi p\bar{n}_F)$ (phase splitting, in-phase/antiphase with linear SdH) into a gate-tunable phase. Unprojected four-band calculations of Eq.~\eqref{eq:app_blg4} at realistic parameters ($v_F=10^6\;\mathrm{m/s}$, $\gamma_1=0.40\;\mathrm{eV}$, $v_3/v_F=0.12$, $\mu=75\;\mathrm{meV}$, $\Gamma=2.2\;\mathrm{meV}$, $U_0=\pm 60\;\mathrm{meV}$, $B_z\approx 5\text{--}7\;\mathrm{T}$; SM Fig.~S6) confirm this gate-switchable $B_z$-even/odd hierarchy~\cite{SM}.

\clearpage
\onecolumngrid
\counterwithout{equation}{section}
\setcounter{secnumdepth}{2}
\setcounter{tocdepth}{2}
\setcounter{section}{0}
\setcounter{subsection}{0}
\setcounter{equation}{0}
\setcounter{figure}{0}
\setcounter{table}{0}
\renewcommand{\thesection}{S\arabic{section}}
\renewcommand{\theHsection}{SM.\arabic{section}}
\renewcommand{\thesubsection}{\Alph{subsection}}
\renewcommand{\theHsubsection}{SM.\arabic{section}.\Alph{subsection}}
\renewcommand{\theequation}{S\arabic{equation}}
\renewcommand{\theHequation}{SM.\arabic{equation}}
\renewcommand{\thefigure}{S\arabic{figure}}
\renewcommand{\theHfigure}{SM.\arabic{figure}}
\renewcommand{\thetable}{S\arabic{table}}
\renewcommand{\theHtable}{SM.\arabic{table}}
\renewcommand{\arraystretch}{1.12}
\renewcommand{\topfraction}{0.95}
\renewcommand{\bottomfraction}{0.95}
\renewcommand{\textfraction}{0.05}
\renewcommand{\floatpagefraction}{0.80}
\setlength{\textfloatsep}{11pt plus 2pt minus 2pt}
\setlength{\floatsep}{10pt plus 2pt minus 2pt}
\setlength{\intextsep}{10pt plus 2pt minus 2pt}
\raggedbottom

\begin{center}
\textbf{\large Supplemental Material for\\[2pt]``Nonlinear Quantum Oscillations from Triangular Landau Orbits''}\\[6pt]
H.~Geng,$^{1,*}$ G.~Y.~Qi,$^{2}$ K.~X.~Jia,$^{3,\dagger}$ L.~Sheng,$^{4}$ and D.~Y.~Xing$^{4}$\\[3pt]
{\small\itshape
$^{1}$College of Physics, Key Laboratory of Aerospace Information Perception and Physics, MIIT, Nanjing University of Aeronautics and Astronautics, Nanjing 211106, China\\
$^{2}$College of Electrical Engineering, Zhejiang University of Water Resources and Electric Power, Hangzhou 310018, China\\
$^{3}$Interdisciplinary Center for Theoretical Physics and Information Sciences (ICTPIS) and State Key Laboratory of Surface Physics, Fudan University, Shanghai 200433, China\\
$^{4}$National Laboratory of Solid State Microstructures, School of Physics, and Collaborative Innovation Center of Advanced Microstructures, Nanjing University, Nanjing 210093, China
}
\end{center}
\vspace{6pt}

\setlength{\textfloatsep}{11pt plus 2pt minus 2pt}
\setlength{\floatsep}{10pt plus 2pt minus 2pt}
\setlength{\intextsep}{10pt plus 2pt minus 2pt}

This Supplemental Material mirrors the structure of the main text and provides self-contained derivations, model calculations, experimental analysis protocols, and numerical settings. Sections~\ref{sec:sm_negf}--\ref{sec:sm_nllk_hall} establish the general microscopic theory (non-equilibrium length/velocity-gauge transport, symmetry selection rules and triangular orbits, nonlinear Lifshitz--Kosevich resummation, disorder protection, semiclassical limit, and Hall-bar conversion); Secs.~\ref{sec:sm_ti_model} and~\ref{sec:sm_blg} apply it to warped topological surface states [$\alpha\text{-Sn}(111)$] and gated Bernal bilayer graphene; and Sec.~\ref{sec:sm_reproducibility} documents the numerical solvers. Tables~\ref{tab:sm_roadmap} and~\ref{tab:sm_notation} summarize the main-text mapping and global conventions.

\vspace{-6pt}
\tableofcontents
\addtocontents{toc}{\protect\fi}

\clearpage
\begin{table}[!htbp]
\caption{\label{tab:sm_roadmap}Mapping from main-text topics and figures to this Supplemental Material.}
\begin{ruledtabular}
\begin{tabular}{p{0.25\textwidth}p{0.45\textwidth}p{0.25\textwidth}}
Main-text topic & Physical content & SM location \\
\hline
Quantum kinetic theory & Cyclotron algebra, Kadanoff--Baym / gauge derivation of $\hat\rho_{\mathcal F}$, survival-time average, polarization--relaxation & Sec.~\ref{sec:sm_negf} (\ref{subsec:sm_kb_gauge}) \\
Linear \& nonlinear $\sigma_\parallel^{(1,2)}$ & 1st/2nd-order expansion (Term~I vs.\ Term~II), $D_{\alpha\beta\gamma}$ loop, velocity-vertex equivalence & Sec.~\ref{sec:sm_negf} (\ref{subsec:sm_expansion_loop}) \\
Table~I \& $\tilde{\mathcal{P}}$ & Generalized Landau parity $\tilde{\mathcal{P}}=U(-1)^{\hat{n}_{\bm{p}_0}}$, magnetic groups, BCD rules & Sec.~\ref{sec:sm_sym_orbits} (\ref{subsec:sm_generalized_parity}), Table~\ref{tab:sm_symmetry_expanded} \\
Triangular orbits \& Figs.~1, 4 & Triangular states $|\Psi_n|^2,\bm{j}_n^{\mathrm{chiral}}$, static octupole vs.\ chiral circulation / $\Tr[x^2\,\delta\hat\rho_{\mathcal F}^{(1)}]$ & Sec.~\ref{sec:sm_sym_orbits} (\ref{subsec:sm_tri_channels}) \\
NLLK formula \& SCBA & Six-path Poisson resummation, all-$\tilde{\Gamma}$ term breakdown, SCBA $3\phi$ protection, Boltzmann limit & Sec.~\ref{sec:sm_nllk_hall} (\ref{subsec:sm_six_path}--\ref{subsec:sm_boltzmann}) \\
Hall-bar $\rho_\parallel^{(2)}$ \& $\theta_H/3$ & Constant-current Hall-bar tensor inversion \& inverse-field decay rate $\Lambda_B^{(2)}$ & Sec.~\ref{sec:sm_nllk_hall} (\ref{subsec:sm_hall_decay}) \\
Warped Dirac states \& Fig.~2(a,b) & Linear-cone BTE limit, $V_{1,2}$, $p=0$ reconciliation, paired-doublet cutoff & Sec.~\ref{sec:sm_ti_model} (\ref{subsec:sm_ti_instantiation}--\ref{subsec:sm_p0_reconciliation}) \\
Fig.~2(c,d) & Experimental $1\omega/2\omega$ analysis protocol \& $\alpha\text{-Sn}(111)$ fits & Sec.~\ref{sec:sm_ti_model} (\ref{subsec:sm_alpha_sn_exp}), Table~\ref{tab:sm_ti_params} \\
Gated BLG, Fig.~3 \& App.~C & BLG Peierls/continuum models, $\Delta n=\pm 3$, gate parity, disk LDOS, torus & Sec.~\ref{sec:sm_blg} (\ref{subsec:sm_blg_ham}--\ref{subsec:sm_blg_torus}), Table~\ref{tab:sm_blg_params} \\
Data availability & Numerical solvers, discretization settings, and reproduction scripts & Sec.~\ref{sec:sm_reproducibility}, Table~\ref{tab:sm_solvers} \\
\end{tabular}
\end{ruledtabular}
\end{table}

\begin{table}[!htbp]
\caption{\label{tab:sm_notation}Global notation and sign conventions shared across the main text and Supplemental Material.}
\begin{ruledtabular}
\begin{tabular}{p{0.20\textwidth}p{0.47\textwidth}p{0.27\textwidth}}
Symbol & Definition / convention & Notes \\
\hline
$-e,\;\Phi_0,\;s_B,\;\ell_B$ & Electron charge ($e>0$), $\Phi_0=2\pi\hbar/e$, $s_B\equiv\operatorname{sgn}B_z$, $\ell_B\equiv\sqrt{\hbar/(e|B_z|)}$ & Uniform $\bm{B}=B_z\bm{e}_z$ \\
$\bm{\Pi},\;\bm{\eta},\;\bm{R}$ & $\bm{\Pi}=-\ii\hbar\bm{\nabla}+e\bm{\mathcal{A}}$, $\bm{\eta}=-s_B(\ell_B^2/\hbar)\bm{e}_z\times\bm{\Pi}$, $\bm{R}=\bm{r}-\bm{\eta}$ & $[\eta_a,\Pi_b]=\ii\hbar\delta_{ab}$, $[\bm{R},\bm{\Pi}]=0$ \\
$a_B,\;\nu,\;n,\;n_0,\;n_F$ & $a_B=\ell_B(\Pi_x-\ii s_B\Pi_y)/(\sqrt{2}\hbar)$; $\nu$: bare oscillator; $n$: LL multiplet & $n_0=k_F^2\ell_B^2/2$, $n_F=n_0-\gamma_B$ \\
$\phi,\;M_x,\;\mathcal{M}_x$ & $\bm{E}=E_\parallel(\cos\phi,\sin\phi)$; $M_x:(x,y)\mapsto(-x,y)$; $\mathcal{M}_x\equiv M_x\Theta$ & BLG: $x\parallel\text{zigzag}$, $y\parallel\text{armchair}$ \\
$\chi_o(B_z),\;\chi_e(B_z)$ & $\sigma_\parallel^{(2)}=\chi_o\cos 3\phi+\chi_e\sin 3\phi$ & $\chi_o$: $B_z$-odd; $\chi_e$: $B_z$-even \\
$\tau,\;\tau_q,\;\Gamma,\;\tilde{\Gamma},\;\beta$ & $\tau\equiv\tau_{\mathrm{tr}}$ vs.\ $\tau_q$; $\Gamma=\hbar/(2\tau_q)$; $\tilde{\Gamma}=\Gamma/(\hbar|\omega_c|)$; $\beta=\omega_c\tau$ & Matched: $\tau_{\mathrm{tr}}=\tau_q\Rightarrow |\beta|=1/(2\tilde{\Gamma})$ \\
$r,\;h,\;\theta_H$ & $r=\rho_{xx}^{(1)}>0$, $h=\rho_{xy}^{(1)}=-\rho_{yx}^{(1)}=E_x/J_y=-E_y/J_x$, $\tan\theta_H=h/r$ & If measuring $R_{yx}=E_y/J_x$, set $h=-R_{yx}$ \\
$\hat\rho_{\mathcal F},\;\hat\rho_{\mathrm{inj}},\;\delta\hat\rho_{\mathcal F}^{(1)}$ & Length-gauge steady state, Hermitian injection source, linear non-equilibrium kernel & $\delta\hat\rho_{\mathcal F}^{(1)}(E)=(\Gamma/\pi)G_0^r\eta_x G_0^a$ \\
$\kappa,\;\zeta$ & $\kappa=k_F\varepsilon_0''(k_F)/\varepsilon_0'(k_F)$; $\zeta=g_z m_c/[4m_e(1-\kappa)]$ & Linear Dirac: $\kappa=0$, $\zeta=\Delta_Z/(\hbar\omega_c)$ \\
\end{tabular}
\end{ruledtabular}
\end{table}

\section{Microscopic non-equilibrium magnetotransport in the length and velocity gauges}
\label{sec:sm_negf}

This section derives the exact steady-state transport formulas used throughout the main text. We begin in a translationally invariant magnetic field with a gauge-invariant local fermionic bath, derive the polarization--relaxation master equation [Eq.~\eqref{eq:sm_pol_relax_master}], and expand it to first and second order in the dc electric field to obtain three equivalent second-order forms, namely the two-term physical decomposition [Eqs.~\eqref{eq:sm_sigma2_I_II}--\eqref{eq:sm_termII}], the eigenbasis three-step dipole loop [Eqs.~\eqref{eq:sigma2_spectral_master}--\eqref{eq:wloop}], and the periodic-lattice velocity-vertex trace [Eq.~\eqref{eq:sm_sigma2_velocity_reduced}].

\subsection{Cyclotron algebra and Kadanoff--Baym derivation of the length-gauge steady state}
\label{subsec:sm_kb_gauge}

Consider a general translationally invariant two-dimensional magnetic Hamiltonian $H_B(\bm\Pi)$ under $\bm B=B_z\bm e_z=\bm\nabla\times\bm{\mathcal A}(\bm r)$, with electron charge $-e$ ($e>0$), field sign $s_B\equiv\operatorname{sgn}B_z$, and magnetic length $\ell_B\equiv\sqrt{\hbar/(e|B_z|)}$. The gauge-invariant kinetic momentum $\bm\Pi=-\ii\hbar\bm\nabla+e\bm{\mathcal A}(\bm r)$ obeys $[\Pi_x,\Pi_y]=-\ii s_B\hbar^2/\ell_B^2$. Defining the cyclotron ladder operators
\begin{equation*}
 a_B=\frac{\ell_B}{\sqrt2\hbar}(\Pi_x-\ii s_B\Pi_y),\qquad
 a_B^\dagger=\frac{\ell_B}{\sqrt2\hbar}(\Pi_x+\ii s_B\Pi_y),\qquad
 [a_B,a_B^\dagger]=1,
\end{equation*}
gives $\Pi_\pm\equiv\Pi_x\pm\ii\Pi_y=(\sqrt2\hbar/\ell_B)[\frac{1\mp s_B}{2}a_B+\frac{1\pm s_B}{2}a_B^\dagger]$.

We separate the position operator as $\bm r=\bm R+\bm\eta$ into the guiding-center coordinate $\bm R=\bm r-\bm\eta$ and the relative cyclotron coordinate $\bm\eta=-s_B(\ell_B^2/\hbar)\bm e_z\times\bm\Pi$~\cite{Kubo1959,Goerbig2011RMP}. In components, $\eta_x=s_B(\ell_B^2/\hbar)\Pi_y=(\ii\ell_B/\sqrt2)(a_B-a_B^\dagger)$ and $\eta_y=-s_B(\ell_B^2/\hbar)\Pi_x=-(s_B\ell_B/\sqrt2)(a_B+a_B^\dagger)$, which obey the decoupled magnetic algebra
\begin{equation}
 [\eta_a,\Pi_b]=\ii\hbar\delta_{ab},\qquad
 [\eta_x,\eta_y]=-\ii s_B\ell_B^2,\qquad
 [R_a,\Pi_b]=[R_a,\eta_b]=0,\qquad
 [R_x,R_y]=\ii s_B\ell_B^2.\label{eq:sm_eta_Pi_comm}
\end{equation}
Since $[R_a,\Pi_b]=0$, $H_B(\bm\Pi)$ factors the Hilbert space into $\mathcal H_{\mathrm{full}}=\mathcal H_{\mathrm{GC}}\otimes\mathcal H_{\mathrm{LL}}$, where the Landau-orbital space $\mathcal H_{\mathrm{LL}}=\mathcal H_{\mathrm{osc}}\otimes\mathbb C^N$ carries areal degeneracy $1/(2\pi\ell_B^2)$ per state.

For a uniform in-plane electric field $\bm E=E_\parallel\bm e_\phi$ along $\bm e_\phi=(\cos\phi,\sin\phi)$, the longitudinal coordinate splits into $r_\parallel(\phi)\equiv\bm e_\phi\cdot\bm r=R_\parallel(\phi)+\eta_\parallel(\phi)$. Writing $\eta_\pm\equiv\eta_x\pm\ii\eta_y=\mp(\ii s_B\ell_B^2/\hbar)\Pi_\pm$, we obtain
\begin{gather}
 \eta_\parallel(\phi)=\frac{1}{2}\big(\ee^{-\ii\phi}\eta_++\ee^{\ii\phi}\eta_-\big)=\frac{\ii\ell_B}{\sqrt2}\big(\ee^{+\ii s_B\phi}a_B-\ee^{-\ii s_B\phi}a_B^\dagger\big),\label{eq:eta_par_ladder}\\
 [R_\parallel(\phi),\eta_\parallel(\phi)]=[R_\parallel(\phi),H_B(\bm\Pi)]=0,\qquad
 [\eta_\parallel(\phi),H_B(\bm\Pi)]=\ii\hbar\,v_\parallel(\phi),\label{eq:v1_comm}
\end{gather}
where $v_\parallel(\phi)=\partial_{\Pi_\parallel}H_B(\bm\Pi)$ is the longitudinal velocity operator of the full magnetic Hamiltonian. In the exact eigenbasis $H_B|\alpha\rangle=E_\alpha|\alpha\rangle$ ($\Delta_{\alpha\beta}\equiv E_\alpha-E_\beta$), Eq.~\eqref{eq:v1_comm} relates velocity and dipole matrix elements via $v_{\parallel,\alpha\beta}(\phi)=(\ii/\hbar)\Delta_{\alpha\beta}\eta_{\parallel,\alpha\beta}(\phi)$, and repeated commutation yields $\partial_{\Pi_\parallel}^3 H_B=-\hbar^{-2}[\eta_\parallel,[\eta_\parallel,v_\parallel]]$.

To construct an exact non-equilibrium steady state with uniform scalar broadening $\Gamma=\hbar/(2\tau_q)$ while preserving energy dissipation and gauge invariance, we adopt the standard local wide-band fermionic-bath model~\cite{Han2013PRB,HanLi2013PRB}. Each spatial point couples to an independent equilibrium fermion reservoir with fixed scalar broadening, distinct from self-consistent zero-current B\"uttiker voltage probes; energy-dependent short-range impurity scattering is analyzed separately within the SCBA in Sec.~\ref{sec:sm_nllk_hall}. In the time-dependent \emph{velocity gauge} with $\bm A(t)=-\bm E t$, the system Hamiltonian is $H_{\mathrm{VG}}(t)=H_B(\bm\Pi-e\bm E t)$, while the local bath remains at zero scalar potential with strictly field-free, time-translation-invariant self-energies,
\begin{equation}
 \Sigma_{\mathrm{VG}}^{r,a}(t_1,t_2)=\mp\ii\Gamma\delta(t_1-t_2),\qquad
 \Sigma_{\mathrm{VG}}^{<}(t_1,t_2)=2\ii\Gamma\int_{-\infty}^{\infty}\frac{\dd E'}{2\pi\hbar}\,\ee^{-\frac{\ii}{\hbar}E'(t_1-t_2)}f_0(E'),\label{eq:sm_sigma_vg}
\end{equation}
where $f_0(E)=[1+\ee^{(E-\mu)/(k_BT)}]^{-1}$. The retarded and lesser Kadanoff--Baym equations read $[\ii\hbar\partial_{t_1}-H_{\mathrm{VG}}(t_1)+\ii\Gamma]G_{\mathrm{VG}}^r(t_1,t_2)=\hbar\delta(t_1-t_2)$ and $G_{\mathrm{VG}}^<=G_{\mathrm{VG}}^r\circ\Sigma_{\mathrm{VG}}^<\circ G_{\mathrm{VG}}^a$.

We pass to the static \emph{length gauge} through the exact unitary transformation $U(t)=\exp[-\frac{\ii}{\hbar}eE_\parallel r_\parallel t]$. For electron charge $-e$ and $A_\parallel(t)=-E_\parallel t$, this maps the Hamiltonian to $U H_{\mathrm{VG}}(t)U^\dagger+\ii\hbar(\partial_t U)U^\dagger=H_B(\bm\Pi)+eE_\parallel r_\parallel\equiv H_{\mathrm{full}}$. Since $U(t)$ is diagonal in real space, the transformed lesser self-energy $\Sigma_{\mathrm{LG}}^<(t_1,t_2)=U(t_1)\Sigma_{\mathrm{VG}}^<(t_1,t_2)U^\dagger(t_2)$ depends solely on $t_1-t_2$. Shifting the integration variable to $E\equiv E'+eE_\parallel r_\parallel$ gives
\begin{align}
 \Sigma_{\mathrm{LG}}^<(t_1,t_2)&=\ee^{-\frac{\ii}{\hbar}eE_\parallel r_\parallel(t_1-t_2)}\,2\ii\Gamma\int_{-\infty}^{\infty}\frac{\dd E'}{2\pi\hbar}\,\ee^{-\frac{\ii}{\hbar}E'(t_1-t_2)}f_0(E')\nonumber\\
 &=2\ii\Gamma\int_{-\infty}^{\infty}\frac{\dd E}{2\pi\hbar}\,\ee^{-\frac{\ii}{\hbar}E(t_1-t_2)}f_0(E-eE_\parallel r_\parallel),\label{eq:sm_sigma_lg_shift}
\end{align}
while $\Sigma_{\mathrm{LG}}^{r,a}(t_1,t_2)=\mp\ii\Gamma\delta(t_1-t_2)$ remain unchanged. The length-gauge steady state is therefore strictly stationary with energy-space self-energy $\Sigma_{\mathrm{LG}}^<(E)=2\ii\Gamma f_0(E-eE_\parallel r_\parallel)$. Physically, the bath distribution shifts together with the local scalar potential, maintaining spatial homogeneity and gauge consistency~\cite{HanLi2013PRB}; enforcing exact charge neutrality at finite $E_\parallel$ would separately require self-consistent electrostatics or an adjusted $\mu(E_\parallel)$~\cite{ChristenButtiker1996EPL}.

Next, we project the Keldysh equation $G_{\mathrm{LG}}^<(E)=G_{\mathrm{LG}}^r(E)\Sigma_{\mathrm{LG}}^<(E)G_{\mathrm{LG}}^a(E)$ onto guiding-center eigenstates $R_\parallel(\phi)|q\rangle_{\mathrm{GC}}=X_q|q\rangle_{\mathrm{GC}}$ [Eq.~\eqref{eq:v1_comm}]. Shifting $E\to E-eE_\parallel X_q$ removes $X_q$ identically from every guiding-center sector, so the full density matrix factors as $\langle q|\hat\rho_{\mathrm{full}}|q'\rangle=\delta_{qq'}\hat\rho_{\mathcal F}$. Writing $\mathcal F\equiv eE_\parallel$, $H_{\mathrm{LG}}\equiv H_B(\bm\Pi)+\mathcal F\eta_\parallel(\phi)$, and $G_{\mathcal F}^{r,a}(E)\equiv(E-H_{\mathrm{LG}}\pm\ii\Gamma)^{-1}$, the equal-time steady-state density matrix $\hat\rho_{\mathcal F}=\frac{1}{2\pi\ii}\int\dd E\,G_{\mathcal F}^<(E)$ on $\mathcal H_{\mathrm{LL}}$ follows directly without further approximation within this bath model~\cite{Han2013PRB,HanLi2013PRB},
\begin{equation}
 \hat\rho_{\mathcal F}=\frac{\Gamma}{\pi}\int_{-\infty}^{\infty}\dd E\,G_{\mathcal F}^r(E)\,f_0\big(E-\mathcal F\eta_\parallel(\phi)\big)\,G_{\mathcal F}^a(E).\label{eq:sm_lg_rho_dc}
\end{equation}
Transforming back to the velocity gauge gives $\hat\rho_{\mathrm{VG}}(t)=U^\dagger(t)(I_{\mathrm{GC}}\otimes\hat\rho_{\mathcal F})U(t)$. At $t=0$, we obtain the exact nonperturbative equalities $\hat\rho_{\mathrm{VG}}(0)=I_{\mathrm{GC}}\otimes\hat\rho_{\mathcal F}$ and $v_{\parallel,\mathrm{VG}}(t=0)=I_{\mathrm{GC}}\otimes v_\parallel(\bm\Pi)$, establishing exact gauge invariance of the dc current at every order in $E_\parallel$~\cite{Han2013PRB,HanLi2013PRB}.

To convert Eq.~\eqref{eq:sm_lg_rho_dc} into a kinetic balance equation, we apply the resolvent identities $H_{\mathrm{LG}}G_{\mathcal F}^r=(E+\ii\Gamma)G_{\mathcal F}^r-1$ and $G_{\mathcal F}^a H_{\mathrm{LG}}=(E-\ii\Gamma)G_{\mathcal F}^a-1$. Together they give $[H_{\mathrm{LG}},G_{\mathcal F}^r f_0 G_{\mathcal F}^a]-2\ii\Gamma G_{\mathcal F}^r f_0 G_{\mathcal F}^a=G_{\mathcal F}^r f_0-f_0 G_{\mathcal F}^a$. Multiplying by $\Gamma/\pi$ and integrating over $E$ shows that $\hat\rho_{\mathcal F}$ is the unique solution of the damped stationary quantum Liouville equation on $\mathcal H_{\mathrm{LL}}$,
\begin{gather}
 \big[H_{\mathrm{LG}},\,\hat\rho_{\mathcal F}\big]-2\ii\Gamma\big(\hat\rho_{\mathcal F}-\hat\rho_{\mathrm{inj}}\big)=0,\label{eq:sm_stationary_liouville}\\
 \hat\rho_{\mathrm{inj}}\equiv\frac{\ii}{2\pi}\int_{-\infty}^{\infty}\dd E\,\Big[G_{\mathcal F}^r(E)\,f_0\big(E-\mathcal F\eta_\parallel\big)-f_0\big(E-\mathcal F\eta_\parallel\big)\,G_{\mathcal F}^a(E)\Big].\label{eq:sm_rho_inj_def}
\end{gather}
Rewriting Eq.~\eqref{eq:sm_stationary_liouville} in the interaction picture $U(t)=\ee^{-\ii H_{\mathrm{LG}}t/\hbar}$ yields the exact survival-time integral
\begin{equation*}
 \hat\rho_{\mathcal F}=\int_0^{\infty}\frac{\dd t}{\tau_q}\,\ee^{-t/\tau_q}\,U(t)\hat\rho_{\mathrm{inj}}U^\dagger(t),\qquad \tau_q=\frac{\hbar}{2\Gamma}.
\end{equation*}
Hence $\hat\rho_{\mathcal F}$ is the survival-time average of the local injection source $\hat\rho_{\mathrm{inj}}$ under collision-free cyclotron evolution by $H_{\mathrm{LG}}$ over lifetime $\tau_q$.

Now trace Eq.~\eqref{eq:sm_stationary_liouville} against $\eta_\parallel(\phi)$. Using trace cyclicity together with Eq.~\eqref{eq:v1_comm} gives $\Tr(\eta_\parallel[H_{\mathrm{LG}},\hat\rho_{\mathcal F}])=\Tr([\eta_\parallel,H_B]\hat\rho_{\mathcal F})=\ii\hbar\Tr(v_\parallel\hat\rho_{\mathcal F})$. Combining this identity with $2\Gamma/\hbar=1/\tau_q$ directly produces the exact polarization--relaxation master identity,
\begin{equation}
 j_\parallel=-\frac{e}{2\pi\ell_B^2}\Tr\big[v_\parallel(\phi)\hat\rho_{\mathcal F}\big]
 =\frac{e}{2\pi\ell_B^2\tau_q}\Tr\!\big[\eta_\parallel(\phi)\big(\hat\rho_{\mathrm{inj}}-\hat\rho_{\mathcal F}\big)\big].\label{eq:sm_pol_relax_master}
\end{equation}
For any function $g(\eta_\parallel)$ that commutes with $f_0(E-\mathcal F\eta_\parallel)$, trace cyclicity simplifies the injection expectation value to $\Tr[g(\eta_\parallel)\hat\rho_{\mathrm{inj}}]=\int\dd E\,\Tr[g(\eta_\parallel)\hat A_{\mathcal F}(E)f_0(E-\mathcal F\eta_\parallel)]$, where $\hat A_{\mathcal F}(E)=\frac{\ii}{2\pi}(G_{\mathcal F}^r-G_{\mathcal F}^a)=(\Gamma/\pi)G_{\mathcal F}^r G_{\mathcal F}^a$ is the field-dressed spectral density.

\textit{Technical notes.}---Three formal properties of Eqs.~\eqref{eq:sm_stationary_liouville}--\eqref{eq:sm_pol_relax_master} merit note. First, $\hat\rho_{\mathrm{inj}}$ is manifestly Hermitian with an energy-convergent integral in any fixed regularized subspace, as the $1/E$ tails of $G_{\mathcal F}^r$ and $G_{\mathcal F}^a$ cancel before integration; at $\mathcal F\neq 0$, however, $G_{\mathcal F}^{r,a}$ and $f_0(E-\mathcal F\eta_\parallel)$ do not commute, so $\hat\rho_{\mathrm{inj}}$ is generally not positive semidefinite and enters strictly as a Hermitian source operator rather than a physical density matrix. Second, all intermediate traces on the unbounded Landau ladder are defined on a finite regularized subspace using the projected coordinate $\eta_\parallel$ and consistent commutator velocity $v_\parallel=(\ii/\hbar)[H_B,\eta_\parallel]$ before taking the cutoff to infinity in the convergent Fermi-surface response (Subsec.~\ref{subsec:sm_ti_landau} details paired-doublet Dirac truncation). Third, dropping $\mathcal F\eta_\parallel$ inside $f_0$ reduces Eq.~\eqref{eq:sm_lg_rho_dc} to $\int\dd E\,f_0(E)\hat A_{\mathcal F}(E)$, for which $\Tr[v_\parallel\hat A_{\mathcal F}(E)]=(\ii/\hbar)\Tr(\eta_\parallel[\hat A_{\mathcal F},H_{\mathrm{LG}}])\equiv 0$ eliminates longitudinal current at \emph{all} orders in $E_\parallel$; the local shift $f_0(E-\mathcal F\eta_\parallel)$ is thus essential for bulk dissipative magnetotransport. By the same commutator identity $\Tr[v_\parallel g(H_B)]\equiv 0$, an isotropic Joule-heating shift $\delta f_{\mathrm{in}}^{(2)}(E)\propto\tau_{\mathrm{in}}E_\parallel^2$ produces identically zero uniform second-order current in a $C_3$-symmetric crystal and enters first at $\mathcal O(E_\parallel^3)$ (which does not exclude inelastic contributions to anisotropic collision rates $\tau_a^{-1}=\tau_{\mathrm{el}}^{-1}+\tau_{\mathrm{in},a}^{-1}$ in higher-order vertex corrections).

\subsection{Order-by-order expansion, two-term physical decomposition, three-step dipole loop, and velocity-vertex equivalence}
\label{subsec:sm_expansion_loop}

We now expand Eq.~\eqref{eq:sm_pol_relax_master} order by order in $\mathcal F=eE_\parallel$. Writing $G_0^{r,a}(E)=(E-H_B\pm\ii\Gamma)^{-1}$ and $\hat A_0(E)=(\Gamma/\pi)G_0^r(E)G_0^a(E)$, we expand the local Fermi distribution as
\begin{equation*}
 f_0(E-\mathcal F\eta_\parallel)=f_0(E)-\mathcal F f_0'(E)\eta_\parallel+\frac{1}{2}\mathcal F^2 f_0''(E)\eta_\parallel^2+\cdots.
\end{equation*}
Since $\hat A_{\mathcal F}(E)=(\Gamma/\pi)G_{\mathcal F}^r(E)G_{\mathcal F}^a(E)$ holds identically, the zeroth-order equilibrium pieces multiplying $f_0(E)$ cancel between $\Tr[\eta_\parallel\hat\rho_{\mathrm{inj}}]$ and $\Tr[\eta_\parallel\hat\rho_{\mathcal F}]$ at every order in $\mathcal F$.

\textit{1. First order (Kubo--Greenwood linear SdH).}---At $\mathcal O(\mathcal F)$, only the $-\mathcal F f_0'(E)\eta_\parallel$ term acting on the zero-field operators survives, yielding the linear longitudinal conductivity
\begin{equation}
 \sigma_{\parallel\parallel}^{(1)}[H_B]=\frac{e^2}{2\pi\ell_B^2\tau_q}\int_{-\infty}^{\infty}\dd E\,\big[-f_0'(E)\big]\left(\Tr\big[\eta_\parallel^2\hat A_0(E)\big]-\frac{\Gamma}{\pi}\Tr\big[\eta_\parallel G_0^r(E) \eta_\parallel G_0^a(E)\big]\right).\label{eq:sm_sigma1_kubo_ando}
\end{equation}
Using the commutator $[G_0^r,\eta_\parallel]=-\ii\hbar G_0^r v_\parallel G_0^r$ together with $\hat A_0=(\Gamma/\pi)G_0^r G_0^a$, Eq.~\eqref{eq:sm_sigma1_kubo_ando} transforms identically into the constant-broadening Kubo--Greenwood formula~\cite{Kubo1957,Greenwood1958} (with $\tau_{\mathrm{tr}}=\tau_q$ in this scalar-bath model),
\begin{equation*}
 \sigma_{\parallel\parallel}^{(1)}=\frac{e^2\hbar}{2\ell_B^2}\int_{-\infty}^{\infty}\dd E\,\big[-f_0'(E)\big]\Tr\big[v_\parallel\hat A_0(E)v_\parallel\hat A_0(E)\big],
\end{equation*}
where the bracketed trace difference in Eq.~\eqref{eq:sm_sigma1_kubo_ando} expands in the eigenbasis as the manifestly non-negative sum $\frac{\Gamma}{2\pi}\sum_{\alpha,\beta}|\eta_{\parallel,\alpha\beta}|^2(E_\alpha-E_\beta)^2/\{[(E-E_\alpha)^2+\Gamma^2][(E-E_\beta)^2+\Gamma^2]\}\ge 0$.

Near an isolated sharp Landau level $|\alpha\rangle$ ($\Gamma\ll\Delta E_{\min}\sim\hbar|\omega_c|$), the resonant contribution simplifies to $A_\alpha(E)(\langle\eta_\parallel^2\rangle_\alpha-\langle\eta_\parallel\rangle_\alpha^2)$ up to $\mathcal O(\tilde\Gamma^2)$ corrections (when $|\alpha\rangle$ is degenerate, the full intra-subspace projection $\sum_{E_\beta=E_\alpha}|\eta_{\parallel,\alpha\beta}|^2$ is subtracted). For parabolic Landau levels, Poisson resummation multiplies both the smooth Drude background and every SdH harmonic $p\ge 1$ by the exact factor $1/(1+4\tilde\Gamma^2)=(\omega_c\tau_q)^2/[1+(\omega_c\tau_q)^2]$. Since local scattering at fixed $r_\parallel=R_\parallel+\eta_\parallel$ shifts the guiding center by $\Delta R_\parallel=\eta_i-\eta_f$ with $\langle(\Delta R_\parallel)^2\rangle_\alpha=2\operatorname{Var}_\alpha(\eta_\parallel)$, Eq.~\eqref{eq:sm_sigma1_kubo_ando} directly recovers the Kubo--Ando Einstein diffusion form~\cite{Kubo1959,AndoUemura1974} $\sigma_{\parallel\parallel,\alpha}^{(1)}\simeq e^2\int\dd E\,[-f_0'(E)]D_\alpha(E)D_\alpha^*$, where $D_\alpha(E)=A_\alpha(E)/(2\pi\ell_B^2)$ and $D_\alpha^*=\langle(\Delta R_\parallel)^2\rangle_\alpha/(2\tau_q)=(\langle\eta_\parallel^2\rangle_\alpha-\langle\eta_\parallel\rangle_\alpha^2)/\tau_q$. On an unbounded Landau ladder, the two separate traces inside Eq.~\eqref{eq:sm_sigma1_kubo_ando} grow individually at large $n$ (logarithmically for parabolic ladders and linearly for linear Dirac cones), whereas their difference cancels the leading high-$n$ terms identically; both traces must therefore be combined within a shared cutoff before taking $N_{\mathrm{cut}}\to\infty$.

\textit{2. Second order (exact two-term physical decomposition).}---At $\mathcal O(\mathcal F^2)$, exactly two pieces survive in Eq.~\eqref{eq:sm_pol_relax_master}. The first comes from expanding the field-dressed operators $G_{\mathcal F}^{r,a}$ and $\hat A_{\mathcal F}$ to first order in $\mathcal F$ while keeping $-\mathcal F f_0'(E)\eta_\parallel$. The second comes from keeping the zero-field operators $G_0^r,G_0^a,\hat A_0$ while taking the quadratic distribution shift $+\frac{1}{2}\mathcal F^2 f_0''(E)\eta_\parallel^2$, followed by a single integration by parts via $\int\dd E\,f_0''(E)h(E)=\int\dd E\,[-f_0'(E)]h'(E)$. Together they give the exact two-term decomposition
\begin{align}
 \sigma_\parallel^{(2)}&=\sigma_{\parallel,\mathrm{I}}^{(2)}+\sigma_{\parallel,\mathrm{II}}^{(2)},\label{eq:sm_sigma2_I_II}\\
 \sigma_{\parallel,\mathrm{I}}^{(2)}&\equiv\left.\frac{\partial\sigma_{\parallel\parallel}^{(1)}[H_B+eE_\parallel\eta_\parallel]}{\partial E_\parallel}\right|_{E_\parallel=0},\label{eq:sm_termI}\\
 \sigma_{\parallel,\mathrm{II}}^{(2)}&\equiv\frac{e^3}{4\pi\ell_B^2\tau_q}\int_{-\infty}^{\infty}\dd E\,\big[-f_0'(E)\big]\,\frac{\dd}{\dd E}\!\left(\Tr\big[\eta_\parallel^3\hat A_0(E)\big]-\frac{\Gamma}{\pi}\Tr\big[\eta_\parallel G_0^r(E) \eta_\parallel^2 G_0^a(E)\big]\right).\label{eq:sm_termII}
\end{align}

Equations~\eqref{eq:sm_sigma2_I_II}--\eqref{eq:sm_termII} separate the two physical mechanisms cleanly. Term~I ($\sigma_{\parallel,\mathrm{I}}^{(2)}$) measures how static first-order orbital deformation inside $H_B+eE_\parallel\eta_\parallel$ modifies the linear conductivity $\sigma_{\parallel\parallel}^{(1)}$. Whenever $C_{3z}$ symmetry enforces zero projected dipole within every energy level ($P_{E_\alpha}\eta_\parallel P_{E_\alpha}=0$, reducing to $\langle\alpha|\eta_\parallel|\alpha\rangle=0$ for nondegenerate states), there is no first-order Stark shift of the Landau energies $E_\alpha$. The derivative $\partial_{E_\parallel}$ therefore acts solely on interlevel wave-function mixing (suppressed by the level spacing $\sim\hbar|\omega_c|$) rather than on the sharp Lorentzian peaks $A_\alpha(E)=(\Gamma/\pi)/[(E-E_\alpha)^2+\Gamma^2]$. In the semiclassical regime ($n_F\gg 1$, $\tilde\Gamma=\Gamma/(\hbar|\omega_c|)\ll 1$) of our warped Dirac and BLG models, Term~I accordingly produces a smooth non-oscillatory background plus a weak in-phase ($\cos 2\pi n_F$) oscillatory correction of relative order $\mathcal O(1/\pi n_F)$ against Term~II.

By contrast, Term~II ($\sigma_{\parallel,\mathrm{II}}^{(2)}$) acts on the undeformed zero-electric-field spectrum and applies $\dd/\dd E$ directly to $A_\alpha(E)$. Differentiating each even Lorentzian peak converts it into an odd derivative waveform ($\cos 2\pi n_F\to\sin 2\pi n_F$), enhances the local peak slope by $\hbar|\omega_c|/\Gamma\gg 1$ over Term~I, and scales as $\pi p n_F\gg 1$ for a fixed $p$th harmonic. Subsection~\ref{subsec:sm_gamma_regimes} derives the exact all-$\tilde\Gamma$ breakdown of Term~I, Term~II-a, and Term~II-b across all broadening regimes.

\textit{3. Consolidation to the Fermi-surface $\hat Q$ trace and eigenbasis three-step dipole loop.}---An equivalent, compact form follows directly from $j_\parallel^{(2)}=-[e/(2\pi\ell_B^2)]\Tr[v_\parallel\hat\rho^{(2)}]$. Expanding Eq.~\eqref{eq:sm_lg_rho_dc} to $\mathcal O(\mathcal F^2)$, dropping the vanishing equilibrium trace $\Tr[v_\parallel\hat A_{\mathcal F}(E)]\equiv 0$, and integrating the $\frac{1}{2}f_0''(E)\Tr(v_\parallel G_0^r \eta_\parallel^2 G_0^a)$ term once by parts via $\partial_E G_0^r=-(G_0^r)^2$ consolidates both Term~I and Term~II into a single Fermi-surface trace,
\begin{align}
 \sigma_\parallel^{(2)}&=-\frac{e^3\Gamma}{2\pi^2\ell_B^2}\int_{-\infty}^{\infty}\dd E\,\big[-f_0'(E)\big]\,2\Re\Tr\big[v_\parallel(\phi)\hat Q(E)\big],\label{eq:sm_sigma2_Q}\\
 \hat Q(E)&=G_0^r(E)\eta_\parallel G_0^r(E)\eta_\parallel G_0^a(E)-\frac{1}{2}\big[G_0^r(E)\big]^2\eta_\parallel^2 G_0^a(E).\label{eq:sm_Q_def}
\end{align}

To express Eq.~\eqref{eq:sm_sigma2_Q} in the warped eigenbasis $\{|\alpha\rangle\}$, we substitute $v_{\parallel,\alpha\beta}=(\ii/\hbar)\Delta_{\alpha\beta}\eta_{\parallel,\alpha\beta}$, introduce $\varepsilon_\lambda(E)\equiv E-E_\lambda+\ii\Gamma=[G_\lambda^r(E)]^{-1}$, and rewrite $(\Gamma/\pi)G_\alpha^a=A_\alpha\varepsilon_\alpha$. Since $\Delta_{\alpha\beta}=\varepsilon_\beta-\varepsilon_\alpha$ and $\varepsilon_\beta G_\beta^r=1$, the energy factors simplify immediately via
\begin{equation*}
 \Delta_{\alpha\beta}G_\beta^r G_\gamma^r = G_\gamma^r - \varepsilon_\alpha G_\beta^r G_\gamma^r,\qquad
 -\frac{1}{2}\Delta_{\alpha\beta}\big[G_\beta^r\big]^2 = -\frac{1}{2}G_\beta^r + \frac{1}{2}\varepsilon_\alpha\big[G_\beta^r\big]^2.
\end{equation*}
Upon taking $-2\Re(\ii Z)=2\Im Z$, the term $\sum_{\alpha,\beta,\gamma}A_\alpha\varepsilon_\alpha G_\beta^r D_{\alpha\beta\gamma}=(\Gamma/\pi)\Tr[G_0^a\eta_\parallel G_0^r\eta_\parallel^2]$ is strictly real and drops out identically. Defining the closed three-step electric-dipole loop $D_{\alpha\beta\gamma}(\phi)\equiv\eta_{\parallel,\alpha\beta}\eta_{\parallel,\beta\gamma}\eta_{\parallel,\gamma\alpha}$ yields
\begin{align}
 \sigma_\parallel^{(2)}&=\frac{e^3\hbar^2}{2\pi\ell_B^2}\int_{-\infty}^\infty\dd E\,\big[-f_0'(E)\big]\sum_\alpha A_\alpha(E)\,\mathcal L_\alpha^{(2)}(E,\phi),\label{eq:sigma2_spectral_master}\\
 \mathcal L_\alpha^{(2)}(E,\phi)&=\frac{1}{\hbar^3}\Im\sum_{\beta,\gamma}w_{\alpha\beta\gamma}(E)D_{\alpha\beta\gamma}(\phi),\label{eq:dipole-loop}\\
 w_{\alpha\beta\gamma}(E)&=2\varepsilon_\alpha G_\gamma^r(E)-2\varepsilon_\alpha^2 G_\beta^r(E)G_\gamma^r(E)+\varepsilon_\alpha^2\big[G_\beta^r(E)\big]^2.\label{eq:wloop}
\end{align}
Symmetrizing under loop reversal $\beta\leftrightarrow\gamma$ ($D_{\alpha\gamma\beta}=D_{\alpha\beta\gamma}^*$) via $w_{\alpha\beta\gamma}^{s,a}\equiv\frac{1}{2}(w_{\alpha\beta\gamma}\pm w_{\alpha\gamma\beta})$ and $G_\gamma^r-G_\beta^r=-\Delta_{\beta\gamma}G_\beta^r G_\gamma^r$ splits $\mathcal L_\alpha^{(2)}$ into circulation-symmetric and circulation-antisymmetric parts,
\begin{gather}
 \mathcal L_\alpha^{(2)}(E,\phi)=\frac{1}{\hbar^3}\sum_{\beta,\gamma}\Big[\Re D_{\alpha\beta\gamma}(\phi)\,\Im w_{\alpha\beta\gamma}^s(E)+\Im D_{\alpha\beta\gamma}(\phi)\,\Re w_{\alpha\beta\gamma}^a(E)\Big],\label{eq:sm_L2_parity_split}\\
 w_{\alpha\beta\gamma}^a(E)=\Delta_{\beta\gamma}G_\beta^r(E)G_\gamma^r(E)\left[-\varepsilon_\alpha+\frac{1}{2}\varepsilon_\alpha^2\big(G_\beta^r(E)+G_\gamma^r(E)\big)\right].\label{eq:sm_wa_factor}
\end{gather}

\textit{4. Equivalent velocity-vertex representation for periodic-torus lattice transport.}---On a disordered periodic system of area $A_{\mathrm{tot}}$, the position operator $r_\parallel=\bm r\cdot\bm e_\phi$ is ill-defined across periodic boundaries, so it is convenient to convert $\hat Q(E)$ into velocity vertices $v_\parallel=(\ii/\hbar)[H,r_\parallel]$ and $v_{\parallel\parallel}=(\ii/\hbar)[v_\parallel,r_\parallel]$. Differentiating $G^r(E-H+\ii\Gamma)=1$ twice yields the exact commutator identities
\begin{equation*}
 [G^r,r_\parallel]=-\ii\hbar G^r v_\parallel G^r,\qquad
 \big[[G^r,r_\parallel],r_\parallel\big]=-\hbar^2 G^r\big(2v_\parallel G^r v_\parallel+v_{\parallel\parallel}\big)G^r.
\end{equation*}
Using the algebraic identity $G^r r_\parallel G^r r_\parallel-\frac{1}{2}(G^r)^2 r_\parallel^2=-\frac{1}{2}G^r[[G^r,r_\parallel],r_\parallel]+\frac{1}{2}G^r r_\parallel^2 G^r$ inside $(2\Gamma/\pi)\hat Q(E)$ gives $(2\Gamma/\pi)\hat Q=\hbar^2(G^r)^2(2v_\parallel G^r v_\parallel+v_{\parallel\parallel})\hat A+G^r r_\parallel^2\hat A$. Multiplying by $v_\parallel$ and taking $\Re\Tr$, the second term vanishes identically since $\Tr[v_\parallel G^r r_\parallel^2\hat A]=(\ii/\hbar)\Tr[\hat A r_\parallel^3]-(\ii\Gamma/\pi\hbar)\Tr[G^a r_\parallel G^r r_\parallel^2]\in\ii\mathbb R$. Equation~\eqref{eq:sm_sigma2_Q} therefore converts identically into the velocity-gauge form used on periodic tight-binding tori~\cite{Parker2019,MichishitaPeters2021,Nakazawa2026},
\begin{equation}
 \sigma_\parallel^{(2)}=-\frac{e^3\hbar^2}{A_{\mathrm{tot}}}\int_{-\infty}^{\infty}\dd E\,\big[-f_0'(E)\big]\,\Re\Tr\!\Big\{\hat{A}(E)\,v_\parallel\big[G^r(E)\big]^2\big[2v_\parallel G^r(E)v_\parallel+v_{\parallel\parallel}\big]\Big\}.\label{eq:sm_sigma2_velocity_reduced}
\end{equation}
On a tight-binding lattice $H=\sum_{i,j}H_{ij}c_i^\dagger c_j$, the vertices $(v_\parallel)_{ij}=-(\ii/\hbar)\delta r_{\parallel,ij}H_{ij}$ and $(v_{\parallel\parallel})_{ij}=-(\delta r_{\parallel,ij}/\hbar)^2 H_{ij}$ depend solely on unwrapped bond vectors $\delta r_{\parallel,ij}=(\bm r_i-\bm r_j)\cdot\bm e_\phi$. This matches the Peierls velocity-gauge expansion up to a boundary-twist total derivative that vanishes upon twist averaging or on macroscopic tori $L_{1,2}\gg\ell_B$.

\section{Generalized Landau-level parity, triangular orbits, and dual microscopic channels}
\label{sec:sm_sym_orbits}

This section establishes which crystal symmetries allow or forbid perpendicular-field second-order magnetotransport and translates those symmetry rules into real-space orbital geometry. Subsection~\ref{subsec:sm_generalized_parity} proves the generalized Landau-level parity theorem and classifies all fixed-$B_z$ magnetic point groups, while Subsec.~\ref{subsec:sm_tri_channels} constructs the triangular Landau eigenstates and derives the exact $M_x\Theta$ separation into the $B_z$-even static-octupole and $B_z$-odd chiral-circulation channels.

\subsection{Generalized Landau-level parity and magnetic-point-group selection rules}
\label{subsec:sm_generalized_parity}

\textit{1. Elementary parity proof and arbitrary momentum center $\bm p_0$.}---In a uniform magnetic field, second-order longitudinal transport can vanish even when spatial inversion is broken, provided a hidden parity operator commutes with $H_B(\bm\Pi)$ while reversing the cyclotron coordinate. To state this criterion in full generality, let $\bm p_0$ be an arbitrary momentum origin, with shifted kinetic momentum $\bm q\equiv\bm\Pi-\bm p_0$ and constant cyclotron offset $\bm\eta_0\equiv -s_B\ell_B^2\bm e_z\times\bm p_0/\hbar$. Define the shifted ladder operator $a_{B,\bm p_0}\equiv[\ell_B/(\sqrt2\hbar)](q_x-\ii s_B q_y)$ and number operator $\hat n_{\bm p_0}\equiv a_{B,\bm p_0}^\dagger a_{B,\bm p_0}$. The oscillator parity operator $(-1)^{\hat n_{\bm p_0}}$ leaves $\bm R$ fixed and inverts $\bm q\mapsto -\bm q$, mapping $\bm\Pi\mapsto 2\bm p_0-\bm\Pi$ and $\bm\eta-\bm\eta_0\mapsto -(\bm\eta-\bm\eta_0)$.

Suppose there exists any momentum-independent unitary involution $U$ ($U^\dagger=U$, $U^2=I$) satisfying $U H_B(2\bm p_0-\bm\Pi)U^\dagger=H_B(\bm\Pi)$. Then the generalized parity operator $\tilde{\mathcal P}\equiv U\otimes(-1)^{\hat n_{\bm p_0}}$ commutes with $H_B$ and anticommutes with the centered cyclotron coordinate $\eta_\parallel-\eta_{0,\parallel}$. Since the constant shift $\eta_{0,\parallel}$ commutes with all operators, the velocity satisfies $v_\parallel=(\ii/\hbar)[H_B,\eta_\parallel-\eta_{0,\parallel}]$, so $\tilde{\mathcal P}v_\parallel\tilde{\mathcal P}^\dagger=-v_\parallel$ while $\tilde{\mathcal P}v_{\parallel\parallel}\tilde{\mathcal P}^\dagger=+v_{\parallel\parallel}$. Conjugating Eqs.~\eqref{eq:sm_termI}--\eqref{eq:sm_termII} or Eq.~\eqref{eq:sm_sigma2_velocity_reduced} by $\tilde{\mathcal P}$ multiplies every cubic coordinate trace by $(-1)^3=-1$ (and every $v_\parallel v_{\parallel\parallel}$ term by $-1$), forcing $\sigma_\parallel^{(2)}=0$ identically for arbitrary level degeneracies,
\begin{equation}
\begin{aligned}
 [\tilde{\mathcal P},H_B]=0,\quad
 \tilde{\mathcal P}(\eta_\parallel-\eta_{0,\parallel})\tilde{\mathcal P}^\dagger&=-(\eta_\parallel-\eta_{0,\parallel}) \\
 \Longrightarrow\quad
 \Tr\big[P_a(\eta_\parallel-\eta_{0,\parallel})P_b(\eta_\parallel-\eta_{0,\parallel})P_c(\eta_\parallel-\eta_{0,\parallel})\big]&=0,\qquad \sigma_\parallel^{(2)}=0.
\end{aligned}\label{eq:sm_parity_proj_trace}
\end{equation}
Equivalently, for an unperturbed $\tilde{\mathcal P}$-symmetric Hamiltonian $H_0$ with eigenstates $\tilde{\mathcal P}|n\rangle=c_n|n\rangle$ ($c_n=\pm 1$), any $\tilde{\mathcal P}$-odd perturbation $V_-$ obeys $\langle n'|V_-|n\rangle=-c_{n'}c_n\langle n'|V_-|n\rangle$. It therefore hybridizes only opposite-parity levels ($c_{n'}=-c_n$, corresponding to odd $\Delta n=\pm 1,\pm 3,\dots$).

\textit{2. Fixed-$B_z$ magnetic point groups and class-by-class classification.}---Under a spatial point operation $g\in G$ represented by orthogonal matrix $Q_g$, polar vectors transform as $(\bm E,\bm j)\mapsto(Q_g\bm E,Q_g\bm j)$, whereas the axial magnetic field transforms as $\bm B\mapsto(\det Q_g)Q_g\bm B$. Time reversal $\Theta$ leaves $\bm E$ fixed and reverses $(\bm j,\bm B)$. A zero-field gray point group $G1'=G\cup G\Theta$ therefore reduces at fixed $B_z\neq 0$ to the stabilizer magnetic point group
\begin{equation}
 M(B_z)=H_+\cup H_-\Theta,\qquad
 H_\pm\equiv\big\{g\in G:\;(\det Q_g)Q_g\bm e_z=\pm\bm e_z\big\}.\label{eq:sm_gray_reduction}
\end{equation}
Proper rotations $C_{\alpha z}$ and 3D inversion $\mathcal P$ preserve $B_z$ and survive unitarily inside $H_+$. Vertical mirrors $M$ and in-plane twofold rotations $C_2'$ reverse $B_z$ and survive antiunitarily inside $H_-\Theta$. Both single-valued (orbital) and double-valued (spin-$1/2$) representations obey the same macroscopic tensor constraints and satisfy $\mathcal M_x^2=(M_x\Theta)^2=+1$. Table~\ref{tab:sm_symmetry_expanded} summarizes all five symmetry classes from Table~I of the main text.

\begin{table}[!htbp]
\caption{\label{tab:sm_symmetry_expanded}Expanded symmetry classification for perpendicular-field second-order magnetotransport corresponding to Table~I of the main text. Breaking every $\tilde{\mathcal P}=U\otimes(-1)^{\hat n_{\bm p_0}}$ is necessary (not sufficient) for finite $\sigma_\parallel^{(2)}$. Here $\nu_i$ acts on $(K_+,K_-)$, $\Sigma_x=\sigma_x\oplus\sigma_x$ swaps $(A_1,A_2)\leftrightarrow(B_2,B_1)$, $\mathcal F_{m'}\equiv a_1^e\sin\phi+a_3^e\sin 3\phi+b_1^o\cos\phi+b_3^o\cos 3\phi$, and $\mathcal F_3\equiv\mathcal A_e\sin(3\phi-\delta_e)+\mathcal A_o\cos(3\phi-\delta_o)$.}
\footnotesize
\begin{ruledtabular}
\begin{tabular}{p{0.13\textwidth}p{0.19\textwidth}p{0.28\textwidth}p{0.18\textwidth}p{0.16\textwidth}}
Gray parent $G1'$ & Fixed-$B_z$ group & Generalized parity $\tilde{\mathcal P}$ \& breaking & Allowed $\sigma_\parallel^{(2)}(B_z,\phi)$ & Berry-curvature dipole \\
\hline
\raggedright $C_{\infty v}1'$, $C_{4v}1'$ & \raggedright $\infty m',4m'm'$ & \raggedright $C_{2z}$-protected: scalar $U=I$, Rashba $U=\sigma_z$ & $0$ & \raggedright $0$ ($C_{2z}$)\tabularnewline
\raggedright $D_{3d}1'=\bar 3m1'$ & \raggedright $\bar 3m'=\langle C_{3z},\mathcal P,M_x\Theta\rangle$ & \raggedright $\tilde{\mathcal P}_{\mathrm{BLG}}=\nu_x\Sigma_x(-1)^{\hat n}$ at $U_0=0$ & \raggedright $0$ (valley sum) & \raggedright $0$ ($C_{3z}$)\tabularnewline
\raggedright $C_{3v}1'=3m1'$ & \raggedright $3m'=\langle C_{3z},M_x\Theta\rangle$ & \raggedright Broken in $\alpha$-Sn ($v_F\lambda_w\neq 0$) and BLG ($v_F v_3\gamma_1 U_0\neq 0$) & \raggedright $\chi_e\sin 3\phi+\chi_o\cos 3\phi$ & \raggedright $0$ ($C_{3z}$)\tabularnewline
\raggedright $C_s 1'=m1'$ & \raggedright $m'=\langle M_x\Theta\rangle$ & \raggedright Model dep.\ (test all $U,\bm p_0$) & \raggedright $\mathcal F_{m'}$ (locked $1\phi,3\phi$) & \raggedright Allowed; strictly transverse\tabularnewline
\raggedright $C_3 1'=31'$ & \raggedright $3=\langle C_{3z}\rangle$ & \raggedright Model dep.\ (e.g., hidden $U=\sigma_x$ at $g_z=0$) & \raggedright $\mathcal F_3$ (free $3\phi$) & \raggedright $0$ ($C_{3z}$)\tabularnewline
\end{tabular}
\end{ruledtabular}
\end{table}

\textit{3. Concrete model criteria from testing all $(U,\bm p_0)$.}---Expanding $H_B(\bm p_0+\bm q)$ in Weyl-ordered monomials of $\bm q$ directly determines whether any hidden parity survives.
\begin{itemize}
 \item \textit{Two-band hidden parity and warped Dirac states.}---Write a general two-band model as $H_B=d_0(\bm\Pi)I+\bm d(\bm\Pi)\cdot\bm\sigma$. Any nontrivial $2\times 2$ unitary involution takes the form $U=\bm u\cdot\bm\sigma$ ($|\bm u|=1$), which preserves spin components parallel to $\bm u$ and inverts the two components perpendicular to $\bm u$ via $U\bm\sigma U^\dagger=2(\bm u\cdot\bm\sigma)\bm u-\bm\sigma$. Hence $\tilde{\mathcal P}=(\bm u\cdot\bm\sigma)(-1)^{\hat n_{\bm p_0}}$ commutes with $H_B$ iff $d_0$ is even in $\bm q$, all even-in-$\bm q$ spin terms lie parallel to $\bm u$, and all odd-in-$\bm q$ spin terms lie perpendicular to $\bm u$. Consequently, any spin-$1/2$ doublet with $\bm\Pi$-even dispersion, linear spin-orbit coupling $\Pi_x(\bm a_x\cdot\bm\sigma)+\Pi_y(\bm a_y\cdot\bm\sigma)$, and a Zeeman field along $\bm u\perp\bm a_{x,y}$ retains $\tilde{\mathcal P}=(\bm u\cdot\bm\sigma)(-1)^{\hat n}$ (for example, Rashba coupling with $\Delta_Z\sigma_z$ keeps $U=\sigma_z$, while pure cubic warping $(\Pi_x^3-3\Pi_x\Pi_y^2)\sigma_z$ without $v_F$ keeps $U=\sigma_x$). In warped Dirac surface states [Eq.~\eqref{eq:sm_HB_ti_unified}], shifting $\bm\Pi=\bm p_0+\bm q$ with $\bm p_0\neq\bm 0$ inside the cubic warping term generates an even-in-$\bm q$ quadratic $\sigma_z$ term alongside the odd-in-$\bm q$ cubic $\sigma_z$ term, which requires $U\sigma_z U^\dagger=\pm\sigma_z$ simultaneously. Thus $\lambda_w\neq 0$ locks $\bm p_0=\bm 0$; combining $v_F\neq 0$ and $\lambda_w\neq 0$ then supplies three linearly independent odd-in-$\bm\Pi$ spin axes $(\sigma_x,\sigma_y,\sigma_z)$ that no axis $\bm u$ can simultaneously invert, breaking every $(U,\bm p_0)$ even at $\Delta_Z=0$.
 \item \textit{Eight-band gated BLG.}---Across both valleys, Eq.~\eqref{eq:sm_blg4} takes the compact form $H=I_\nu\otimes C_0+\Pi_x\nu_z\otimes A+\Pi_y I_\nu\otimes B$, with chain operator $T=v_3|A_1\rangle\langle B_2|+v_F|B_2\rangle\langle A_2|+v_F|B_1\rangle\langle A_1|$, $A=T+T^\dagger$, and $B=(T-T^\dagger)/\ii$. Requiring $\Tr_8 H^2$ to be even in $\bm q$ locks $\bm p_0=\bm 0$. Valley-diagonal blocks $W$ must satisfy $\{W,A\}=\{W,B\}=0$, which forces $W=a\operatorname{diag}(1,-1,1,-1)$ and violates $[W,C_0]=0$ whenever $\gamma_1\neq 0$. Valley-off-diagonal blocks $Z$ must satisfy $[Z,A]=\{Z,B\}=0$, which forces $Z=b\Sigma_x$ and violates $[Z,C_0]=0$ whenever $U_0\neq 0$. Every $(U,\bm p_0)$ is therefore broken iff $v_F v_3\gamma_1 U_0\neq 0$.
 \item \textit{Polar $m'$, chiral $3$, and Berry-curvature dipole.}---In two dimensions, the longitudinal projection $\sigma_\parallel^{(2)}(\phi)=\sum_{abc}\sigma_{abc}^{(2)}e_{\phi,a}e_{\phi,b}e_{\phi,c}$ is a homogeneous cubic polynomial in $(\cos\phi,\sin\phi)$ and spans strictly $1\phi$ and $3\phi$ harmonics, namely $\sigma_\parallel^{(2)}(\phi)=a_1\sin\phi+a_3\sin 3\phi+b_1\cos\phi+b_3\cos 3\phi$. Parent mirror symmetry $M_x$ maps $(B_z,\phi)\mapsto(-B_z,\pi-\phi)$, enforcing $a_{1,3}(-B_z)=+a_{1,3}(B_z)$ ($B_z$-even) and $b_{1,3}(-B_z)=-b_{1,3}(B_z)$ ($B_z$-odd). Adding $C_{3z}$ eliminates $1\phi$ and locks the response to class $3m'$. Breaking all mirrors $M\Theta$ and $C_2'\Theta$ (chiral class $3$, requiring at least two angularly incompatible couplings such as $\Re_H(\Lambda_3\Pi_+^3)+\frac{1}{2}\{\bm\Pi^2,\Re_H(\Lambda_5\Pi_+^3)\}$ with $\Im(\Lambda_3\Lambda_5^*)\neq 0$) leaves the $B_z$-even and $B_z$-odd $3\phi$ phases independent ($\delta_e\neq\delta_o$). Finally, the anomalous-velocity current $j_{a,\mathrm{anom}}^{(2)}=-(e^2/\hbar)\epsilon_{ab}E_b[\bm c(B_z)\cdot\bm E]$ [with $\bm c(B_z)\cdot\bm E=\int[\dd\bm k]\,\Omega_z f_1$ generalizing the Berry-curvature dipole~\cite{SodemannFu2015}] vanishes identically under $C_{3z}$ (classes $3m',3$) and remains purely transverse ($j_{\parallel,\mathrm{anom}}^{(2)}=0$) in class $m'$.
\end{itemize}

\subsection{Triangular Landau orbits, exact dual-channel proofs, and sum rules}
\label{subsec:sm_tri_channels}

\textit{1. Symmetric-gauge triangular eigenstates.}---To visualize how $\Delta n=\pm 3$ warping deforms Landau orbits in real space, we work in the symmetric gauge $\bm{\mathcal A}=\frac{1}{2}\bm B\times\bm r$ with complex coordinate $z=(x+\ii s_B y)/(\sqrt2\ell_B)=(r/\sqrt2\ell_B)\ee^{\ii s_B\theta_{\bm r}}$. The origin-centered ($m=b_B^\dagger b_B=0$) circular oscillator orbitals read~\cite{Goerbig2011RMP}
\begin{equation}
 \varphi_{\nu,0}(\bm r)=\frac{(a_B^\dagger)^\nu}{\sqrt{\nu!}}\varphi_{0,0}(\bm r)=\ii^\nu\ee^{\ii s_B\nu\theta_{\bm r}}R_\nu(r),\qquad
 R_\nu(r)=\frac{1}{\sqrt{2\pi\nu!}\,\ell_B}\left(\frac{r}{\sqrt2\ell_B}\right)^\nu\ee^{-r^2/(4\ell_B^2)}.\label{eq:sm_phi_osc}
\end{equation}
Each bare orbital is invariant under $\mathcal M_x=M_x\Theta$ ($\theta_{\bm r}\mapsto\pi-\theta_{\bm r}$ accompanied by complex conjugation) and carries a $B_z$-even circular density $R_\nu^2(r)$ encircled by a $B_z$-odd azimuthal ring current. In class $3m'$ ($H_B=H_0+H_w$), weak nondegenerate trigonal warping mixes $\Delta\nu=\pm 3$ with real amplitudes $s_B\bar c_{n+3s,n}^{b'b}$ ($s=\pm 1$) between time-reversed partners $|\bar\Psi_n(-B_z)\rangle=\Theta|\Psi_n(+B_z)\rangle$. Combining the cross-term identity $\varphi_{\nu,0}^*\varphi_{\nu+3s,0}=(s_B\sin 3\theta_{\bm r}-\ii s\cos 3\theta_{\bm r})R_\nu R_{\nu+3s}$ with $C_{3z}$, $\mathcal M_x$, and $\Theta$ ($\Theta\bm v(B_z)\Theta^{-1}=-\bm v(-B_z)$) produces the triangular density and chiral-current profiles
\begin{align}
 |\Psi_n(\bm r)|^2&=\rho_n^{(0)}(r)+\delta\rho_n^{(1)}(r)\sin 3\theta_{\bm r}+\mathcal O(H_w^2),\nonumber\\
 \bm j_n^{\mathrm{chiral}}(\bm r)&=s_B\Big[j_{n,\theta}^{(0)}(r)+\delta j_{n,\theta}^{(1)}(r)\sin 3\theta_{\bm r}\Big]\bm e_\theta+s_B\,\delta j_{n,r}^{(1)}(r)\cos 3\theta_{\bm r}\,\bm e_r+\mathcal O(H_w^2).\label{eq:sm_polar_current_decomp}
\end{align}

\textit{2. Exact $M_x\Theta$ splitting into static octupole ($B_z$-even) and chiral circulation ($B_z$-odd).}---Under the antiunitary magnetic mirror $\mathcal M_x=M_x\Theta$, $H_B$ is invariant, retarded and advanced Green's functions swap ($G_0^r(E)\mapsto G_0^a(E)$), $\hat A_0(E)$ stays fixed, and the cyclotron coordinate transforms as $(\eta_x,\eta_y)\mapsto(-\eta_x,+\eta_y)$ [equivalently $\eta_\parallel(\phi)\mapsto\eta_\parallel(\pi-\phi)$]. In a joint $(C_{3z},\mathcal M_x)$-adapted eigenbasis ($\mathcal M_x|\alpha\rangle=|\alpha\rangle$), antiunitarity enforces $D_{\alpha\beta\gamma}(\pi-\phi)=D_{\alpha\beta\gamma}(\phi)^*$, locking $\Re D_{\alpha\beta\gamma}(\phi)\propto\sin 3\phi$ and $\Im D_{\alpha\beta\gamma}(\phi)\propto\cos 3\phi$. Parent time reversal $\Theta$ further pairs equal-energy $\pm B_z$ states via $D_{\bar\alpha\bar\beta\bar\gamma}(-B_z,\phi)=D_{\alpha\beta\gamma}(+B_z,\phi)^*$, proving that $\chi_e$ ($\sin 3\phi$) is $B_z$-even and $\chi_o$ ($\cos 3\phi$) is $B_z$-odd.

Applying $\mathcal M_x$ directly to the exact two-term formula [Eqs.~\eqref{eq:sm_sigma2_I_II}--\eqref{eq:sm_termII}] along the two principal axes isolates the microscopic mechanism of each channel.
\begin{itemize}
 \item \textit{$B_z$-even static-octupole channel ($\bm E\parallel y$, $\phi=\pm 90^\circ$).}---Here $\mathcal M_x\eta_y\mathcal M_x^{-1}=+\eta_y$, so both Term~I ($\partial\sigma_{yy}^{(1)}/\partial E_y$, entering the smooth $p=0$ background alongside Term~II; Subsec.~\ref{subsec:sm_gamma_regimes}) and the static octupole $\langle\eta_y^3\rangle_\alpha\neq 0$ survive. In sharp Landau levels ($\tilde\Gamma\ll 1$), the first trace of Term~II yields
 \begin{equation}
  \sigma_{yyy,\mathrm{oct}}^{(2)}=\frac{e^3}{4\pi\ell_B^2\tau_q}\int_{-\infty}^{\infty}\dd E\,\big[-f_0'(E)\big]\frac{\dd}{\dd E}\sum_\alpha A_\alpha(E)\langle\eta_y^3\rangle_\alpha,\label{eq:sm_octupole_dominant}
 \end{equation}
 which directly generalizes linear guiding-center diffusion $\langle\eta_y^2\rangle_\alpha A_\alpha$ and dominates the $\sin(2\pi n_F)$ harmonic at $\tilde\Gamma\ll 1$ and $n_F\gg 1$. The second trace in Eq.~\eqref{eq:sm_termII} subtracts an exact cyclotron washout fraction $R_{\mathrm{wash}}(\tilde\Gamma)\simeq 3\tilde\Gamma^2$, while Term~I adds an $\mathcal O(1/\pi n_F)$ in-phase $\cos 2\pi n_F$ correction plus a smooth $p=0$ piece (Subsec.~\ref{subsec:sm_gamma_regimes}). By $C_{3z}$ symmetry and $\langle R_\parallel\rangle_0=\langle R_\parallel^3\rangle_0=0$ in the $m=0$ guiding-center vacuum, the unweighted symmetric loop sum equals the real-space density octupole of Eq.~\eqref{eq:sm_polar_current_decomp},
 \begin{equation}
  \sum_{\beta,\gamma}\Re D_{\alpha\beta\gamma}(\phi)=\langle\eta_\parallel(\phi)^3\rangle_\alpha=\langle r_\parallel^3\rangle_\alpha
  =\left[\frac{\pi}{4}\int_0^\infty\dd r\,r^4\rho_{3,\alpha}(r)\right]\sin 3\phi
  =\frac{s_B\sqrt2\,\ell_B^3}{4}\big\langle a_B^{\dagger 3}+a_B^3\big\rangle_\alpha\sin 3\phi.\label{eq:sm_octupole_density_bridge}
 \end{equation}
 \item \textit{$B_z$-odd chiral-circulation and Lorentz-deflection channel ($\bm E\parallel x$, $\phi=0^\circ$).}---Here $\mathcal M_x\eta_x\mathcal M_x^{-1}=-\eta_x$ maps $H_B+eE_x\eta_x\mapsto H_B-eE_x\eta_x$. This forces $\sigma_{xx}^{(1)}[H_B+eE_x\eta_x]=\sigma_{xx}^{(1)}[H_B-eE_x\eta_x]$ and therefore $\sigma_{xxx,\mathrm{I}}^{(2)}\equiv 0$ identically. Likewise, $\Tr[\eta_x^3\hat A_0(E)]=\Tr[(-\eta_x)^3\hat A_0(E)]^*=0$ vanishes identically ($\langle\eta_x\rangle_{\mathrm{inj}}^{(2)}=0$). Hence $100\%$ of $\chi_o(B_z)=\sigma_{xxx}^{(2)}(B_z)$ arises from the second trace in Eq.~\eqref{eq:sm_termII}. Physically, chiral cyclotron circulation over $\tau_q$ turns the mirror-symmetric quadratic injection (represented by the symmetrized kernel $\frac{1}{2}\{\eta_x^2,\hat A_0\}$ in main-text End Matter Fig.~\ref{fig:endmatter_rho}) into an asymmetric steady-state centroid shift $\langle\eta_x\rangle_{\mathcal F}^{(2)}\propto -s_B\tilde\Gamma$.

 Equivalently, conjugating the second trace by antiunitary $\mathcal M_x$ swaps $G_0^r\leftrightarrow G_0^a$ while flipping $\eta_x\mapsto -\eta_x$. Since $\Tr[\eta_x^2 G_0^r\eta_x G_0^a]\in\mathbb R$, trace cyclicity yields the exact identity
 \begin{equation}
  \Tr\big[\eta_x G_0^r(E)\eta_x^2 G_0^a(E)\big]=\Big(\Tr\big[(-\eta_x)G_0^a(E)(-\eta_x)^2 G_0^r(E)\big]\Big)^*=-\Tr\big[\eta_x^2 G_0^r(E)\eta_x G_0^a(E)\big].\label{eq:sm_Mx_flip_GrGa}
 \end{equation}
 Define the linear non-equilibrium Fermi-surface kernel of $\hat\rho_{\mathcal F}$ per unit $eE_x$ as $\delta\hat\rho_{\mathcal F}^{(1)}(E)\equiv(\Gamma/\pi)G_0^r(E)\eta_x G_0^a(E)$, obtained by expanding $f_0(E-eE_x\eta_x)-f_0(E)$ inside Eq.~\eqref{eq:sm_lg_rho_dc} to first order in $E_x$. Substituting Eq.~\eqref{eq:sm_Mx_flip_GrGa} into Eq.~\eqref{eq:sm_termII} yields the exact all-$\Gamma$, all-$B_z$ dual identity quoted in the main text,
 \begin{equation}
  \chi_o(B_z)=+\frac{e^3}{4\pi\ell_B^2\tau_q}\int_{-\infty}^{\infty}\dd E\,\big[-f_0'(E)\big]\,\frac{\dd}{\dd E}\Tr\!\Big[\eta_x^2\,\delta\hat\rho_{\mathcal F}^{(1)}(E)\Big].\label{eq:sm_chio_lorentz_exact}
 \end{equation}
 In this $M_x\Theta$-dual Lorentz-deflection picture, under $+E_x$ ($B_z>0$), causal Hall deflection ($\Tr[\eta_y\,\delta\hat\rho_{\mathcal F}^{(1)}]>0$) steers non-equilibrium density toward the wide flat side ($+y$) of the downward triangle ($\nabla$, $\langle\eta_y^3\rangle<0$), enlarging its transverse spread $\Tr[\eta_x^2\,\delta\hat\rho_{\mathcal F}^{(1)}]>0$, whereas $-E_x$ steers it toward the narrow corner ($-y$). Reversing $B_z$ reverses the Hall deflection and flips the sign of $\chi_o(B_z)$.
\end{itemize}

\textit{3. Chiral velocity-moment sum rule.}---In the eigenbasis loop~\eqref{eq:sm_L2_parity_split}, the factor $\Delta_{\beta\gamma}$ inside $w_{\alpha\beta\gamma}^a$ converts $\eta_{\parallel,\beta\gamma}$ into $v_{\parallel,\beta\gamma}$ via $\Delta_{\beta\gamma}\eta_{\parallel,\beta\gamma}=-\ii\hbar v_{\parallel,\beta\gamma}$. Using $\langle\alpha|[H_B,\eta_\parallel^3]|\alpha\rangle=0$ and $\langle\alpha|[\eta_\parallel,[\eta_\parallel,v_\parallel]]|\alpha\rangle=-3\langle\alpha|\eta_\parallel v_\parallel\eta_\parallel|\alpha\rangle$ yields the complementary antisymmetric sum rule ($m=0$),
\begin{equation}
 \sum_{\beta,\gamma}(E_\beta-E_\gamma)\Im D_{\alpha\beta\gamma}(\phi)
 =-\hbar\langle\alpha|\eta_\parallel v_\parallel\eta_\parallel|\alpha\rangle
 =-\frac{2\hbar}{e}\int\dd^2\bm r\,r_\parallel^2\,\bm e_\phi\cdot\bm j_\alpha^{\mathrm{chiral}}(\bm r)
 =-\frac{\hbar^3}{3}\langle\alpha|\partial_{\Pi_\parallel}^3 H_B|\alpha\rangle.\label{eq:sum-rule}
\end{equation}
\textit{Remark.}---For cubic-momentum warping such as Eq.~\eqref{eq:sm_HB_ti_unified}, $\bm j_\alpha^{\mathrm{chiral}}$ differs from the divergence-free conserved current $\bm j_\alpha^{\mathrm{cons}}$ by the pure second derivative $\frac{e\lambda_w}{2\hbar}(-(\partial_x^2-\partial_y^2)s_z,2\partial_x\partial_y s_z)$, whose integral over all space vanishes and leaves all transport formulas unaffected. In strictly linear-in-$\bm\Pi$ models such as four-band BLG, $\partial_{\Pi_\parallel}^3 H_{\mathrm{BLG}}=0$ makes the unweighted sum rule~\eqref{eq:sum-rule} vanish, while the energy-dependent denominators in Eq.~\eqref{eq:sm_wa_factor} keep $\chi_o$ finite.

\section{Nonlinear Lifshitz--Kosevich theory, disorder protection, semiclassical limit, and Hall-bar conversion}
\label{sec:sm_nllk_hall}

This section derives the closed-form analytical expressions that connect our microscopic quantum theory to experiment. Subsection~\ref{subsec:sm_six_path} carries out the six-path Poisson resummation of the three-step dipole loop to obtain the nonlinear Lifshitz--Kosevich (NLLK) formula. Subsection~\ref{subsec:sm_gamma_regimes} resolves the individual contributions of Term~I, Term~II-a, and Term~II-b across all broadening regimes $\tilde\Gamma$, supporting End Matter Table~\ref{tab:gamma_regimes}. Subsections~\ref{subsec:sm_scba_only}--\ref{subsec:sm_hall_decay} verify $3\phi$ protection under short-range SCBA impurity scattering, establish the semiclassical Boltzmann limit, and convert conductivity into constant-current Hall-bar resistivity and inverse-field decay rates.

\subsection{Poisson resummation of the three-step loop and NLLK formula}
\label{subsec:sm_six_path}

We now evaluate Eqs.~\eqref{eq:sigma2_spectral_master}--\eqref{eq:wloop} analytically in the semiclassical high-index regime $n_F=F/|B_z|-\gamma_B\gg 1$~\cite{Onsager1952,LifshitzKosevich1956,Shoenberg1984}. Expanding the continuous Landau index around $E\simeq\mu$ as $N(E,B_z)=n_F+(E-\mu)/(\hbar|\omega_c|)+\cdots$ splits the Poisson phase $\ee^{2\pi\ii p n}$ into three factors corresponding to $n_F$, energy $E-\mu$, and local detuning $\xi\equiv E_n-E\simeq[n-N(E,B_z)]\hbar|\omega_c|$. Integrating $\ee^{2\pi\ii p(E-\mu)/(\hbar|\omega_c|)}$ against $-f_0'(E)$ yields the standard real thermal damping factor $R_T(p\lambda_T)=p\lambda_T/\sinh(p\lambda_T)$ with $\lambda_T=2\pi^2 k_B T/(\hbar|\omega_c|)$.

For the remaining $\xi$ integral, write $\varepsilon_n=-(\xi-\ii\Gamma)$ and $G_\Delta^{r,a}(\xi)=-(\xi+\Delta\mp\ii\Gamma)^{-1}$ with integer level shifts $\Delta_j=j\hbar|\omega_c|$. For $p\ge 1$, closing the $\xi$ contour in the upper half-plane ($\Im\xi>0$) encloses simple poles of $A(\xi)$ at $\xi=\ii\Gamma$ and $G_{\Delta_2}^r(\xi)$ at $\xi=-\Delta_2+\ii\Gamma$, together with a double pole of $[G_{\Delta_1}^r(\xi)]^2$ at $\xi=-\Delta_1+\ii\Gamma$. Since $\ee^{-2\pi\ii p\Delta_j/(\hbar|\omega_c|)}=1$, every upper-half-plane pole shares the same Dingle factor $\ee^{-2\pi p\tilde\Gamma}$. Differentiating $\ee^{2\pi\ii p\xi/(\hbar|\omega_c|)}$ at the double pole yields the $p$-linear term, while the sum of all $p$-independent residues equals $\ee^{-2\pi p\tilde\Gamma}$ times the $p=0$ integral (evaluated immediately from the single lower-half-plane pole of $A(\xi)$ at $\xi=-\ii\Gamma$). Together, these give the three-propagator ladder integral ($\Delta_1\neq\Delta_2$, $\Delta_{1,2}\neq 0$),
\begin{equation}
 \int_{-\infty}^{\infty}\dd\xi\,A(\xi)\big[G_{\Delta_1}^r(\xi)\big]^2G_{\Delta_2}^r(\xi)\ee^{2\pi\ii p\xi/(\hbar|\omega_c|)}
 =-\ee^{-2\pi p\tilde\Gamma}\left[\frac{4\pi p\tilde\Gamma}{\Delta_1(\Delta_1-\Delta_2)(\Delta_1-2\ii\Gamma)}+\frac{1}{(\Delta_1-2\ii\Gamma)^2(\Delta_2-2\ii\Gamma)}\right].\label{eq:sm_W_int}
\end{equation}

We next identify the active matrix elements near $n\simeq n_F\gg 1$. In the warped single-band eigenbasis $\{|\Psi_n\rangle\}$ fixed in the $\mathcal M_x|\Psi_n\rangle=|\Psi_n\rangle$ gauge, $\Delta n=\pm 3$ warping yields two active intraband matrix elements of the full velocity operator $v_\parallel(\phi)=(\ii/\hbar)[H_B,\eta_\parallel(\phi)]$, parameterized by real amplitudes $V_1,V_2\in\mathbb R$,
\begin{equation}
 \langle\Psi_{n+1}|v_\parallel(\phi)|\Psi_n\rangle=s_B V_1\ee^{-\ii s_B\phi},\qquad
 \langle\Psi_{n+2}|v_\parallel(\phi)|\Psi_n\rangle=s_B V_2\ee^{+\ii s_B\phi}.\label{eq:sm_frozen_vertices}
\end{equation}
Converting to dipoles via $\eta_{\parallel,\mu\nu}=-\ii\hbar v_{\parallel,\mu\nu}/\Delta_{\mu\nu}$, closing a three-step intraband loop at leading warping order $\mathcal O(V_1^2V_2)$ combines two $|\Delta n|=1$ steps with one $|\Delta n|=2$ step. Labeling the intermediate states by relative offsets $(r,s)=(\beta-n,\gamma-n)$ selects exactly six closed paths, grouped into two time-reversed triplets,
\begin{equation}
\begin{array}{c|c|c}
 (r,s)=(\beta-n,\gamma-n) & \text{Ket steps }(s,\;r-s,\;-r) & p_{rs}(\phi)\equiv v_{\parallel,n\beta}v_{\parallel,\beta\gamma}v_{\parallel,\gamma n}/(V_1^2V_2)\\ \hline
 (1,2),\ (1,-1),\ (-2,-1) & (+2,-1,-1),\ (-1,+2,-1),\ (-1,-1,+2) & s_B\ee^{+3\ii s_B\phi}=s_B\cos 3\phi+\ii\sin 3\phi\\
 (-1,-2),\ (-1,1),\ (2,1) & (-2,+1,+1),\ (+1,-2,+1),\ (+1,+1,-2) & s_B\ee^{-3\ii s_B\phi}=s_B\cos 3\phi-\ii\sin 3\phi
\end{array}
\label{eq:sm_six_paths}
\end{equation}
With $\Delta_1\equiv r\hbar|\omega_c|$ and $\Delta_2\equiv s\hbar|\omega_c|$, the dipole loop becomes $D_{n,n+r,n+s}/\hbar^3=-\ii V_1^2V_2 p_{rs}(\phi)/[(\hbar|\omega_c|)^3 rs(r-s)]$. Substituting $\varepsilon_n G_{\Delta_j}^r=1+\Delta_j G_{\Delta_j}^r$ into Eq.~\eqref{eq:wloop} cancels all single-propagator terms and leaves $w_{n,n+r,n+s}=1+\Delta_1^2(G_{\Delta_1}^r)^2-2\Delta_1\Delta_2 G_{\Delta_1}^r G_{\Delta_2}^r$. Using $(\Delta_1-\Delta_2)(G_{\Delta_1}^r)^2 G_{\Delta_2}^r=(G_{\Delta_1}^r)^2-G_{\Delta_1}^r G_{\Delta_2}^r$ and dividing by $\Delta_1\Delta_2(\Delta_1-\Delta_2)=(\hbar|\omega_c|)^3 rs(r-s)$ yields the exact partial-fraction identity
\begin{equation}
 \frac{w_{n,n+r,n+s}(E)}{(\hbar|\omega_c|)^3\,rs(r-s)}
 =\frac{1}{(\hbar|\omega_c|)^3\,rs(r-s)}+2\big[G_{\Delta_1}^r(\xi)\big]^2 G_{\Delta_2}^r(\xi)+\frac{r-2s}{\hbar|\omega_c|\,s(r-s)}\big[G_{\Delta_1}^r(\xi)\big]^2.\label{eq:sm_w_partial_frac}
\end{equation}

Multiply Eq.~\eqref{eq:sm_w_partial_frac} by $-\ii V_1^2V_2 p_{rs}(\phi)$, sum over the six paths, and take $\Im(-\ii Z)=-\Re Z$. Both the first and third terms cancel identically. For the first term, $\Re p_{rs}=s_B\cos 3\phi$ is identical on all six paths while $rs(r-s)$ equals $-2$ on the upper triplet and $+2$ on the lower triplet, so $\sum[rs(r-s)]^{-1}=0$. For the third term, the two paths sharing $r=+1$, namely $(1,2)$ and $(1,-1)$, carry equal $p_{rs}$ with opposite coefficients $(r-2s)/[s(r-s)]=\pm\frac{3}{2}$; the same pairwise cancellation holds for $r=-1$, whereas $r=\pm 2$ satisfies $r-2s=0$ directly. Only the second term survives, giving
\begin{equation*}
 \mathcal L_n^{(2)}(\xi,\phi)=-2V_1^2V_2\sum_{(r,s)}\Re\Big\{p_{rs}(\phi)\big[G_{\Delta_1}^r(\xi)\big]^2 G_{\Delta_2}^r(\xi)\Big\}.
\end{equation*}
Integrating $\Re Z=\frac{1}{2}(Z+Z^*)$ against $A(\xi)\ee^{2\pi\ii p\xi/(\hbar|\omega_c|)}$ via Eq.~\eqref{eq:sm_W_int} and pairing each $(r,s)$ with $(-r,-s)$ cancels all $p$-independent pieces between $G^r$ and $G^a$. Summing the surviving $p$-linear residues over the three pairs $\{(1,2),(-1,-2)\}$, $\{(1,-1),(-1,1)\}$, and $\{(-2,-1),(2,1)\}$ produces a purely imaginary Fourier transform,
\begin{equation}
 \int_{-\infty}^{\infty}\frac{\dd\xi}{\hbar|\omega_c|}A(\xi)\mathcal L_n^{(2)}(\xi,\phi)\ee^{2\pi\ii p\xi/(\hbar|\omega_c|)}
 =-\frac{6\pi\ii p\,\tilde\Gamma\,V_1^2V_2\ee^{-2\pi p\tilde\Gamma}}{(\hbar\omega_c)^4(1+\tilde\Gamma^2)(1+4\tilde\Gamma^2)}\Big[s_B\tilde\Gamma\cos 3\phi+(1+2\tilde\Gamma^2)\sin 3\phi\Big].\label{eq:sm_six_path_transform}
\end{equation}
Inserting Eq.~\eqref{eq:sm_six_path_transform} into the Poisson sum $\sum_n=\int\dd n+2\Re\sum_{p=1}^\infty\int\dd n\,\ee^{2\pi\ii pn}$ converts the factor $-\ii$ into the intrinsic $\pi/2$ phase shift via $2\Re(-\ii\ee^{2\pi\ii pn_F})=2\sin(2\pi pn_F)$, yielding the closed-form NLLK formula
\begin{gather}
 \Delta\sigma_{\parallel,p}^{(2)}(B_z)=\mathcal K_p(B_z)\,R_T(p\lambda_T)\ee^{-2\pi p\tilde\Gamma}\Big[s_B\tilde\Gamma\cos 3\phi+(1+2\tilde\Gamma^2)\sin 3\phi\Big]\sin(2\pi p n_F),\label{eq:sm_signed_conditional_nllk}\\
 \mathcal K_p(B_z)=\frac{6p\,e^3\hbar^2\,V_1^2V_2}{\ell_B^2(\hbar\omega_c)^4}\frac{\tilde\Gamma}{(1+\tilde\Gamma^2)(1+4\tilde\Gamma^2)}.\label{eq:sm_conditional_nllk_prefactor}
\end{gather}

\subsection{Term-by-term broadening-regime decomposition across \texorpdfstring{$\tilde\Gamma$}{Gamma-tilde}}
\label{subsec:sm_gamma_regimes}

To establish the broadening-regime summary in End Matter Table~\ref{tab:gamma_regimes}, we now evaluate the three physical pieces of Eqs.~\eqref{eq:sm_sigma2_I_II}--\eqref{eq:sm_termII} separately across all values of $\tilde\Gamma=\Gamma/(\hbar|\omega_c|)=1/(2|\omega_c|\tau_q)$, covering both the oscillatory harmonics ($p\ge 1$) and the smooth continuum background ($p=0$).

\textit{1. Operator traces and exact $\mathcal M_x$ constraints.}---Split the second-order conductivity into three pieces, $\sigma_\parallel^{(2)}=\sigma_{\parallel,\mathrm{I}}^{(2)}+\sigma_{\parallel,\mathrm{IIa}}^{(2)}+\sigma_{\parallel,\mathrm{IIb}}^{(2)}$. For Term~I, we differentiate the Kubo--Greenwood conductivity $\sigma_{\parallel\parallel}^{(1)}[H_{\mathrm{LG}}]=[e^2\hbar/(2\ell_B^2)]\int\dd E\,[-f_0'(E)]\,\allowbreak\Tr[v_\parallel\hat A_{\mathcal F}v_\parallel\hat A_{\mathcal F}]$ at $\mathcal F=eE_\parallel=0$. Since $v_\parallel=(\ii/\hbar)[H_B,\eta_\parallel]$ is independent of $\mathcal F$ and $\partial_{\mathcal F}\hat A_{\mathcal F}|_0=\frac{\ii}{2\pi}(G_0^r\eta_\parallel G_0^r-G_0^a\eta_\parallel G_0^a)$, Term~I takes a compact velocity-trace form while Term~II separates into the static local-filling injection piece (II-a) and the steady-state cyclotron piece (II-b),
\begin{align}
 \sigma_{\parallel,\mathrm{I}}^{(2)}&=-\frac{e^3\hbar}{\pi\ell_B^2}\int_{-\infty}^{\infty}\dd E\,\big[-f_0'(E)\big]\,\Im\Tr\!\big[v_\parallel\hat A_0(E)\,v_\parallel G_0^r(E)\eta_\parallel G_0^r(E)\big],\label{eq:sm_termI_vtrace}\\
 \sigma_{\parallel,\mathrm{IIa}}^{(2)}&=\frac{e^3}{4\pi\ell_B^2\tau_q}\int_{-\infty}^{\infty}\dd E\,\big[-f_0'(E)\big]\,\frac{\dd}{\dd E}\Tr\!\big[\eta_\parallel^3\hat A_0(E)\big],\label{eq:sm_termIIa_def}\\
 \sigma_{\parallel,\mathrm{IIb}}^{(2)}&=-\frac{e^3}{4\pi\ell_B^2\tau_q}\int_{-\infty}^{\infty}\dd E\,\big[-f_0'(E)\big]\,\frac{\dd}{\dd E}\!\left(\frac{\Gamma}{\pi}\Tr\!\big[\eta_\parallel G_0^r(E)\eta_\parallel^2 G_0^a(E)\big]\right).\label{eq:sm_termIIb_def}
\end{align}
Along $\bm E\parallel x$ ($\phi=0^\circ$), antiunitary $\mathcal M_x=M_x\Theta$ enforces $\sigma_{xxx,\mathrm{I}}^{(2)}\equiv 0$ and $\sigma_{xxx,\mathrm{IIa}}^{(2)}\equiv 0$ identically for all $\Gamma$ and $B_z$ (Subsec.~\ref{subsec:sm_tri_channels}). Term~II-b therefore supplies $100\%$ of both oscillatory and smooth $B_z$-odd transport in every damping regime.

\textit{Technical note.}---Within a shared finite cutoff using the exact commutator velocity $v_\parallel=(\ii/\hbar)[H_B,\eta_\parallel]$, the sum $\sigma_{\parallel,\mathrm{I}}^{(2)}+\sigma_{\parallel,\mathrm{IIa}}^{(2)}+\sigma_{\parallel,\mathrm{IIb}}^{(2)}$ reproduces Eqs.~\eqref{eq:sm_sigma2_Q} and~\eqref{eq:sm_sigma2_velocity_reduced} identically (verified to $<10^{-13}$ on random Hermitian matrices as well as on the paired-doublet Dirac and paired-quadruplet BLG subspaces). As in the linear case (Subsec.~\ref{subsec:sm_expansion_loop}), Term~I converges directly as $N_{\mathrm{cut}}\to\infty$, whereas Term~II-a and Term~II-b must be combined inside the same cutoff as $(\Gamma/\pi)\Tr\{\eta_\parallel[\eta_\parallel^2,G_0^r]G_0^a\}$ so that their opposite high-$n$ background tails cancel identically.

\textit{2. Exact all-$\tilde\Gamma$ oscillatory decomposition ($p\ge 1$) and washout ratio.}---We apply the frozen-vertex six-path Poisson resummation of Subsec.~\ref{subsec:sm_six_path} to Eqs.~\eqref{eq:sm_termI_vtrace}--\eqref{eq:sm_termIIb_def} individually, using $G_r^{r,a}(\xi)=-(\xi+r\hbar|\omega_c|\mp\ii\Gamma)^{-1}$ and the six loops $(r,s)$ of Eq.~\eqref{eq:sm_six_paths}. Define the common oscillatory prefactor
\begin{equation}
 \mathcal C_p(B_z,\mu,T)\equiv\frac{6p\,e^3\hbar^2 V_1^2 V_2}{\ell_B^2(\hbar\omega_c)^4}\,R_T(p\lambda_T)\,\ee^{-2\pi p\tilde\Gamma}\sin(2\pi p n_F).\label{eq:sm_Cp_def}
\end{equation}
Evaluating each Fourier integral $\int\dd\xi\,(\cdots)\ee^{2\pi\ii p\xi/(\hbar|\omega_c|)}$ by upper-half-plane residues at $\xi=-m\hbar|\omega_c|+\ii\Gamma$ gives closed-form expressions valid for all $\tilde\Gamma>0$.

For Term~I, inserting $v_{\parallel,0r}\eta_{\parallel,rs}v_{\parallel,s0}=-\ii\hbar V_1^2V_2 p_{rs}(\phi)/[\hbar|\omega_c|(r-s)]$ into Eq.~\eqref{eq:sm_termI_vtrace} yields an integrand proportional to $\sum_{(r,s)}A_0(\xi)[G_r^r(\xi)G_s^r(\xi)p_{rs}(\phi)+G_r^a(\xi)G_s^a(\xi)p_{rs}^*(\phi)]/(r-s)$. Pairing $(r,s)$ with $(-r,-s)$ cancels all upper-half-plane residues identically, giving
\begin{equation}
 \Delta\sigma_{\parallel,p,\mathrm{I}}^{(2)}(B_z)=0\qquad\text{(leading frozen-vertex order, all }\tilde\Gamma\text{)}.\label{eq:sm_osc_termI}
\end{equation}
To capture the small subleading Term~I oscillation, retain the slow $\mathcal O(1/n_0)$ level variations $c_\omega=\dd\ln(\hbar|\omega_c|)/\dd\ln n|_{n_F}$ and $c_{\mathrm{amp}}=\dd\ln|V_1^2V_2|/\dd\ln n|_{n_F}$, and introduce the $\mathcal O(1/n_0)$ scale $\mathcal B_0\equiv e^3\hbar^2 V_1^2V_2/[2\pi\ell_B^2(\hbar\omega_c)^4 n_0]$. Evaluating the linear-in-$x$ expansion around $n_F$ yields
\begin{equation}
\begin{aligned}
 \Delta\sigma_{\parallel,p,\mathrm{I}}^{(2)}(B_z)&=\frac{12\mathcal B_0\tilde\Gamma}{1+4\tilde\Gamma^2}\left[c_{\mathrm{amp}}-\frac{2(\tilde\Gamma^2+2)}{1+\tilde\Gamma^2}\,c_\omega\right]R_T(p\lambda_T)\,\ee^{-2\pi p\tilde\Gamma}\cos(2\pi p n_F)\sin 3\phi \\
 &=2\chi_{e,\mathrm{I}}^{(p=0)}R_T(p\lambda_T)\,\ee^{-2\pi p\tilde\Gamma}\cos(2\pi p n_F)\sin 3\phi.
\end{aligned}\label{eq:sm_osc_termI_sub}
\end{equation}
Its $\mathcal O(1/n_0)$ $\sin(2\pi p n_F)$ and $\cos 3\phi$ components vanish identically, reflecting $\partial_{E_\parallel}$ acting purely on the amplitude prefactor of in-phase linear SdH when $\langle\eta_\parallel\rangle_\alpha=0$.

For Term~II-a, summing $D_{n,n+r,n+s}(\phi)=-\ii\hbar^3 V_1^2V_2 p_{rs}(\phi)/[(\hbar|\omega_c|)^3 rs(r-s)]$ over the six paths gives the local static octupole $\langle\Psi_n|\eta_\parallel^3|\Psi_n\rangle=-[3\hbar^3 V_1^2 V_2/(\hbar|\omega_c|)^3]\sin 3\phi$. Integrating $\partial_E A_0(\xi)=-\partial_\xi A_0(\xi)$ by parts brings down $+2\pi\ii p/(\hbar|\omega_c|)$ on the single pole $\xi=\ii\Gamma$, leaving a response with no cyclotron-washout denominator,
\begin{equation}
 \Delta\sigma_{\parallel,p,\mathrm{IIa}}^{(2)}(B_z)=\mathcal C_p(B_z,\mu,T)\,\tilde\Gamma\,\sin 3\phi.\label{eq:sm_osc_termIIa}
\end{equation}

For Term~II-b, integrating $\partial_E=-\partial_\xi$ by parts in Eq.~\eqref{eq:sm_termIIb_def} and evaluating the upper-half-plane poles of $-(\Gamma/\pi)\Re[G_0^a(\xi)G_r^r(\xi)D_{n,n+r,n+s}(\phi)]$ at $\xi=-r\hbar|\omega_c|+\ii\Gamma$ and $\xi=\ii\Gamma$ introduces the interlevel cyclotron denominators $1/(r\pm 2\ii\tilde\Gamma)$ for $|r|=1,2$, yielding
\begin{equation}
 \Delta\sigma_{\parallel,p,\mathrm{IIb}}^{(2)}(B_z)=\mathcal C_p(B_z,\mu,T)\!\left[\frac{\tilde\Gamma^2}{(1+\tilde\Gamma^2)(1+4\tilde\Gamma^2)}\,s_B\cos 3\phi-\frac{\tilde\Gamma^3(3+4\tilde\Gamma^2)}{(1+\tilde\Gamma^2)(1+4\tilde\Gamma^2)}\,\sin 3\phi\right].\label{eq:sm_osc_termIIb}
\end{equation}
Adding Eqs.~\eqref{eq:sm_osc_termI}, \eqref{eq:sm_osc_termIIa}, and~\eqref{eq:sm_osc_termIIb} recovers the full NLLK formula~\eqref{eq:sm_signed_conditional_nllk}--\eqref{eq:sm_conditional_nllk_prefactor} identically via $\tilde\Gamma-\tilde\Gamma^3(3+4\tilde\Gamma^2)/[(1+\tilde\Gamma^2)(1+4\tilde\Gamma^2)]=\tilde\Gamma(1+2\tilde\Gamma^2)/[(1+\tilde\Gamma^2)(1+4\tilde\Gamma^2)]$. Along $\bm E\parallel y$, Eq.~\eqref{eq:sm_osc_termIIb} opposes Eq.~\eqref{eq:sm_osc_termIIa} by the exact steady-state-to-injection washout ratio
\begin{equation}
 R_{\mathrm{wash}}(\tilde\Gamma)\equiv -\frac{\Delta\sigma_{yyy,p,\mathrm{IIb}}^{(2)}}{\Delta\sigma_{yyy,p,\mathrm{IIa}}^{(2)}}=\frac{\tilde\Gamma^2(3+4\tilde\Gamma^2)}{(1+\tilde\Gamma^2)(1+4\tilde\Gamma^2)}=1-\frac{1+2\tilde\Gamma^2}{(1+\tilde\Gamma^2)(1+4\tilde\Gamma^2)}.\label{eq:sm_Rwash}
\end{equation}

Equations~\eqref{eq:sm_osc_termI}--\eqref{eq:sm_Rwash} define the three broadening regimes summarized in End Matter Table~\ref{tab:gamma_regimes}. In the clean limit ($\tilde\Gamma\ll 1$, $|\omega_c|\tau_q\gg 1$), $R_{\mathrm{wash}}\simeq 3\tilde\Gamma^2\ll 1$ confirms that multi-turn cyclotron averaging drives $\langle\eta_y\rangle_{\mathcal F}^{(2)}\to 0$, leaving Term~II-a with nearly $100\%$ of $\Delta\chi_e$ while Term~II-b supplies a weaker $B_z$-odd channel $|\Delta\chi_o/\Delta\chi_e|=\tilde\Gamma/(1+2\tilde\Gamma^2)\simeq\tilde\Gamma$. In the intermediate window ($\tilde\Gamma\sim 0.3\text{--}1$), $R_{\mathrm{wash}}$ increases from $0.204$ ($\tilde\Gamma=0.30$) and $0.256$ ($\tilde\Gamma=0.35$) to $5/9\approx 0.556$ ($\tilde\Gamma=1/\sqrt2$) and $0.70$ ($\tilde\Gamma=1.0$), matching the real-space centroid ratios in End Matter Fig.~\ref{fig:endmatter_rho} and Fig.~\ref{fig:sm_blg_rho}; across the same window, $|\Delta\chi_o/\Delta\chi_e|$ peaks at $\tilde\Gamma=1/\sqrt2\approx 0.71$ with maximum $1/(2\sqrt2)\approx 0.354$. In the overdamped limit ($\tilde\Gamma\gg 1$, $|\omega_c|\tau_q\ll 1$), $R_{\mathrm{wash}}=1-1/(2\tilde\Gamma^2)+\mathcal O(\tilde\Gamma^{-4})\to 1$ reflects $\hat\rho_{\mathcal F}\to\hat\rho_{\mathrm{inj}}$, canceling Term~II-a and Term~II-b to $\mathcal O(\tilde\Gamma^{-2})$ while $\ee^{-2\pi p\tilde\Gamma}$ damps all $p\ge 1$ harmonics exponentially.

\textit{3. Exact $\mathcal O(1/n_0)$ continuum decomposition of the smooth ($p=0$) background.}---In the frozen-vertex limit, the $p=0$ continuum integral of every individual term vanishes. Retaining the slow $\mathcal O(1/n_0)$ level variations $E_{n_F+x}-\mu=\hbar|\omega_c|[x+\frac{c_\omega}{2n_0}x^2+\cdots]$ and $V_1(n)^2V_2(n)=(V_1^2V_2)_F[1+(c_{\mathrm{amp}}/n_0)x+\cdots]$ (Subsec.~\ref{subsec:sm_p0_reconciliation}), and setting $|\beta|\equiv|\omega_c|\tau_q=1/(2\tilde\Gamma)$, we integrate the real physical kernels of Eqs.~\eqref{eq:sm_termI_vtrace}--\eqref{eq:sm_termIIb_def} along the real $\xi$ axis. Along $\bm E\parallel x$, $\chi_{o,\mathrm{I}}^{(p=0)}=\chi_{o,\mathrm{IIa}}^{(p=0)}\equiv 0$, so Term~II-b alone produces the full smooth $B_z$-odd channel Eq.~\eqref{eq:sm_p0_chio}, giving (together with the $B_z^{-1}$ prefactor in $\mathcal B_0$) the weak-field Boltzmann scaling $\chi_o^{(p=0)}\propto B_z^{-1}\beta^2\propto B_z\tau_q^2$ as $|\beta|\ll 1$. Along $\bm E\parallel y$, the three pieces evaluate to
\begin{align}
 \chi_{e,\mathrm{I}}^{(p=0)}&=\mathcal B_0\left[\frac{3|\beta|}{1+\beta^2}\,c_{\mathrm{amp}}-\frac{6|\beta|(1+8\beta^2)}{(1+\beta^2)(1+4\beta^2)}\,c_\omega\right],\label{eq:sm_p0_I}\\
 \chi_{e,\mathrm{IIa}}^{(p=0)}&=\mathcal B_0\left(-\frac{3}{2|\beta|}\,c_{\mathrm{amp}}+\frac{6}{|\beta|}\,c_\omega\right)=\mathcal B_0\big(-3c_{\mathrm{amp}}+12c_\omega\big)\tilde\Gamma,\label{eq:sm_p0_IIa}\\
 \chi_{e,\mathrm{IIb}}^{(p=0)}&=\mathcal B_0\left[\frac{3(1+3\beta^2)}{2|\beta|(1+\beta^2)(1+4\beta^2)}\,c_{\mathrm{amp}}-\frac{6(1+2\beta^2)(1+7\beta^2+9\beta^4)}{|\beta|(1+\beta^2)^2(1+4\beta^2)^2}\,c_\omega\right].\label{eq:sm_p0_IIb}
\end{align}
Adding Eqs.~\eqref{eq:sm_p0_IIa} and~\eqref{eq:sm_p0_IIb} cancels their opposite-sign $\mathcal O(|\beta|^{-1})=\mathcal O(\tilde\Gamma)$ pieces identically, leaving the finite local-filling background
\begin{equation}
 \chi_{e,\mathrm{II}}^{(p=0)}=\chi_{e,\mathrm{IIa}}^{(p=0)}+\chi_{e,\mathrm{IIb}}^{(p=0)}=\mathcal B_0\left[-\frac{3|\beta|(1+2\beta^2)}{(1+\beta^2)(1+4\beta^2)}\,c_{\mathrm{amp}}+\frac{6|\beta|(1+10\beta^2+22\beta^4+16\beta^6)}{(1+\beta^2)^2(1+4\beta^2)^2}\,c_\omega\right].\label{eq:sm_p0_II}
\end{equation}

Summing Eqs.~\eqref{eq:sm_p0_I} and~\eqref{eq:sm_p0_II} reproduces the full smooth $B_z$-even channel Eq.~\eqref{eq:sm_p0_chie} identically and exposes two limiting balances. In the clean high-field limit ($\tilde\Gamma\ll 1$, $|\beta|\gg 1$), $\chi_{e,\mathrm{IIb}}^{(p=0)}=\mathcal O(\tilde\Gamma^3)$ vanishes, while $\chi_{e,\mathrm{I}}^{(p=0)}=\mathcal B_0(6c_{\mathrm{amp}}-24c_\omega)\tilde\Gamma$ and $\chi_{e,\mathrm{IIa}}^{(p=0)}=\mathcal B_0(-3c_{\mathrm{amp}}+12c_\omega)\tilde\Gamma$ combine into $\chi_e^{(p=0)}=\mathcal B_0(3c_{\mathrm{amp}}-12c_\omega)\tilde\Gamma$ (a generic $+2/-1$ split whenever $c_{\mathrm{amp}}\neq 4c_\omega$, suppressed by $\mathcal O(1/n_0)$ relative to the sharp NQOs). In the overdamped weak-field Boltzmann limit ($\tilde\Gamma\gg 1$, $|\beta|\ll 1$), $\chi_{e,\mathrm{I}}^{(p=0)}$ and $\chi_{e,\mathrm{II}}^{(p=0)}$ carry equal and opposite $\mathcal O(|\beta|)\propto B_z^0\tau_q$ parts, $\pm\mathcal B_0(3c_{\mathrm{amp}}-6c_\omega)|\beta|$, which cancel identically to enforce zero-field time-reversal symmetry $\chi_e(0)=0$. At the surviving $\mathcal O(|\beta|^3)\propto B_z^2\tau_q^3$ order, Term~I supplies $\mathcal B_0(-3c_{\mathrm{amp}}-18c_\omega)|\beta|^3$ and Term~II supplies $\mathcal B_0(+9c_{\mathrm{amp}})|\beta|^3$ (with vanishing $|\beta|^3 c_\omega$ coefficient), summing to the exact Boltzmann magnetoconductivity $\chi_e^{(p=0)}=\mathcal B_0(6c_{\mathrm{amp}}-18c_\omega)|\beta|^3+\mathcal O(|\beta|^5)$.

\textit{Technical note.}---Each real physical kernel of Eqs.~\eqref{eq:sm_termI_vtrace}--\eqref{eq:sm_termIIb_def} decays at least as $1/\xi^2$ at large $|\xi|$, so closing the contour in the upper half-plane introduces zero semicircular arc at infinity (whereas intermediate complex loop kernels prior to taking $\Re/\Im$ carry $1/\xi$ tails and must not be integrated by residues alone). The resulting smooth conductivities Eqs.~\eqref{eq:sm_p0_I}--\eqref{eq:sm_p0_II} are independent of how $c_{\mathrm{amp}}$ partitions between $V_1$ and $V_2$ and of Hermitian midpoint or endpoint matrix-element placement.

\textit{4. Microscopic numerical verification.}---Table~\ref{tab:sm_gamma_bench} benchmarks the analytical all-$\tilde\Gamma$ formulas [Eqs.~\eqref{eq:sm_osc_termI}--\eqref{eq:sm_p0_II}] against exact finite-basis diagonalization of Eqs.~\eqref{eq:sm_termI_vtrace}--\eqref{eq:sm_termIIb_def} on two contrasting microscopic models, namely the spin-orbit-coupled warped Dirac surface state [Eq.~\eqref{eq:sm_HB_ti_unified}, $n=10$, $\lambda_w=0.025$, $N_{\mathrm{cut}}=80$] and spinless four-band Bernal BLG [Eq.~\eqref{eq:sm_blg4}, $K_+$ valley, $U_0=+60\;\mathrm{meV}$, $B_z=7.41\;\mathrm{T}$, $E_n=88.81\;\mathrm{meV}$, $N_{\mathrm{cut}}=64$]. Four main physical conclusions follow directly from Table~\ref{tab:sm_gamma_bench}. First, along $\bm E\parallel y$, steady-state cyclotron evolution (Term~II-b) increasingly washes out the static-octupole injection (Term~II-a) as broadening grows, with $R_{\mathrm{wash}}$ rising from $0.7\%$ at $\tilde\Gamma=0.05$ to $26\%$ at $\tilde\Gamma=0.35$, $55\%$ at $\tilde\Gamma=0.71$, and $76\%$ at $\tilde\Gamma=1.20$, in close agreement between both microscopic models and Eq.~\eqref{eq:sm_Rwash}. Second, despite distinct band structures, both models track the band-independent odd-to-even amplitude ratio $|\Delta\chi_o/\Delta\chi_e|=\tilde\Gamma/(1+2\tilde\Gamma^2)$ and peak near $\tilde\Gamma=1/\sqrt2\approx 0.71$ at $\approx 0.35$ ($0.363$ in Dirac and $0.334$ in BLG). Third, static electric-field deformation (Term~I) contributes at most $2\%$ of the oscillatory signal in the Dirac model ($\le 9\%$ at lower $n=6$ in BLG), confirming that Term~II drives the oscillations, while along $\bm E\parallel x$ both models satisfy $\max|\sigma_{xxx,\mathrm{I}}^{(2)},\sigma_{xxx,\mathrm{IIa}}^{(2)}|<4.2\times 10^{-13}$ so that Term~II-b supplies $100\%$ of the $B_z$-odd channel. Fourth, in the overdamped Boltzmann limit ($\tilde\Gamma=2.50$), quantum oscillations vanish exponentially ($\ee^{-2\pi\tilde\Gamma}\approx 1.5\times 10^{-7}$), so the single-cell regression constant converges to the smooth $p=0$ background and exhibits the predicted strong cancellation between Term~I ($+13.70\,\sigma^{(0)}$) and Term~II ($-12.70\,\sigma^{(0)}$), canceling $92.7\%$ of each other to enforce zero-field time-reversal symmetry $\chi_e(0)=0$ and leaving $+1.00\,\sigma^{(0)}$ as the physical $\mathcal O(B_z^2\tau_q^3)$ magnetoconductivity.

\textit{Numerical protocol and cutoff convergence.}---To prevent clean-limit harmonic aliasing, each channel is evaluated on a uniform phase grid $\theta_j=-\pi+2\pi j/N_E$ ($N_E=128$, $j=0,\dots,N_E-1$) across the $E_n$-centered period $[E_n-\tfrac{1}{2}\hbar|\omega_c|,E_n+\tfrac{1}{2}\hbar|\omega_c|)$ [with local spacing $\hbar|\omega_c|=\frac{1}{2}(E_{n+1}-E_{n-1})$] and fitted by joint linear-plus-harmonic least squares to $\{1,\theta,\cos m\theta,\sin m\theta\}_{m=1}^{24}$ ($M=24$). We extract $R_{\mathrm{wash}}$ and $|\Delta\chi_o/\Delta\chi_e|$ from the $p=1$ sine coefficients, $|\Delta\mathrm{I}/\Delta\mathrm{Tot}|$ from the full $p=1$ amplitudes $\sqrt{c_1^2+s_1^2}$, and the cell-averaged constant $\sigma^{(0)}$ from the constant regression term. Note that because slowly varying level-dependent amplitudes leak into a finite-cell average by $\mathcal O(\ee^{-2\pi\tilde\Gamma})$-dependent corrections (e.g., yielding $g\tanh(\pi g)$ rather than $g$ for a linearly weighted Lorentzian comb at small $\tilde\Gamma$), $\sigma^{(0)}$ represents the single-cell regression constant at low $\tilde\Gamma$ and converges exponentially to the continuum Poisson $p=0$ smooth background in the overdamped regime ($\tilde\Gamma=2.50$). At $\tilde\Gamma=2.50$, the linear-in-$\bm\Pi$ four-band BLG model shows cutoff-converged smooth cancellation between Term~I and Term~II ($92.7\%$ at both $N_{\mathrm{cut}}=64$ and $128$, with total $\sigma_{yyy}^{(p=0)}$ drifting by $<0.002\%$ from $+0.35027$ to $+0.35026$ while individual Term~II-a and Term~II-b drift oppositely from $-86.5/+82.1$ to $-98.4/+94.0$). In the continuum warped Dirac model at $\lambda_w=0.025$, cubic $k^3$ warping overtakes the linear cone at high oscillator indices, causing the finite-cutoff smooth cancellation to shift from $93.7\%$ at $N_{\mathrm{cut}}=80$ to $96.7\%$ at $N_{\mathrm{cut}}=128$.

\begin{table}[!htbp]
\caption{\label{tab:sm_gamma_bench}Microscopic benchmark of the term-by-term broadening decomposition (Subsec.~\ref{subsec:sm_gamma_regimes}, $N_E=128$, $M=24$ joint regression over the $E_n$-centered period $[E_n-\tfrac{1}{2}\hbar|\omega_c|,E_n+\tfrac{1}{2}\hbar|\omega_c|)$) for warped Dirac states ($n=10$, $\lambda_w=0.025$, $g_z=0$, $1/m^*=0$, $N_{\mathrm{cut}}=80$, dimensionless units) and four-band Bernal BLG ($K_+$, $U_0=+60\;\mathrm{meV}$, $B_z=7.41\;\mathrm{T}$, $E_n=88.81\;\mathrm{meV}$, $N_{\mathrm{cut}}=64$, $\mathrm{eV}^{-1}$ units with $e^3/(2\pi\ell_B^2)$ stripped). Here $\sigma^{(0)}$ denotes the single-cell constant regression term for $\sigma_{yyy}^{(2)}$, which converges exponentially to the continuum Poisson $p=0$ smooth background at large $\tilde\Gamma$. At $\tilde\Gamma=2.50$, $p=1$ oscillations are damped by $\ee^{-2\pi\tilde\Gamma}\approx 1.5\times 10^{-7}$ and only the converged smooth $p=0$ ratios are listed. Along $\bm E\parallel x$, $\max|\sigma_{xxx,\mathrm{I}}^{(2)},\sigma_{xxx,\mathrm{IIa}}^{(2)}|<4.2\times 10^{-13}$ for every entry.}
\footnotesize
\begin{ruledtabular}
\begin{tabular}{cccccccccc}
 & \multicolumn{2}{c}{Analytical NLLK} & \multicolumn{3}{c}{Warped Dirac ($n=10$)} & \multicolumn{4}{c}{Four-band BLG ($K_+$, $n=6$)} \\
\cline{2-3}\cline{4-6}\cline{7-10}
$\tilde\Gamma$ & $R_{\mathrm{wash}}$ & $|\Delta\chi_o/\Delta\chi_e|$ & $R_{\mathrm{wash}}$ & $|\Delta\chi_o/\Delta\chi_e|$ & $|\Delta\mathrm{I}/\Delta\mathrm{Tot}|$ & $R_{\mathrm{wash}}$ & $|\Delta\chi_o/\Delta\chi_e|$ & $\sigma_{\mathrm{I}}^{(0)}/\sigma^{(0)}$ & $\sigma_{\mathrm{II}}^{(0)}/\sigma^{(0)}$ \\
\hline
$0.05$ & $0.0074$ & $0.0498$ & $0.0075$ & $0.0505$ & $0.017$ & $0.0074$ & $0.0485$ & $+1.10$ & $-0.10$ \\
$0.20$ & $0.1048$ & $0.1852$ & $0.1065$ & $0.1882$ & $0.015$ & $0.1042$ & $0.1798$ & $+1.46$ & $-0.46$ \\
$0.35$ & $0.2556$ & $0.2811$ & $0.2601$ & $0.2865$ & $0.013$ & $0.2529$ & $0.2715$ & $+1.85$ & $-0.85$ \\
$0.71$ & $0.5574$ & $0.3536$ & $0.5677$ & $0.3633$ & $0.007$ & $0.5491$ & $0.3344$ & $+2.71$ & $-1.71$ \\
$1.20$ & $0.7648$ & $0.3093$ & $0.7793$ & $0.3218$ & $0.020$ & $0.7566$ & $0.2781$ & $+4.32$ & $-3.32$ \\
$2.50$ & $0.9284$ & $0.1852$ & --- & --- & --- & --- & --- & $+13.70$ & $-12.70$ \\
\end{tabular}
\end{ruledtabular}
\end{table}
\FloatBarrier

\subsection{Short-range impurity scattering in the SCBA and ensemble \texorpdfstring{$3\phi$}{3phi} protection}
\label{subsec:sm_scba_only}

To verify that our constant-$\Gamma$ findings survive realistic energy-dependent impurity broadening, we examine a static scalar white-noise disorder potential $U(\bm r)\sigma_0$ with $\langle U(\bm r)U(\bm r')\rangle_{\mathrm{imp}}=\gamma_{\mathrm{imp}}\delta^{(2)}(\bm r-\bm r')$~\cite{AndoUemura1974}. Guiding-center completeness $\sum_q\varphi_{nq}(\bm r)\varphi_{n'q}^*(\bm r)=\delta_{nn'}/(2\pi\ell_B^2)$ makes the SCBA rainbow self-energy independent of the oscillator index even when $H_w$ mixes $\Delta n=\pm 3$, giving $\Sigma^r_{nqs,n'q's'}(E)=\delta_{nn'}\delta_{qq'}[\gamma_{\mathrm{imp}}/(2\pi\ell_B^2)]\sum_{n''}G^r_{n''s,n''s'}(E)$. Under $C_{3z}$ symmetry in a two-component Dirac doublet, $\Sigma^r(E)=I_{\mathrm{osc}}\otimes[\Sigma_0^r(E)\sigma_0+\Sigma_z^r(E)\sigma_z]$ is spin-diagonal, and locality $[\eta_b,\Sigma^r]=0$ ensures the exact equal-branch Ward identity $\Lambda_b^{rr}(E,E)=v_b$.

Nonperturbatively, consider any statistically $C_{3z}$-invariant disorder ensemble ($U\circ\mathsf R_3^{-1}\stackrel{d}{=}U$). Conjugating $H_{\mathrm{dis}}(\bm E;U)=H_B(\bm\Pi)+e\bm E\cdot\bm r+U(\bm r)\sigma_0$ by full-space $U_{C_3}$ enforces $\langle j_+^{(2)}(\mathsf R_3\bm E)\rangle_{\mathrm{imp}}=\ee^{\ii 2\pi/3}\langle j_+^{(2)}(\bm E)\rangle_{\mathrm{imp}}$, which protects the pure $3\phi$ angular harmonic across all SCBA self-energies and Bethe--Salpeter vertex ladders. Direct numerical checks in main Fig.~2(b) and Fig.~\ref{fig:sm_hall_scba}(c,d) confirm close waveform agreement with projected scalar SCBA broadening [$\Sigma^r(E)=\Gamma_0\cot\{\pi[(E+\ii 0^+-\Sigma^r(E))/(\hbar|\omega_c|)+\gamma_B]\}$ with frozen six-path vertices]. They also show that the Dirac short-range spinor-overlap kernel $K_{\mathrm{imp}}(\theta,\theta')=[1+\cos(\theta-\theta')]/(2\pi)$ leaves a $3\phi$ source unmixed ($K_{\mathrm{BS};\ell,\pm 3}=0$) while splitting the harmonic relaxation times into $\tau_1=2\tau_q\neq\tau_2=\tau_3=\tau_q$.

\subsection{Semiclassical Boltzmann magnetotransport under Lorentz deflection}
\label{subsec:sm_boltzmann}

In the semiclassical regime ($k_F\ell_B\gg 1$), we treat an isolated Bloch band with bare dispersion $\varepsilon_{\bm k}$, Berry curvature $\Omega_z(\bm k)=\ii(\langle\partial_x u_{\bm k}|\partial_y u_{\bm k}\rangle-\text{c.c.})$, and orbital magnetic moment $m_{\mathrm{orb},z}(\bm k)=-\frac{\ii e}{2\hbar}(\langle\partial_x u_{\bm k}|[H(\bm k)-\varepsilon_{\bm k}]|\partial_y u_{\bm k}\rangle-\text{c.c.})$. The magnetic field modifies the band energy to $\tilde\varepsilon_{\bm k}=\varepsilon_{\bm k}-m_{\mathrm{orb},z}(\bm k)B_z+\varepsilon_{\bm k}^{\mathrm{Z}}$ (velocity $\tilde{\bm v}_{\bm k}=\hbar^{-1}\bm\nabla_{\bm k}\tilde\varepsilon_{\bm k}$) and introduces the phase-space factor $\mathcal D_{\bm k}=1+(eB_z/\hbar)\Omega_z(\bm k)$~\cite{SundaramNiu1999,XiaoChangNiu2010}. Substituting the Sundaram--Niu equations of motion,
\begin{equation*}
 \dot{\bm r}=\mathcal D_{\bm k}^{-1}\!\left(\tilde{\bm v}_{\bm k}+\frac{e}{\hbar}\bm E\times\bm\Omega_{\bm k}\right),\qquad
 \hbar\dot{\bm k}=\mathcal D_{\bm k}^{-1}\big(-e\bm E-e\tilde{\bm v}_{\bm k}\times\bm B\big),
\end{equation*}
into the constant-$\tau$ ($\tau\equiv\tau_{\mathrm{tr}}$) steady-state Boltzmann equation yields $(1+\tau\hat L_B-\tau\hat{\mathcal D}_E)f=\tilde f_0$, where $\hat{\mathcal D}_E=(e/\hbar)\mathcal D_{\bm k}^{-1}E_b\partial_b$ and $\hat L_B=-(eB_z/\hbar)\mathcal D_{\bm k}^{-1}\epsilon_{ab}\tilde v_b\partial_a$ ($\hat L_B\tilde f_0=0$). Introducing the cyclotron resolvent $\hat R\equiv(1+\tau\hat L_B)^{-1}$, the formal Zener--Jones solution reads $f_1=\hat R\tau\hat{\mathcal D}_E\tilde f_0$ at first order and
\begin{equation}
 f_2=\hat R\tau\hat{\mathcal D}_E\hat R\tau\hat{\mathcal D}_E\tilde f_0
 =\underbrace{\tau^2\hat{\mathcal D}_E^2\tilde f_0}_{f_2^{(0)}}
 \underbrace{-\tau^3\big(\hat L_B\hat{\mathcal D}_E^2+\hat{\mathcal D}_E\hat L_B\hat{\mathcal D}_E\big)\tilde f_0}_{f_2^{(1)}}+\mathcal O(\tau^4\hat L_B^2\hat{\mathcal D}_E^2\tilde f_0)\label{eq:sm_f2_lorentz_expansion}
\end{equation}
at second order. Since $C_{3z}$ eliminates the anomalous-velocity dipole (Subsec.~\ref{subsec:sm_generalized_parity}), the second-order current $j_a^{(2)}=-e\int[\dd\bm k]\,\tilde v_a f_2$ ($[\dd\bm k]\equiv\dd^2 k/(2\pi)^2$) simplifies in circular components $j_+^{(2)}=j_x^{(2)}+\ii j_y^{(2)}$ and $E_\pm=E_x\pm\ii E_y$ under $C_{3z}$ and $M_x$~\cite{Birss1964} to
\begin{equation}
 j_+^{(2)}=K(B_z)E_-^2=\big[\chi_o(B_z)+\ii\chi_e(B_z)\big](E_x-\ii E_y)^2,\qquad
 \sigma_\parallel^{(2)}(B_z,\phi)=\chi_o(B_z)\cos 3\phi+\chi_e(B_z)\sin 3\phi,\label{eq:sm_sigma_par_3phi}
\end{equation}
with $K(-B_z)=-K^*(B_z)$, so that $\chi_o(-B_z)=-\chi_o(B_z)$ and $\chi_e(-B_z)=+\chi_e(B_z)$.

Now take an isotropic parent dispersion $\varepsilon_0(k)$ with radial velocity $v_*(k)=\hbar^{-1}\varepsilon_0'(k)>0$ and curvature $\kappa\equiv k_F\varepsilon_0''(k_F)/\varepsilon_0'(k_F)$ ($\varepsilon_0'''=0$), perturbed by weak $B_z$-odd trigonal terms $\delta\varepsilon_{\bm k}=w(k)\cos 3\theta$ and $\mathcal D_{\bm k}-1=\delta_D\cos 3\theta$. Integrating $f_2^{(0)}$ and $f_2^{(1)}$ by parts at $T\to 0$ with $\beta=\omega_c\tau=eB_z\tau v_*(k_F)/(\hbar k_F)$ and $w_F\equiv w(k_F)$ gives the leading weak-field channels,
\begin{align}
 \chi_o^{(0)}(B_z)&=-\frac{e^3\tau^2}{16\pi\hbar^3}\left[k_F w_F''+5w_F'+3\kappa\frac{w_F}{k_F}-(1+2\kappa)\hbar v_*(k_F)\delta_D\right]=\mathcal O(\tau^2 B_z),\label{eq:sm_chio_explicit}\\
 \chi_e^{(1)}(B_z)&=\frac{e^3\tau^2\beta}{8\pi\hbar^3}\left[k_F w_F''+(7-2\kappa)w_F'+(2\kappa^2-7\kappa+8)\frac{w_F}{k_F}-3\hbar v_*(k_F)\delta_D\right]=\mathcal O(\tau^3 B_z^2).\label{eq:sm_chie_L1_explicit}
\end{align}
At arbitrary $\beta$, expanding $\hat R$ around the circular resolvent $\hat R_0\,\ee^{\ii m\theta}=[1+\ii m\beta(k)]^{-1}\ee^{\ii m\theta}$ ($m=\pm 1,\pm 2$) generates the universal denominator $(1+\beta^2)^2(1+4\beta^2)^2$. At matched $\tau_{\mathrm{tr}}=\tau_q$, this semiclassical result agrees identically with the intraband $p=0$ continuum limit of Eq.~\eqref{eq:sigma2_spectral_master} (Subsec.~\ref{subsec:sm_p0_reconciliation}).

\subsection{Constant-current Hall-bar conversion and inverse-field envelope decay}
\label{subsec:sm_hall_decay}

In a constant-current Hall bar ($\bm J=J\bm e_\phi$, open-circuit $\bm J^{(2)}=\bm 0$), expanding $J_a=\sigma^{(1)}_{ab}E_b+\sigma^{(2)}_{abc}E_bE_c$ and inverting yields $\rho^{(2)}_{abc}=-\rho^{(1)}_{ad}\sigma^{(2)}_{def}\rho^{(1)}_{eb}\rho^{(1)}_{fc}$. Under $C_{3z}$, the linear resistivity tensor $\bm\rho^{(1)}=(\bm\sigma^{(1)})^{-1}=\begin{psmallmatrix}r&h\\-h&r\end{psmallmatrix}$ ($r=\rho_{xx}^{(1)}>0$, $h=\rho_{xy}^{(1)}=-\rho_{yx}^{(1)}$) defines the linear Hall angle $\theta_H(B_z)\equiv\arctan(h/r)\in(-\frac{\pi}{2},\frac{\pi}{2})$. In circular coordinates, $E_-^{(1)}=(r+\ii h)J_-=\sqrt{r^2+h^2}\,\ee^{+\ii\theta_H}J_-$ and $E_+^{(2)}=-(r-\ii h)j_+^{(2)}=-\sqrt{r^2+h^2}\,\ee^{-\ii\theta_H}j_+^{(2)}$. Combining two input legs ($+2\theta_H$) with one compensating output leg ($-\theta_H$) gives a net phase shift $+\theta_H$, yielding the constant-current Hall-bar resistivity
\begin{gather}
 \rho_\parallel^{(2)}(B_z,\phi)=-\big(r^2+h^2\big)^{3/2}\Big[\chi_o(B_z)\cos\big(3\phi-\theta_H(B_z)\big)+\chi_e(B_z)\sin\big(3\phi-\theta_H(B_z)\big)\Big],\label{eq:sm_rho2_phase}\\
 \phi_{\mathrm{node}}^{\rho}(B_z)-\phi_{\mathrm{node}}^{\sigma}(B_z)=\frac{\theta_H(B_z)}{3}\pmod{\frac{\pi}{3}}.\label{eq:sm_node_shift}
\end{gather}
In weak fields ($|\beta|\ll 1$, $\sigma_0\propto\tau$), dividing by $\sigma_0^3$ yields $\rho_\parallel^{(2)}(B_z,0)\propto\tau^{-1}B_z$ and $\rho_\parallel^{(2)}(B_z,\pi/6)\propto\tau^0 B_z^2$.

Finally, consider the fundamental ($p=1$) constant-current oscillation envelope $A_2=C_\phi(\bar r^2+\bar h^2)^{3/2}|\mathcal K_1|R_T(\lambda_T)\ee^{-2\pi\tilde\Gamma}$, with angular factor $C_\phi=|s_B\tilde\Gamma\cos(3\phi-\theta_H)+(1+2\tilde\Gamma^2)\sin(3\phi-\theta_H)|$. Whenever $|V_2|\propto\ell_B^{-2}\propto\tilde\Gamma^{-1}$ (as in warped Dirac states), the prefactor $|\mathcal K_1|\propto\tilde\Gamma^3/[(1+\tilde\Gamma^2)(1+4\tilde\Gamma^2)]$ grows algebraically with $1/|B_z|$ throughout $\tilde\Gamma<1.30$. The local inverse-field decay rate $\Lambda_B^{(2)}\equiv -\dd\ln A_2/\dd(1/|B_z|)$ therefore satisfies
\begin{gather}
 \Lambda_B^{(2)}=2\pi\frac{\Gamma m_c}{\hbar e}-\frac{\dd\ln R_T(\lambda_T)}{\dd(1/|B_z|)}-\frac{\dd\ln|\mathcal K_1|}{\dd(1/|B_z|)}-\frac{3}{2}\frac{\dd\ln(\bar r^2+\bar h^2)}{\dd(1/|B_z|)}-\frac{\dd\ln C_\phi}{\dd(1/|B_z|)},\label{eq:sm_Lambda2_loc}\\
 \frac{\dd\ln|\mathcal K_1|}{\dd(1/|B_z|)}=|B_z|\,\frac{3+5\tilde\Gamma^2-4\tilde\Gamma^4}{(1+\tilde\Gamma^2)(1+4\tilde\Gamma^2)},\qquad
 \frac{\dd\ln R_T(\lambda_T)}{\dd(1/|B_z|)}=|B_z|\big(1-\lambda_T\coth\lambda_T\big).\label{eq:sm_local_decay}
\end{gather}
Subtracting $\dd\ln|\mathcal K_1|/\dd(1/|B_z|)>0$ (up to $+3|B_z|$ at $\tilde\Gamma\ll 1$) from the Dingle slope $2\pi\Gamma m_c/(\hbar e)$ lowers $\Lambda_B^{(2)}$ substantially below the linear SdH slope $\Lambda_B^{(1)}$, even after partial restoration by $(\bar r^2+\bar h^2)^{3/2}$ in parallel-shunted films.

\section{Microscopic instantiation I: Hexagonally warped topological surface states}
\label{sec:sm_ti_model}

We instantiate the framework of Secs.~\ref{sec:sm_negf}--\ref{sec:sm_nllk_hall} for the $C_{3v}$-symmetric topological Dirac surface state~\cite{Fu2009}:
\begin{equation}
 H_B(\bm\Pi)=\frac{\bm\Pi^2}{2m^*}\sigma_0+v_F\big(\Pi_x\sigma_y-\Pi_y\sigma_x\big)+\left[\frac{\lambda_w}{2\hbar^3}\big(\Pi_+^3+\Pi_-^3\big)+\Delta_Z\right]\sigma_z,\label{eq:sm_HB_ti_unified}
\end{equation}
with $\Delta_Z=g_z\mu_B B_z/2$. As proved in Subsec.~\ref{subsec:sm_generalized_parity}, $\lambda_w=0$ preserves $\tilde{\mathcal P}=\sigma_z(-1)^{\hat n}$ and $v_F=\Delta_Z=0$ preserves $\sigma_x(-1)^{\hat n}$; combining $v_F\neq 0$ with $\lambda_w\neq 0$ breaks every generalized parity $\tilde{\mathcal P}$ even at $\Delta_Z=0$.

\subsection{Semiclassical Boltzmann background for the warped Dirac cone}
\label{subsec:sm_ti_instantiation}

\begin{figure}[!htbp]
 \centering
 \SupplementFigure{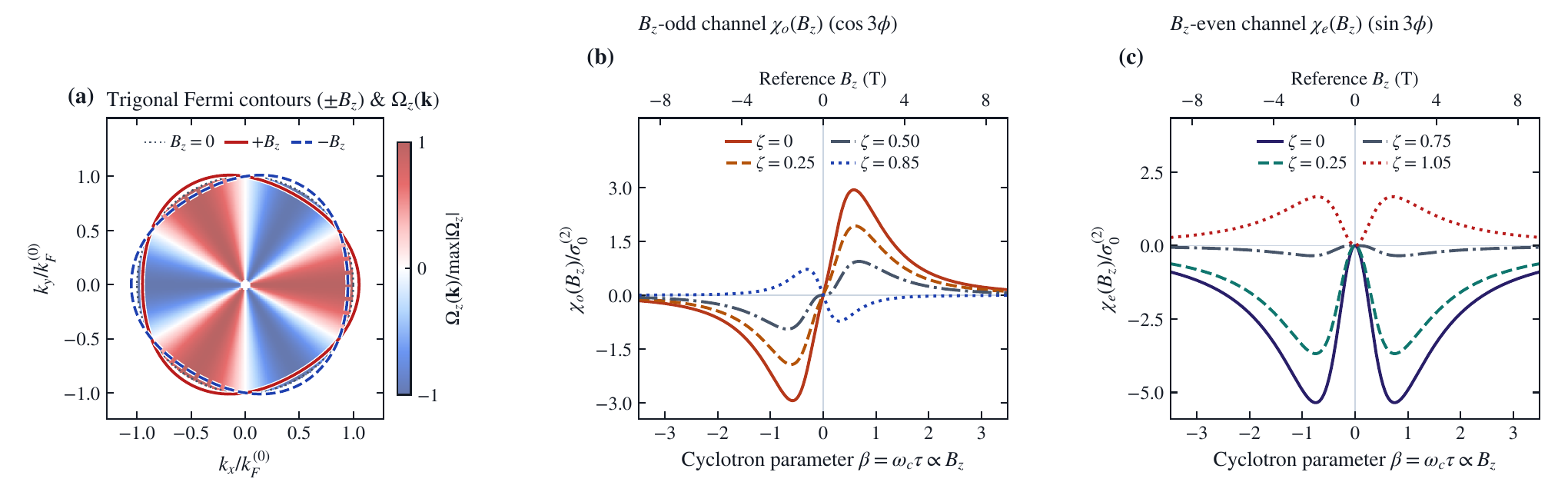}
 \caption{Semiclassical Boltzmann background for the warped Dirac surface state~\eqref{eq:sm_HB_ti_unified} supporting main Fig.~2(a) (parameters in Table~\ref{tab:sm_ti_params}). (a)~Field-skewed Fermi contours $\tilde\varepsilon_{\bm k}=\mu$ for $\pm B_z$ and $B_z=0$ over normalized Berry curvature $\Omega_{+,z}(\bm k)$. (b),(c)~Linear-cone ($\kappa=0$) bipolar curves of $\chi_o/\sigma_0^{(2)}$ and $\chi_e/\sigma_0^{(2)}$ vs.\ $\beta=\omega_c\tau$ from Eq.~\eqref{eq:sm_linear_dirac_bte} across $\zeta$, bracketing the weak-field sign-reversal thresholds ($\zeta=1/2,3/4$).}
 \label{fig:sm_boltzmann}
\end{figure}

In the semiclassical limit $\bm\Pi\to\hbar\bm k$, $H(\bm k)=ak^2\sigma_0+\bm d(\bm k)\cdot\bm\sigma$ with $a\equiv\hbar^2/(2m^*)$, $b\equiv\hbar v_F$, $\bm d=(-bk_y,bk_x,M(\bm k))$, $M(\bm k)=\lambda_w k^3\cos 3\theta+\Delta_Z$, and $\varepsilon_s(\bm k)=ak^2+s\epsilon_{\bm k}$ ($\epsilon_{\bm k}=\sqrt{b^2k^2+M^2}$). Here $v_*(k)=(b+2ak)/\hbar$ and $\kappa=2ak_F/(b+2ak_F)\in[0,1)$. The conduction-band Berry curvature and orbital magnetic moment read
\begin{equation}
 \Omega_{+,z}(\bm k)=\frac{b^2\big(2\lambda_w k^3\cos 3\theta-\Delta_Z\big)}{2\epsilon_{\bm k}^3},\qquad
 m_{\mathrm{orb},z}^{(+)}(\bm k)=\frac{e\epsilon_{\bm k}}{\hbar}\Omega_{+,z}(\bm k)=\frac{eb^2\big(2\lambda_w k^3\cos 3\theta-\Delta_Z\big)}{2\hbar\epsilon_{\bm k}^2}.\label{eq:sm_berry_moment}
\end{equation}
To bilinear order $\mathcal O(\lambda_w B_z)$, $\tilde\varepsilon_{\bm k}=\varepsilon_0(k)+w(k)\cos 3\theta$ and $\mathcal D_{\bm k}=1+\delta_D\cos 3\theta$ with $w(k)=-\frac{e\lambda_w B_z}{\hbar}k+\frac{\lambda_w\Delta_Z}{b}k^2$, $\delta_D=\frac{e\lambda_w B_z}{\hbar b}$, and $\zeta\equiv\frac{g_z m_c}{4m_e(1-\kappa)}$. Solving Eq.~\eqref{eq:sm_f2_lorentz_expansion} to all orders in $\beta=\omega_c\tau$ in the linear-Dirac limit ($\kappa=0$, $\sigma_0^{(2)}\equiv e^3\tau\lambda_w k_F/[16\pi\hbar^3 v_F]$, $Q_\beta\equiv(1+\beta^2)^2(1+4\beta^2)^2$) yields the smooth background curves plotted in main Fig.~2(a) and Fig.~\ref{fig:sm_boltzmann}:
\begin{align}
 \frac{\chi_o(B_z)}{\sigma_0^{(2)}}&=\frac{6\beta\big[(1-2\zeta)+(15-20\zeta)\beta^2+(20-24\zeta)\beta^4\big]}{Q_\beta},\nonumber\\
 \frac{\chi_e(B_z)}{\sigma_0^{(2)}}&=-\frac{12\beta^2\big[(3-4\zeta)+(22-27\zeta)\beta^2+(16-20\zeta)\beta^4\big]}{Q_\beta}.\label{eq:sm_linear_dirac_bte}
\end{align}
Note that the weak-field background channels invert sign at $\zeta=1/2$ ($\chi_o$) and $\zeta=3/4$ ($\chi_e$), distinct from the Landau-level NLLK threshold $\zeta=1$ [Fig.~\ref{fig:sm_nllk_zeeman}(d)].

\subsection{Landau-quantized regime: triangular spinors, \texorpdfstring{$V_1,V_2$}{V1, V2}, and paired-doublet cutoff}
\label{subsec:sm_ti_landau}

For $B_z>0$ ($s_B=+1$; $-B_z$ follows from $\Theta$), defining $\hbar\omega_D\equiv\sqrt2\hbar v_F/\ell_B$, $\Omega_M\equiv\hbar eB_z/m^*$, $w_B\equiv\sqrt2\lambda_w/\ell_B^3$, and $\delta_B\equiv\Delta_Z-\frac{1}{2}\Omega_M$, Eq.~\eqref{eq:sm_HB_ti_unified} splits into $H_B=H_0+H_w$:
\begin{equation}
 H_0=\hbar\omega_D\begin{pmatrix}0&-\ii a_B\\ \ii a_B^\dagger&0\end{pmatrix}+\Omega_M\!\left(a_B^\dagger a_B+\frac{1}{2}\right)\sigma_0+\Delta_Z\sigma_z,\qquad
 H_w=w_B\big(a_B^3+a_B^{\dagger 3}\big)\sigma_z.\label{eq:sm_H0_Hw_ti}
\end{equation}
The unwarped Hamiltonian $H_0$ closes within each Dirac doublet $\{-\ii|n-1,\uparrow\rangle,|n,\downarrow\rangle\}$ ($n\ge 1$), giving $E_{n,s}^{(0)}=n\Omega_M+s\epsilon_n$ ($\epsilon_n=\sqrt{n(\hbar\omega_D)^2+\delta_B^2}$) and $|n,s\rangle=-\ii u_{n,s}|n-1,\uparrow\rangle+v_{n,s}|n,\downarrow\rangle$ with $u_{n,s}=\sqrt{(1+s\delta_B/\epsilon_n)/2}$, $v_{n,s}=s\sqrt{(1-s\delta_B/\epsilon_n)/2}$, plus the unpaired $n=0$ mode $|0,\downarrow\rangle$ at $E_0^{(0)}=-\delta_B$.

\textit{Paired Dirac doublet truncation for finite-matrix numerical evaluation.}---When diagonalizing Eq.~\eqref{eq:sm_H0_Hw_ti} on a finite oscillator cutoff $0\le\nu\le N_{\mathrm{cut}}-1$, the rectangular basis $\{|\nu,\uparrow\rangle,|\nu,\downarrow\rangle\}_{\nu=0}^{N_{\mathrm{cut}}-1}$ leaves one unpaired top-edge orphan state ($|N_{\mathrm{cut}}-1,\uparrow\rangle$ for $B_z>0$, or $|N_{\mathrm{cut}}-1,\downarrow\rangle$ for $B_z<0$) whose missing Dirac partner breaks the continuum commutator $[\eta_\parallel,H_B]=\ii\hbar\partial_{\Pi_\parallel}H_B$ at the UV edge. Dropping that single orphan state so that the truncated space consists solely of $|0,\downarrow\rangle$ (for $B_z>0$, or $|0,\uparrow\rangle$ for $B_z<0$) plus $N_{\mathrm{cut}}-1$ complete doublets $\{|n-1,\uparrow\rangle,|n,\downarrow\rangle\}_{n=1}^{N_{\mathrm{cut}}-1}$ removes $H_0$ edge-mismatch artifacts, yielding exact finite-$N_{\mathrm{cut}}$ machine-precision equivalence between length-gauge [Eq.~\eqref{eq:sigma2_spectral_master}] and commutator-vertex velocity-gauge [Eq.~\eqref{eq:sm_sigma2_velocity_reduced}] traces, and $\mathcal O(1/N_{\mathrm{cut}})$ convergence when physical momentum-derivative vertices are used (as $H_w\propto a_B^3+a_B^{\dagger 3}$ still crosses the UV boundary).

\begin{figure}[!htbp]
 \centering
 \SupplementFigure{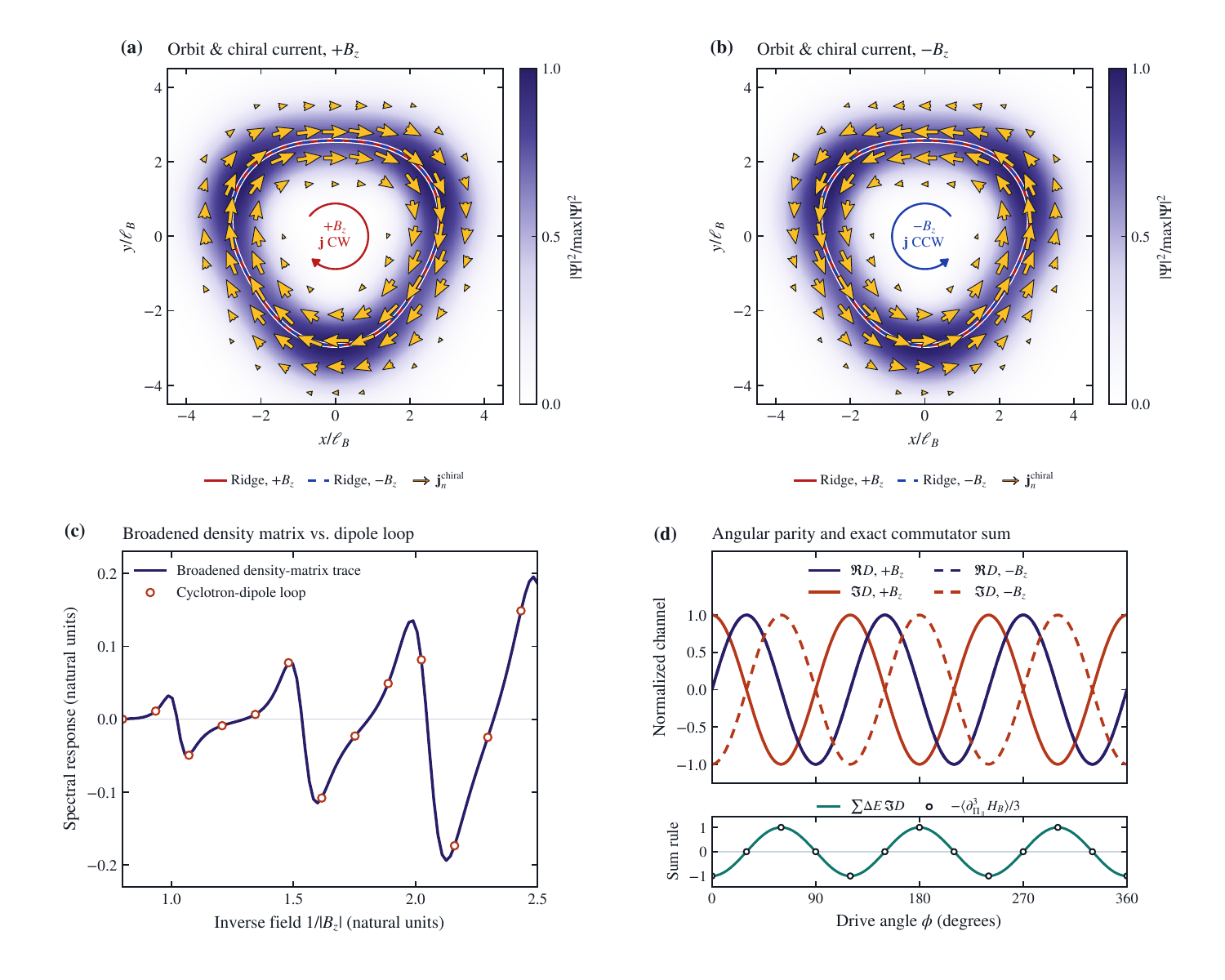}
 \caption{Real-space triangular Dirac Landau eigenstates, chiral charge current, dipole-loop verification, and magnetic-field parity (parameters in Table~\ref{tab:sm_ti_params}). (a),(b)~Normalized symmetric-gauge densities $|\Psi_{n,s}(\bm r)|^2/\max|\Psi_{n,s}|^2$ of $n=4,s=+1$ ($m=0$) at $\lambda_w=+0.025$ for (a)~$+B_z$ and (b)~$-B_z$, showing identical $B_z$-even triangular density ridges (solid red $+B_z$, dashed blue $-B_z$) encircled by the $B_z$-odd conventional charge current $\bm j_n^{\mathrm{chiral}}(\bm r)=-e\Re[\Psi_n^\dagger(\bm v\Psi_n)]$ (gold arrows; clockwise for $+B_z$ and counterclockwise for $-B_z$, opposite to the electron cyclotron velocity in main Fig.~1(c) owing to electron charge $-e$; reversing $\lambda_w\to-\lambda_w$ inverts $\bm r\to-\bm r$). (c)~Exact $T=0$ agreement between Eq.~\eqref{eq:sm_sigma2_Q} and Eq.~\eqref{eq:dipole-loop} at $\lambda_w=0.006,\Delta_Z=0.15 B_z$. (d)~Angular $B_z$ parities of both channels and the sum rule~\eqref{eq:sum-rule}.}
 \label{fig:sm_dirac_orbit}
\end{figure}

In first-order perturbation theory in $H_w$, $W_{n+3,n}^{s's}\equiv\langle n+3,s'|H_w|n,s\rangle=w_B[u_{n+3,s'}u_{n,s}\sqrt{n(n+1)(n+2)}-v_{n+3,s'}v_{n,s}\sqrt{(n+1)(n+2)(n+3)}]$ mixes $|n,s\rangle$ with $|n\pm 3,s'\rangle$ ($c_{\delta s'}\equiv W_{n+\delta,n}^{s's}/(E_{n,s}^{(0)}-E_{n+\delta,s'}^{(0)})$). Expanding in symmetric-gauge orbitals~\eqref{eq:sm_phi_osc} yields the triangular density and velocity-current profiles of Eq.~\eqref{eq:sm_polar_current_decomp} [verified in Fig.~\ref{fig:sm_dirac_orbit} alongside the exact sum rule $\sum(E_\beta-E_\gamma)\Im D_{\alpha\beta\gamma}=-2\lambda_w\cos 3\phi\,\langle\alpha|\sigma_z|\alpha\rangle$]:
\begin{equation}
 |\Psi_{n,s}(\bm r)|^2=\rho_{n,s}^{(0)}(r)+2\sum_{\delta=\pm 3,\,s'=\pm 1}c_{\delta s'}\big(u_{n,s}u_{n+\delta,s'}R_{n-1}R_{n+\delta-1}+v_{n,s}v_{n+\delta,s'}R_n R_{n+\delta}\big)\sin 3\theta_{\bm r}+\mathcal O(\lambda_w^2).\label{eq:triangular-density}
\end{equation}

\begin{figure}[!htbp]
 \centering
 \SupplementFigure{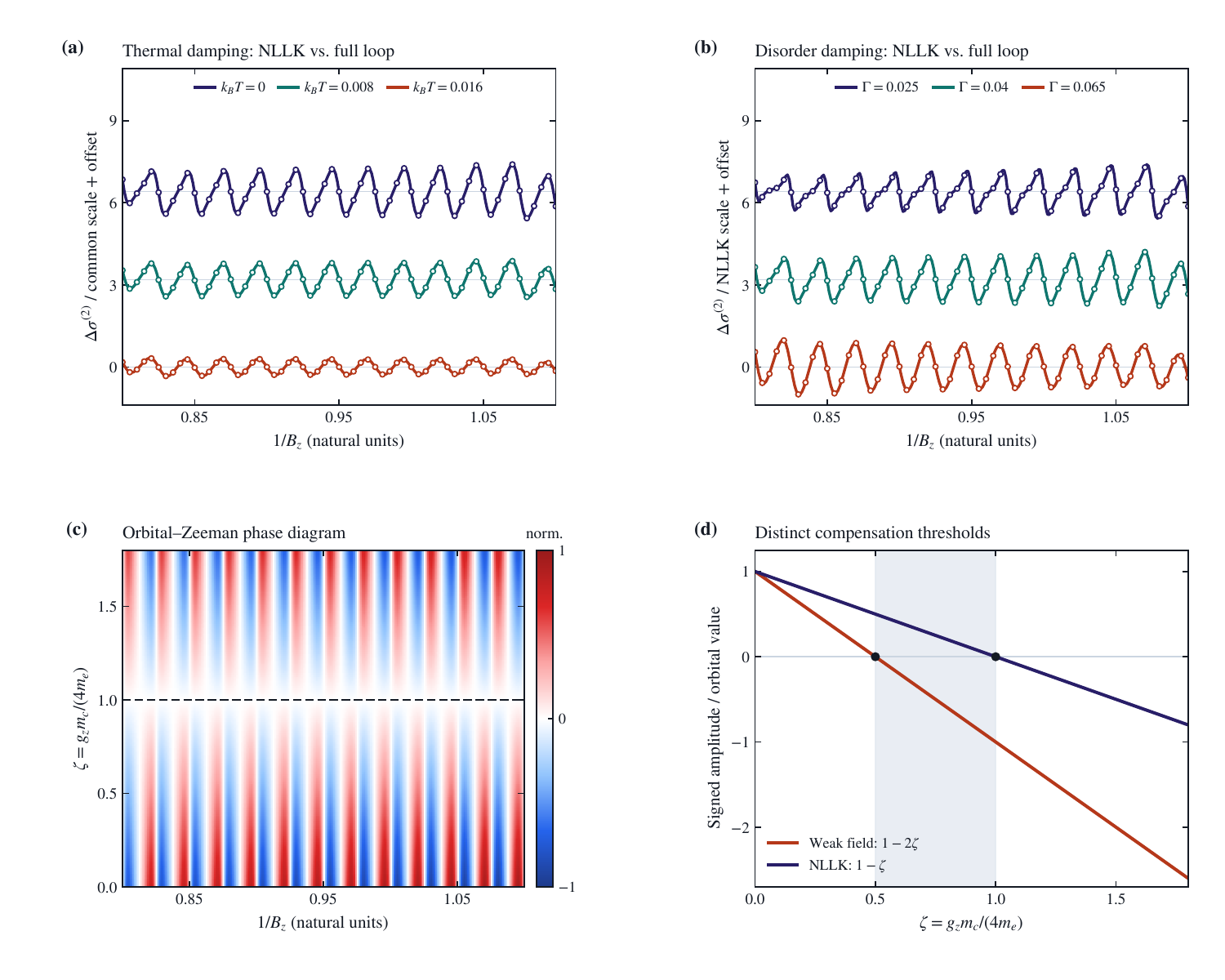}
 \caption{\label{fig:sm_nllk_zeeman}NLLK benchmark and analytical Zeeman phase diagram for a linear Dirac cone ($\kappa=0$; parameters in Table~\ref{tab:sm_ti_params}). (a),(b)~Parameter-free comparison of Eq.~\eqref{eq:sm_signed_conditional_nllk} (solid) against the full two-band loop [Eq.~\eqref{eq:sigma2_spectral_master}, circles] at $g_z=0$ vs.\ temperature ($k_BT=0,0.008,0.016$; $\Gamma=0.04$) and linewidth ($\Gamma=0.025,0.04,0.065$; $T=0$). (c)~Analytical normalized phase map $\Delta\sigma_\parallel^{(2)}(1/B_z,\zeta)\propto(1-\zeta)$ from Eq.~\eqref{eq:sm_signed_conditional_nllk} and $V_2\propto(1-\zeta)$ showing $180^\circ$ phase inversion across $\zeta=1$ (dashed). (d)~Analytical weak-field background factor $1-2\zeta$ vs.\ NLLK factor $1-\zeta$ (opposite-sign window $1/2<\zeta<1$ shaded).}
\end{figure}

Decomposing $v_\parallel(\phi)=v_{0,\parallel}(\phi)+v_{w,\parallel}(\phi)$ with $v_{D,\parallel}=-\ii v_F(\ee^{-\ii\phi}\sigma_+-\ee^{\ii\phi}\sigma_-)$ and $v_{w,\parallel}=[3\lambda_w/(\hbar\ell_B^2)](\ee^{\ii\phi}a_B^{\dagger 2}+\ee^{-\ii\phi}a_B^2)\sigma_z$, large-$n$ projection onto the conduction band ($n\simeq n_0=k_F^2\ell_B^2/2\gg 1$, $\delta_B/\epsilon_n=[2(1-\kappa)\zeta-\kappa]/[4(1-\kappa)n]$) yields $V_1=v_*(k_F)/2=v_F/[2(1-\kappa)]$ and $V_2=-[\lambda_w/(\hbar\ell_B^2)]\mathcal G(\kappa,\zeta)$, combining bare warping ($\mathcal G_{\mathrm{dir}}$) with intraband ($\mathcal G_{\mathrm{intra}}$) and interband ($\mathcal G_{\mathrm{inter}}$) mixing:
\begin{align}
 \mathcal G_{\mathrm{dir}}&=\frac{6-3\kappa}{4(1-\kappa)}-\frac{3}{2}\zeta,\qquad
 \mathcal G_{\mathrm{intra}}=-\frac{6+3\kappa-4\kappa^2}{24(1-\kappa)}+\frac{3+2\kappa}{12}\zeta,\qquad
 \mathcal G_{\mathrm{inter}}=-\frac{2+\kappa}{8(1-\kappa)}+\frac{1}{4}\zeta,\nonumber\\
 \mathcal G(\kappa,\zeta)&=\mathcal G_{\mathrm{dir}}+\mathcal G_{\mathrm{intra}}+\mathcal G_{\mathrm{inter}}=1-\zeta+\frac{\kappa^2+\kappa(1-\kappa)\zeta}{6(1-\kappa)}.\label{eq:sm_finite_curvature_G}
\end{align}
At $\kappa=0$, $\mathcal G(0,\zeta)=1-\zeta$ gives $V_1=v_F/2$ and $V_2=-(\lambda_w/\hbar\ell_B^2)(1-\zeta)$ [benchmarked without fit parameters against the full two-band loop at $g_z=0$ in Fig.~\ref{fig:sm_nllk_zeeman}(a,b), with the resulting analytical $\zeta$ phase map and thresholds plotted in Fig.~\ref{fig:sm_nllk_zeeman}(c,d)].

\subsection{Intraband \texorpdfstring{$p=0$}{p=0} continuum limit: Boltzmann correspondence and multiband boundaries}
\label{subsec:sm_p0_reconciliation}

Retaining the slow $\mathcal O(n_0^{-1})$ variations $E_{n_F+x}-\mu=\hbar|\omega_c|[x+\frac{c_\omega}{2n_0}x^2+\cdots]$ ($c_\omega\equiv\dd\ln(\hbar|\omega_c|)/\dd\ln n|_{n_F}$) and $c_{\mathrm{amp}}\equiv\dd\ln|V_1^2 V_2|/\dd\ln n|_{n_F}$ inside the $p=0$ continuum integral of the six intraband paths~\eqref{eq:sm_six_paths} at matched $\tau\equiv\tau_{\mathrm{tr}}=\tau_q=\hbar/(2\Gamma)$ ($|\beta|=1/(2\tilde\Gamma)$) yields closed expressions for both smooth channels:
\begin{align}
 \chi_o^{(p=0)}(B_z)&=-\frac{3s_B\,e^3\hbar^2V_1^2V_2}{2\pi\ell_B^2(\hbar\omega_c)^4n_0}\left[\frac{\beta^2\,c_{\mathrm{amp}}}{(1+\beta^2)(1+4\beta^2)}-\frac{\beta^2(1+15\beta^2+20\beta^4)}{(1+\beta^2)^2(1+4\beta^2)^2}\,c_\omega\right],\label{eq:sm_p0_chio}\\
 \chi_e^{(p=0)}(B_z)&=+\frac{6e^3\hbar^2V_1^2V_2}{2\pi\ell_B^2(\hbar\omega_c)^4n_0}\left[\frac{|\beta|^3\,c_{\mathrm{amp}}}{(1+\beta^2)(1+4\beta^2)}-\frac{|\beta|^3(3+22\beta^2+16\beta^4)}{(1+\beta^2)^2(1+4\beta^2)^2}\,c_\omega\right].\label{eq:sm_p0_chie}
\end{align}
Inserting Eq.~\eqref{eq:sm_finite_curvature_G} at $\kappa=0$ (accounting for $\dd\zeta/\dd\ln n=\zeta/2$) reproduces the linear-Dirac Boltzmann expressions~\eqref{eq:sm_linear_dirac_bte} identically with zero symbolic residual (shifting the smooth weak-field compensation zeros to $\zeta=1/2$ for $\chi_o$ and $3/4$ for $\chi_e$, Fig.~\ref{fig:sm_nllk_zeeman}(d)), while Eqs.~\eqref{eq:sm_p0_I}--\eqref{eq:sm_p0_II} resolve how Term~I, Term~II-a, and Term~II-b combine to form Eq.~\eqref{eq:sm_p0_chie}. Beyond the projected intraband $p=0$ loop, the full multiband trace~\eqref{eq:sm_sigma2_velocity_reduced} retains a small $\Gamma$-odd interband residual $\chi_e^{\mathrm{Full}}(0,\Gamma)\simeq -3e^3\lambda_w\Gamma/(4\pi\hbar^3 v_F^2\mu)$ at $B_z=0$ ($0.28\%$ of peak $|\chi_e|$ in Fig.~2(a), Lorentz-suppressed at finite $B_z$).

\subsection{Experimental analysis protocol and \texorpdfstring{$\alpha\text{-Sn}(111)$}{alpha-Sn(111)} comparisons}
\label{subsec:sm_alpha_sn_exp}

For a Hall bar of length $L$ and width $W$ driven by $I(t)=I_0\sin(\omega t)=\sqrt2\,I_{\mathrm{rms}}\sin(\omega t)$, the $1\omega$ resistances give $r=(W/L)R_{xx}^{(1)}$, $h=R_{xy}^{(1)}$, and $\theta_H=\arctan(h/r)$ (set $h=-R_{yx}$ if measuring $E_y/J_x$), while the $2\omega$ resistance obeys $R_{2\omega}=\mathcal C_{2\omega}\rho_\parallel^{(2)}$ with $\mathcal C_{2\omega}=LI_0/(2W^2)=LI_{\mathrm{rms}}/(\sqrt2 W^2)$. Analysis proceeds in three steps: (1)~extract $(F,m_c,\tau_q)$ and $(r,h,\theta_H)$ from $1\omega$ SdH/Hall traces; (2)~jointly fit background-subtracted $\Delta R_{2\omega}(B_z,\phi)$ across all angles via Eqs.~\eqref{eq:sm_signed_conditional_nllk} and~\eqref{eq:sm_rho2_phase} with zero per-angle parameters; and (3)~extract inverse-field envelope slopes $\Lambda_B^{(1,2)}$ on a common window via joint polynomial-plus-harmonic detrending ($d=3,4$, sliding window $w_{\mathrm{cyc}}\simeq 1.6\text{--}2.2$ cycles).

\begin{table}[!htbp]
\caption{\label{tab:sm_ti_params}Parameter sets for warped Dirac benchmarks and $\alpha\text{-Sn}(111)$ comparisons across main Fig.~2 and Figs.~\ref{fig:sm_boltzmann}--\ref{fig:sm_hall_scba}.}
\footnotesize
\begin{ruledtabular}
\begin{tabular}{p{0.15\textwidth}p{0.13\textwidth}p{0.27\textwidth}p{0.23\textwidth}p{0.15\textwidth}}
Figure / panels & Role & Model \& band parameters & Field / thermal / damping & Linear Hall / shunt \\
\hline
Fig.~\ref{fig:sm_boltzmann}(a) & Schematic contour & $e=\hbar=b=k_F^{(0)}=1$, $a=0.12$ ($\kappa\simeq 0.194$), $\lambda_w=0.065$ & $\mu=1.12$, $|B_z|=1$, $g_z=0$ & Intrinsic $\sigma^{(2)}$ \\
Fig.~\ref{fig:sm_boltzmann}(b,c) & $\kappa=0$ analytical & $\kappa=0$; $\zeta\in\{0,0.25,0.50,0.75,0.85,1.05\}$ & Top axis: $m_c=0.210\,m_e$, $\tau_0=0.46\;\mathrm{ps}$ & Intrinsic $\sigma^{(2)}$ \\
Fig.~\ref{fig:sm_dirac_orbit}(a,b) & Exact orbit & $e=\hbar=v_F=1$, $m^*=6$, $\lambda_w=+0.025$ & $n=4,s=+1$, $|B_z|=1$, $\Delta_Z=-2B_z$ & Pure eigenstate \\
Fig.~\ref{fig:sm_dirac_orbit}(c,d) & Loop \& sum rule & $e=\hbar=v_F=1$, $m^*=6$, $\lambda_w=0.006$ & $E=2.3$, $\Gamma=0.08$, $\Delta_Z=0.15B_z$, $T=0$ & Intrinsic $\sigma^{(2)}$ \\
Fig.~\ref{fig:sm_nllk_zeeman}(a--d) & NLLK benchmark & $e=\hbar=v_F=1$, $1/m^*=0$, $\lambda_w=2\times 10^{-5}$ & $\mu=\sqrt{80}$ ($F=40$), $\Gamma=0.025\text{--}0.065$ & Intrinsic $\sigma^{(2)}$ \\
Main Fig.~4 (End Matter) & Real-space $\delta\rho_{\mathrm{inj,sym}}^{(2)},\delta\rho_{\mathcal F}^{(2)}$ & $e=\hbar=v_F=1$, $1/m^*=0$, $\lambda_w=+0.042$ & $n=4$, $|B_z|=1$, $\tilde\Gamma=0.06,0.35,1.20$ & Intrinsic $\sigma^{(2)}$ \\
Main Fig.~2(a,b) & BTE / NLLK / SCBA & $E_F=31.53\;\mathrm{meV}$, $v_F=1.625\times 10^5\;\mathrm{m/s}$, $\lambda_w=250\;\mathrm{eV\,\text{\AA}^3}$ & $F=28.1\;\mathrm{T}$ (warped), $T=5\;\mathrm{K}$, $\Gamma=0.715\;\mathrm{meV}$, $g_z=0$ & Intrinsic $\sigma^{(2)}$ \\
Fig.~\ref{fig:sm_hall_scba}(a) & Generic $\theta_H/3$ demo & $F=28.6\;\mathrm{T}$, $m_c=0.126\,m_e$, $2\pi\Gamma m_c/(\hbar e)=10.22\;\mathrm{T}$ & $T=2\;\mathrm{K}$, Eq.~(7) $p=1$ NLLK & Generic high-$\mu$ 2-carrier \\
Fig.~\ref{fig:sm_hall_scba}(c,d) & Proj.\ SCBA \& BS & $F=28.6\;\mathrm{T}$, $\Gamma_0=0.72\;\mathrm{meV}$, $\phi=30^\circ$ & $T=0$; kernel $[1+\cos(\theta-\theta')]/(2\pi)$ & Intrinsic $\sigma^{(2)}$ \\
Main Fig.~2(c) & $\alpha$-Sn angular fit & Exp.\ $1\omega$: $m_c=0.23\,m_e$, $\tau_q=0.46\;\mathrm{ps}$ ($T_D=2.64\;\mathrm{K}$)~\cite{Kalappattil2026} & $T=5\;\mathrm{K}$; fitted $F=19.13\;\mathrm{T}$, $\gamma_B=0.083$ & Rep.\ comp.\ Hall ($R_0=352\,\Omega$) \\
Main Fig.~2(d) & $\alpha$-Sn decay ratio & Bare cone: $E_F=31.53\;\mathrm{meV}$, $m_c=0.21\,m_e$, $\lambda_w=250\;\mathrm{eV\,\text{\AA}^3}$ & $F=28.6\;\mathrm{T}$, $T=5\;\mathrm{K}$, $\Gamma=0.72\;\mathrm{meV}$ & Shunt $\sigma_{b0}\simeq 3.5\bar\sigma_{xx}$ \\
\end{tabular}
\end{ruledtabular}
\end{table}

\begin{figure}[!htbp]
\centering
\SupplementFigure{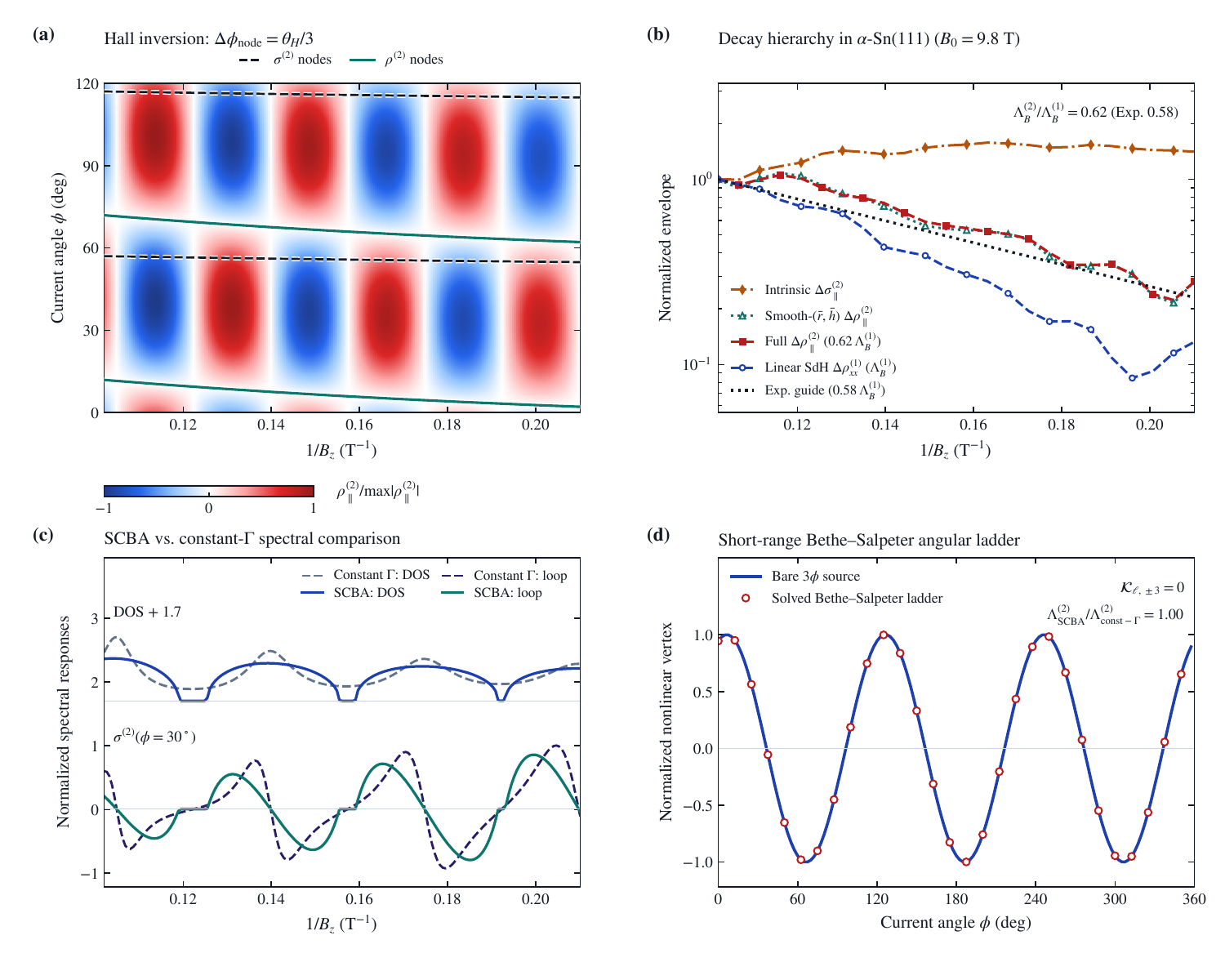}
\caption{Hall transformation, inverse-field decay hierarchy in $\alpha\text{-Sn}(111)$, and short-range impurity checks (parameters in Table~\ref{tab:sm_ti_params}). (a)~Generic high-mobility two-carrier Hall-bar illustration ($n_1=2.99\times 10^{17}\;\mathrm{m}^{-2}$, $\mu_1=0.1198\;\mathrm{m^2\,V^{-1}\,s^{-1}}$; $n_2=2.31\times 10^{16}\;\mathrm{m}^{-2}$, $\mu_2=0.3813\;\mathrm{m^2\,V^{-1}\,s^{-1}}$; distinct from the $\alpha$-Sn fits below) demonstrating $\Delta\phi_{\mathrm{node}}=\theta_H/3$ for $p=1$ Eq.~(7). (b)~Semilogarithmic $1/B_z$ decay hierarchy in $\alpha\text{-Sn}(111)$: intrinsic $\Delta\sigma_\parallel^{(2)}$ (amber), smooth-$(\bar r,\bar h)$-converted $\Delta\rho_\parallel^{(2)}$ envelope (teal, $0.63\,\Lambda_B^{(1)}$), full $\Delta\rho_\parallel^{(2)}$ (red, $\Lambda_B^{(2)}/\Lambda_B^{(1)}=0.62$), linear SdH (blue), and experimental ratio guide ($0.58\,\Lambda_B^{(1)}$, dotted). (c)~Projected scalar SCBA with frozen six-path vertices (solid) vs.\ constant-$\Gamma$ broadening (dashed) at $\phi=30^\circ, T=0$. (d)~Dirac short-range spinor-overlap kernel confirming $K_{\mathrm{BS};\ell,\pm 3}=0$.}
\label{fig:sm_hall_scba}
\end{figure}

For $\alpha\text{-Sn}(111)/\mathrm{Si}(111)$ Hall bars ($L=100\,\mu\mathrm{m}$, $W=40\,\mu\mathrm{m}$, $T=5\;\mathrm{K}$)~\cite{Kalappattil2026}, we adopt a representative compensated Hall-bar model with $\sigma_{xx}^{(1)}=\sigma_{\mathrm{sh}}+n_be\mu_b/[1+(\mu_b B_z)^2]+n_se\mu_s/[1+(\mu_s B_z)^2]$ and $\sigma_{xy}^{(1)}=n_be\mu_b^2 B_z/[1+(\mu_b B_z)^2]-n_se\mu_s^2 B_z/[1+(\mu_s B_z)^2]$ ($\sigma_{\mathrm{sh}}^{-1}=210\,\Omega$, $n_b=1.15\times 10^{13}\;\mathrm{cm}^{-2}$, $\mu_b=850\;\mathrm{cm^2\,V^{-1}\,s^{-1}}$, $n_s=2.31\times 10^{12}\;\mathrm{cm}^{-2}$, $\mu_s=2100\;\mathrm{cm^2\,V^{-1}\,s^{-1}}$; inverted via $h=-\sigma_{xy}^{(1)}/[(\sigma_{xx}^{(1)})^2+(\sigma_{xy}^{(1)})^2]$), matching $R_0=352\,\Omega$ with $|\theta_H|\le 5.6^\circ$ up to $14\;\mathrm{T}$ ($4.6^\circ$ up to $10\;\mathrm{T}$); note that these effective high-field compensated parameters differ from the low-field ($B_z\le 2\;\mathrm{T}$) unshunted two-carrier fit in Table~2 of Ref.~\cite{Kalappattil2026} SM. Dividing the theoretical $\Delta\rho_\parallel^{(2)}$ by an empirical smooth background envelope $R_{2\omega,\mathrm{bg}}(B_z)=-3.6+5.96 B_z-0.27 B_z^2\;\Omega$ and fitting $(F=19.13\;\mathrm{T},\gamma_B=0.083,C_{\mathrm{cal}}=1.34\times 10^4\;\mathrm{A\,m^{-1}})$ ($p_{\max}=3$, joint $R^2=0.757$; per-angle RMSEs $0.0153$ at $30^\circ$, $0.0071$ at $60^\circ$, and $0.0109$ at $90^\circ$, where the suppressed $60^\circ$ trace has standard deviation $0.0018$ near the noise floor; note that $C_{\mathrm{cal}}$ absorbs sample-specific geometry, shunting, and scattering-model amplitude factors, so Fig.~2(c) tests the normalized angular and phase waveform rather than absolute second-harmonic magnitude) under the standard $C_{3v}$ mirror-axis assignment ($\bar\Gamma\bar M$ at $\phi=30^\circ$, $\bar\Gamma\bar K$ at $60^\circ$, $\bar\Gamma\bar M'$ at $90^\circ$) captures the antiphase $B_z>0$ $\bar\Gamma\bar M$ / $\bar\Gamma\bar M'$ oscillations and strongly suppressed $\bar\Gamma\bar K$ response [Fig.~2(c)]. (For background sensitivity, note that Ref.~\cite{Kalappattil2026} SM p.~14 prints $-0.7\,B_z^2$; replacing $-0.27\,B_z^2$ by a linear or weak-quadratic $-0.007\,B_z^2$ background shifts the fit modestly to $F\simeq 19.28\;\mathrm{T},\gamma_B\simeq 0.104,R^2\simeq 0.712$ without altering the angular antiphase structure.) Because $F=19.13\;\mathrm{T}$ differs from the $1\omega$ SdH frequency ($\approx 28\;\mathrm{T}$, which remains unexplained within the single-pocket model), $\gamma_B$ is fitted freely, and Ref.~\cite{Kalappattil2026} recorded $\Delta R_{2\omega}$ only for $B_z>0$, reversing $B_z\to -B_z$ and measuring $1\omega/2\omega$ at a shared frequency provide direct tests of $B_z$-even dominance and the $\pi/2$ phase shift over mechanisms beyond scalar disorder~\cite{Zhang2022SciAdv,Kalappattil2026}. Evaluating the full quantum response at $F=28.6\;\mathrm{T}$ over $4.76\text{--}9.80\;\mathrm{T}$ ($\sigma_{b0}=2.05\times 10^{-4}\;\mathrm{S}\simeq 3.5\bar\sigma_{xx}$, $\mu_b=0.08\;\mathrm{m^2\,V^{-1}\,s^{-1}}$) gives $\Lambda_B^{(2)}=14.5\;\mathrm{T}$ vs.\ $\Lambda_B^{(1)}=23.4\;\mathrm{T}$, i.e., a decay-rate ratio $\Lambda_B^{(2)}/\Lambda_B^{(1)}=0.62$ ($0.55\text{--}0.65$ across realistic bulk-shunt ratios $2.75\text{--}7.5$ and all four detrending protocols, arising from the slowly growing intrinsic $\Delta\sigma_\parallel^{(2)}$ envelope converted by $(r^2+h^2)^{3/2}$ as shown in Fig.~\ref{fig:sm_hall_scba}(b), whereas lower unphysical shunts $0.5\text{--}1.5$ near compensation crossover transiently raise the ratio to $0.80\text{--}1.02$ before returning to $0.58$ at zero shunt; $0.67$ via asymptotic NLLK), consistent with the measured ratio $21.3/36.5=0.58$ [Fig.~2(d), where the dotted line shows the ratio guide $0.58\,\Lambda_B^{(1)}$, and Fig.~\ref{fig:sm_hall_scba}(b)], while noting that the theoretical absolute slopes lie $32\%\text{--}36\%$ below the measured values (reflecting an overall effective broadening scale).

\FloatBarrier
\section{Microscopic instantiation II: Purely orbital mechanism in gated Bernal bilayer graphene}
\label{sec:sm_blg}

\subsection{Peierls tight-binding lattice, four-band continuum model, and \texorpdfstring{$\Delta n=\pm 3$}{Delta n = +-3} selection rule}
\label{subsec:sm_blg_ham}

\begin{figure}[!htbp]
 \centering
 \SupplementFigure{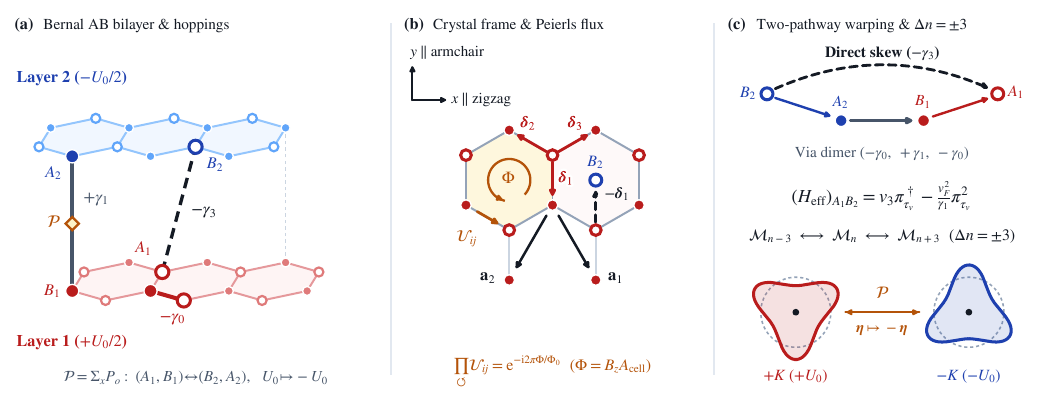}
 \caption{\label{fig:sm_blg_lattice}Peierls tight-binding lattice and orbital trigonal warping in gated Bernal BLG (parameters in Table~\ref{tab:sm_blg_params}). (a)~AB-stacked bilayer showing $(A_1,B_2)$, $(B_1,A_2)$, $(-\gamma_0,+\gamma_1,-\gamma_3)$, $\pm U_0/2$, and inversion center $\mathcal P$. (b)~Plan view of bonds and Peierls flux $-2\pi\Phi/\Phi_0$. (c)~Interference of $-\gamma_3$ and virtual dimer hopping yielding opposite triangular orbits for $(+K,+U_0)$ and $(-K,-U_0)$ at $B_z>0$.}
\end{figure}

Setting $g_s=0$ ($N_s=2$) in the crystal frame $x\parallel\text{zigzag}$, $y\parallel\text{armchair}$ ($d_{\mathrm{nn}}=a_{\mathrm{lat}}/\sqrt3$, $A_{\mathrm{cell}}=\sqrt3 a_{\mathrm{lat}}^2/2$, primitive lattice vectors $\bm a_{1,2}=a_{\mathrm{lat}}(\pm\frac{1}{2},-\frac{\sqrt3}{2})$), the four-sublattice Peierls tight-binding Hamiltonian [geometry in Table~\ref{tab:sm_blg_params} and Fig.~\ref{fig:sm_blg_lattice}] reads~\cite{McCannKoshino2013RPP}
\begin{equation}
 H=\sum_i\Big[\tfrac{1}{2}\xi_i U_0+V_{\mathrm{imp}}(\bm r_i)\Big]c_i^\dagger c_i+\Bigg[\!-\gamma_0\!\sum_{\langle i,j\rangle}\mathcal U_{ij}c_i^\dagger c_j+\gamma_1\!\sum_{\langle i,j\rangle_1}\mathcal U_{ij}c_i^\dagger c_j-\gamma_3\!\sum_{\langle i,j\rangle_3}\mathcal U_{ij}c_i^\dagger c_j+\mathrm{H.c.}\Bigg],\label{eq:sm_blg_tb}
\end{equation}
with $\mathcal U_{ij}=\exp[-(\ii e/\hbar)\int_{\bm r_j}^{\bm r_i}\bm{\mathcal A}\cdot\dd\bm r]$ threading flux $\Phi=B_z A_{\mathrm{cell}}$.

\textit{1. Tight-binding $\bm k$-space Bloch Hamiltonian.}---In the AB-stacked unit cell, the four basis sites lie at $\bm\rho_{A_1}=(0,+d_{\mathrm{nn}},-d_\perp/2)$, $\bm\rho_{B_2}=(0,-d_{\mathrm{nn}},+d_\perp/2)$, and $\bm\rho_{B_1,A_2}=(0,0,\mp d_\perp/2)$. The three intralayer nearest-neighbor vectors pointing from $A_1\to B_1$ (and identically from $A_2\to B_2$) are
\begin{equation}
 \bm\delta_1=(0,-d_{\mathrm{nn}}),\qquad
 \bm\delta_2=\bm\delta_1-\bm a_1=\Big(-\frac{a_{\mathrm{lat}}}{2},+\frac{d_{\mathrm{nn}}}{2}\Big),\qquad
 \bm\delta_3=\bm\delta_1-\bm a_2=\Big(+\frac{a_{\mathrm{lat}}}{2},+\frac{d_{\mathrm{nn}}}{2}\Big),\label{eq:sm_blg_bonds}
\end{equation}
while the vertical dimer bond $A_2\to B_1$ has zero in-plane displacement and the three oblique skew bonds $A_1\to B_2$ project onto $-\bm\delta_\nu$ ($\nu=1,2,3$; Table~\ref{tab:sm_blg_params}). At $B_z=0$ and $V_{\mathrm{imp}}=0$, introducing the physical-position Bloch operators $c_{\alpha,\bm k}=N_{\mathrm{cell}}^{-1/2}\sum_{\bm R}\ee^{-\ii\bm k\cdot(\bm R+\bm\rho_\alpha)}c_{\alpha,\bm R}$ (so that every hopping matrix element carries its exact physical displacement factor $\ee^{\ii\bm k\cdot(\bm r_j-\bm r_i)}$) maps Eq.~\eqref{eq:sm_blg_tb} onto the full-Brillouin-zone Bloch matrix in basis $(A_1,B_2,A_2,B_1)^{\mathsf T}$:
\begin{equation}
 H_{\mathrm{BLG}}(\bm k,U_0)=\begin{pmatrix}U_0/2&-\gamma_3 f^*(\bm k)&0&-\gamma_0 f(\bm k)\\ -\gamma_3 f(\bm k)&-U_0/2&-\gamma_0 f^*(\bm k)&0\\ 0&-\gamma_0 f(\bm k)&-U_0/2&\gamma_1\\ -\gamma_0 f^*(\bm k)&0&\gamma_1&U_0/2\end{pmatrix},\label{eq:sm_blg_bloch}
\end{equation}
where the lattice structure factor is
\begin{equation}
 f(\bm k)=\sum_{\nu=1}^3\ee^{\ii\bm k\cdot\bm\delta_\nu}=\ee^{-\ii k_y d_{\mathrm{nn}}}+2\ee^{\ii k_y d_{\mathrm{nn}}/2}\cos\!\left(\frac{k_x a_{\mathrm{lat}}}{2}\right).\label{eq:sm_blg_fk}
\end{equation}
Because $f(-\bm k)=f^*(\bm k)$, Eq.~\eqref{eq:sm_blg_bloch} obeys the exact inversion identity $\Sigma_x H_{\mathrm{BLG}}(\bm k,U_0)\Sigma_x^{-1}=H_{\mathrm{BLG}}(-\bm k,-U_0)$ ($\Sigma_x=I_2\otimes\sigma_x$ swapping $A_1\leftrightarrow B_2,A_2\leftrightarrow B_1$). Under Peierls substitution $\hbar\bm k\to\bm\Pi$, this becomes $\tilde{\mathcal P}_0 H_{\mathrm{BLG}}(\bm\Pi/\hbar,U_0)\tilde{\mathcal P}_0^{-1}=H_{\mathrm{BLG}}(\bm\Pi/\hbar,-U_0)$ with $\tilde{\mathcal P}_0\equiv\Sigma_x(-1)^{\hat n}$.

\textit{2. Low-energy expansion around $\bm K_\pm$ to the four-band continuum model.}---The Dirac corners lie at $\bm K_{\tau_v}=(\tau_v\frac{4\pi}{3a_{\mathrm{lat}}},0)$ ($\tau_v=\pm 1$), where $\cos(\tau_v\frac{2\pi}{3})=-\frac{1}{2}$ gives $f(\bm K_{\tau_v})=1+2(-\frac{1}{2})=0$. Evaluating the first partial derivatives of Eq.~\eqref{eq:sm_blg_fk} at $\bm K_{\tau_v}$ using $d_{\mathrm{nn}}=a_{\mathrm{lat}}/\sqrt3$ yields
\begin{equation}
 \left.\frac{\partial f}{\partial k_x}\right|_{\bm K_{\tau_v}}=-a_{\mathrm{lat}}\sin\!\left(\tau_v\frac{2\pi}{3}\right)=-\tau_v\frac{\sqrt3\,a_{\mathrm{lat}}}{2},\qquad
 \left.\frac{\partial f}{\partial k_y}\right|_{\bm K_{\tau_v}}=-\ii d_{\mathrm{nn}}+\ii d_{\mathrm{nn}}\cos\!\left(\tau_v\frac{2\pi}{3}\right)=-\ii\frac{\sqrt3\,a_{\mathrm{lat}}}{2},\label{eq:sm_blg_fk_deriv}
\end{equation}
so expanding $\bm k=\bm K_{\tau_v}+\bm q$ to linear order in $\bm q$ gives
\begin{align}
 f(\bm K_{\tau_v}+\bm q)&=-\frac{\sqrt3\,a_{\mathrm{lat}}}{2}(\tau_v q_x+\ii q_y)+\mathcal O(q^2),\nonumber\\
 -\gamma_0 f(\bm K_{\tau_v}+\bm q)&\simeq v_F(\tau_v\hbar q_x+\ii\hbar q_y),\qquad
 -\gamma_3 f^*(\bm K_{\tau_v}+\bm q)\simeq v_3(\tau_v\hbar q_x-\ii\hbar q_y),\label{eq:sm_blg_linearize}
\end{align}
with $v_{F,3}\equiv\sqrt3 a_{\mathrm{lat}}\gamma_{0,3}/(2\hbar)$. Replacing $\hbar\bm q\to\bm\Pi$ and defining $\pi_{\tau_v}\equiv\tau_v\Pi_x+\ii\Pi_y$ ($\pi_{\tau_v}^\dagger=\tau_v\Pi_x-\ii\Pi_y$) converts Eq.~\eqref{eq:sm_blg_bloch} into the four-band magnetic continuum Hamiltonian~\cite{McCannFalko2006,McCannKoshino2013RPP}:
\begin{equation}
 H_{\mathrm{BLG}}^{(\tau_v)}(\bm\Pi,U_0)=\begin{pmatrix}U_0/2&v_3\pi_{\tau_v}^\dagger&0&v_F\pi_{\tau_v}\\ v_3\pi_{\tau_v}&-U_0/2&v_F\pi_{\tau_v}^\dagger&0\\ 0&v_F\pi_{\tau_v}&-U_0/2&\gamma_1\\ v_F\pi_{\tau_v}^\dagger&0&\gamma_1&U_0/2\end{pmatrix}.\label{eq:sm_blg4}
\end{equation}

\textit{3. Schur-complement reduction to the two-band effective model and $\mathcal O(U_0/\gamma_1^2)$ valley splitting.}---Partitioning $H_{\mathrm{BLG}}^{(\tau_v)}(\psi_N,\psi_D)^{\mathsf T}=E(\psi_N,\psi_D)^{\mathsf T}$ into non-dimer $\psi_N=(A_1,B_2)^{\mathsf T}$ and high-energy dimer $\psi_D=(A_2,B_1)^{\mathsf T}$ subspaces defines the $2\times 2$ blocks
\begin{equation}
 H_{NN}=\begin{pmatrix}U_0/2&v_3\pi_{\tau_v}^\dagger\\ v_3\pi_{\tau_v}&-U_0/2\end{pmatrix},\qquad
 H_{ND}=H_{DN}=\begin{pmatrix}0&v_F\pi_{\tau_v}\\ v_F\pi_{\tau_v}^\dagger&0\end{pmatrix},\qquad
 H_{DD}=\begin{pmatrix}-U_0/2&\gamma_1\\ \gamma_1&U_0/2\end{pmatrix}.\label{eq:sm_blg_blocks}
\end{equation}
Exact elimination $\psi_D=(E-H_{DD})^{-1}H_{DN}\psi_N$ yields $H_{\mathrm{eff}}^{(\tau_v)}(E)=H_{NN}+H_{ND}(E-H_{DD})^{-1}H_{DN}$. Expanding $(E-H_{DD})^{-1}=-\frac{1}{\gamma_1}\sigma_x-\frac{E}{\gamma_1^2}I_2+\frac{U_0}{2\gamma_1^2}\sigma_z+\mathcal O(\gamma_1^{-3})$ for $|E|,|U_0|\ll\gamma_1$ separates the leading $\mathcal O(1/\gamma_1)$ two-band Hamiltonian from the next-order $\mathcal O(1/\gamma_1^2)$ dimer diagonal correction:
\begin{align}
 H_{\mathrm{eff}}^{(\tau_v)}&=\begin{pmatrix}U_0/2&v_3\pi_{\tau_v}^\dagger-\frac{v_F^2}{\gamma_1}\pi_{\tau_v}^2\\ v_3\pi_{\tau_v}-\frac{v_F^2}{\gamma_1}(\pi_{\tau_v}^\dagger)^2&-U_0/2\end{pmatrix},\nonumber\\
 \delta H_{\mathrm{dim}}^{(\tau_v)}&=\operatorname{diag}\!\left(-\frac{E+U_0/2}{\gamma_1^2}v_F^2\pi_{\tau_v}\pi_{\tau_v}^\dagger,\;-\frac{E-U_0/2}{\gamma_1^2}v_F^2\pi_{\tau_v}^\dagger\pi_{\tau_v}\right).\label{eq:sm_blg2_schur}
\end{align}
Whereas the bare $\mathcal O(1/\gamma_1)$ two-band model $H_{\mathrm{eff}}^{(\tau_v)}$ retains valley-degenerate $n\ge 2$ Landau levels at $v_3=0$, with energies $E_n=\pm\sqrt{(U_0/2)^2+(\hbar\omega_0)^2 n(n-1)}$ ($\hbar\omega_0\equiv 2\hbar^2 v_F^2/(\gamma_1\ell_B^2)=\hbar e|B_z|/m_{\mathrm{BLG}}$, $m_{\mathrm{BLG}}=\gamma_1/(2v_F^2)$), the $\mathcal O(1/\gamma_1^2)$ dimer correction $\delta H_{\mathrm{dim}}^{(\tau_v)}$ couples $U_0$ to the non-commutativity $[\pi_{\tau_v},\pi_{\tau_v}^\dagger]=-2\tau_v s_B\hbar^2/\ell_B^2$ and shifts $(A_1,B_2)$ asymmetrically, directly generating the $B_z$-odd valley Landau-level splitting ($\Delta\gamma_B(-B_z)=-\Delta\gamma_B(B_z)$).

\textit{4. $\Delta n=\pm 3$ selection rule, velocity amplitudes $V_{1,2}$, and two-valley NLLK sum.}---For $s_B=+1,\tau_v=+1$ ($\pi_+=\sqrt2\hbar a_B^\dagger/\ell_B$), the unwarped four-band multiplet $|\Phi_{n,\alpha}^+\rangle=(c_1|n\rangle,c_2|n-2\rangle,c_3|n-1\rangle,c_4|n-1\rangle)^{\mathsf T}$ ($n\ge 2$) offsets the non-dimer $(A_1,B_2)$ oscillator indices by two. Restoring $(H_{v_3})_{12}=(\sqrt2\hbar v_3/\ell_B)a_B$ adds $\Delta\nu=\pm 1$, enforcing $2+1=3$, i.e., strict $\Delta n=\pm 3$ hybridization:
\begin{equation}
 \langle\Phi_{n',\alpha'}^+|H_{v_3}|\Phi_{n,\alpha}^+\rangle=\frac{\sqrt2\hbar v_3}{\ell_B}\Big[\sqrt{n-2}\,c_{1,n'\alpha'}^*c_{2,n\alpha}\,\delta_{n',n-3}+\sqrt{n+1}\,c_{2,n'\alpha'}^*c_{1,n\alpha}\,\delta_{n',n+3}\Big].\label{eq:sm_blg_warp_selection}
\end{equation}
In the low-energy two-band limit ($U_0=0,n\gg 1$), the unwarped eigenstates of $H_{\mathrm{eff}}^{(+)}$ read $|n,s\rangle=\frac{1}{\sqrt2}(|n,A_1\rangle-s|n-2,B_2\rangle)$ with $E_{n,s}^{(0)}=s\hbar\omega_0\sqrt{n(n-1)}$ ($s=\pm 1$), and $v_x=\partial_{\Pi_x}H_{\mathrm{eff}}^{(+)}=v_{0,x}+v_{w,x}$ splits into $v_{0,x}=-(2v_F^2/\gamma_1)(\pi_+\sigma_+ + \pi_+^\dagger\sigma_-)$ and $v_{w,x}=v_3\sigma_x$. Evaluating $\langle n+1,+|v_{0,x}|n,+\rangle$ yields $V_{1,+}\simeq\frac{1}{2}\sqrt{2n}\,\hbar\omega_0\ell_B/\hbar=v_*/2$ ($v_*=\hbar k_F/m_{\mathrm{BLG}}$). For $V_{2,+}=\langle\Psi_{n+2,+}|v_x|\Psi_{n,+}\rangle$ to first order in $H_{v_3}$, combining the bare skew matrix element $\langle n+2,+|v_{w,x}|n,+\rangle=-\frac{1}{2}v_3$ with $\Delta n=\pm 3$ intraband ($s'=+1$, $+\frac{1}{3}v_3$) and interband ($s'=-1$, $-\frac{1}{2}v_3$) perturbation mixing yields
\begin{equation}
 V_{1,+}\simeq\frac{v_*}{2},\qquad
 V_{2,+}(U_0=0)\simeq\left(-\frac{1}{2}+\frac{1}{3}-\frac{1}{2}\right)v_3=-\frac{2}{3}\,v_3,\qquad
 V_{2,-}(U_0=0)=-V_{2,+}(U_0=0).\label{eq:sm_blg_v12}
\end{equation}
At $U_0\neq 0$, four-band dimer coupling in Eq.~\eqref{eq:sm_blg4} produces both the $B_z$-odd valley phase splitting $\Delta\gamma_B\equiv\gamma_{B,+}-\gamma_{B,-}$ [$\Delta\gamma_B(-B_z)=-\Delta\gamma_B(B_z)$ by $M_x:(K_+,+B_z)\leftrightarrow(K_-,-B_z)$] and unequal valley amplitudes. Summing Eq.~\eqref{eq:sm_signed_conditional_nllk} over both valleys in the leading common-spacing approximation ($g_\tau\equiv\Gamma_\tau/(\hbar|\omega_{c,\tau}|)\simeq\tilde\Gamma$) with $s_B$ factored from $\chi_o$ gives valley weights obeying $W_{+p}(-B_z,U_0)=W_{-p}(+B_z,U_0)$ and $W_{\tau p}(B_z,-U_0)=-W_{-\tau,p}(B_z,+U_0)$, yielding $\sum_{\tau=\pm}W_{\tau p}\sin(2\pi p n_{F,\tau})=(W_{+p}+W_{-p})\cos(\pi p\,\Delta\gamma_B)\sin(2\pi p\,\bar n_F)-(W_{+p}-W_{-p})\sin(\pi p\,\Delta\gamma_B)\cos(2\pi p\,\bar n_F)$. Both terms are strictly $B_z$-even and $U_0$-odd (vanishing identically at $U_0=0$), mixing the $\pi/2$-shifted amplitude-imbalance term with the in-phase/antiphase phase-splitting term into a gate-tunable phase.

\begin{table}[!htbp]
\caption{\label{tab:sm_blg_params}Bernal BLG lattice geometry ($x\parallel\text{zigzag}$, $y\parallel\text{armchair}$; $d_{\mathrm{nn}}=a_{\mathrm{lat}}/\sqrt3$) and simulation parameter sets across main Fig.~3 and Figs.~\ref{fig:sm_blg_lattice}--\ref{fig:sm_blg_impurity}.}
\footnotesize
\begin{ruledtabular}
\begin{tabular}{p{0.27\textwidth}p{0.68\textwidth}}
\multicolumn{2}{l}{\textit{(a) Four-sublattice unit-cell geometry and hopping bonds}} \\
\hline
Basis sites & $\bm\rho_{A_1}=(0,+d_{\mathrm{nn}},-d_\perp/2)$, $\bm\rho_{B_2}=(0,-d_{\mathrm{nn}},+d_\perp/2)$, $\bm\rho_{B_1,A_2}=(0,0,\mp d_\perp/2)$ \\
Intralayer / skew bonds & $\bm\delta_1=(0,-d_{\mathrm{nn}})$, $\bm\delta_{2,3}=(\mp a_{\mathrm{lat}}/2,+d_{\mathrm{nn}}/2)$; skew $A_1\to B_2$ ($-\gamma_3$) projects onto $-\bm\delta_\nu$ \\
\hline
\multicolumn{2}{l}{\textit{(b) Model hierarchy and simulation parameter sets}} \\
\hline
Fig.~\ref{fig:sm_blg_lattice}(c) (orbit schematic) & 4-band Eq.~\eqref{eq:sm_blg4}: $\gamma_1=0.40\;\mathrm{eV}$, $U_0=60\;\mathrm{meV}$, $E=75\;\mathrm{meV}$, $v_3/v_F=0.22$ \\
Fig.~\ref{fig:sm_blg_valleys}(a--d) (realistic continuum) & 4-band Eq.~\eqref{eq:sm_blg4}: $v_F=10^6\;\mathrm{m/s}$, $\gamma_1=0.40\;\mathrm{eV}$, $v_3/v_F=0.12$, $\mu=75\;\mathrm{meV}$, $\Gamma=2.2\;\mathrm{meV}$, $U_0=\pm 60\;\mathrm{meV}$ \\
Main Fig.~3(b) (1-imp.\ disk) & Scaled TB disk ($61{,}268$ sites): $\gamma_0=0.60$, $\gamma_1=0.40$, $\gamma_3=0.18\;\mathrm{eV}$; $U_0=+60\;\mathrm{meV}$, $\Phi/\Phi_0=1/180$, $E=99.19\;\mathrm{meV}$ \\
Fig.~\ref{fig:sm_blg_impurity}(a--h) (1-/3-imp.\ disks) & Scaled TB ($61{,}268$ / $92{,}846$ sites): $\gamma_0=0.60$, $\gamma_1=0.38$, $\gamma_3=0.18\;\mathrm{eV}$; $U_0=+80,0\;\mathrm{meV}$ \\
Main Fig.~3(c,d) (disordered torus) & Scaled TB torus ($\gamma_0=0.60$, $\gamma_1=0.38$, $\gamma_3=0.18\;\mathrm{eV}$; $\mu=0.24\;\mathrm{eV}$, $U_0=80\;\mathrm{meV}$, $q\in[62,156]$): $\Gamma_{\mathrm{clean}}=4.80$, $\Gamma_0=2.50\;\mathrm{meV}$ \\
\end{tabular}
\end{ruledtabular}
\end{table}
\FloatBarrier

\subsection{Inversion gate parity, valley splitting, open-disk LDOS, and disordered torus transport}
\label{subsec:sm_blg_torus}

\begin{figure}[!t]
\centering
\SupplementFigure{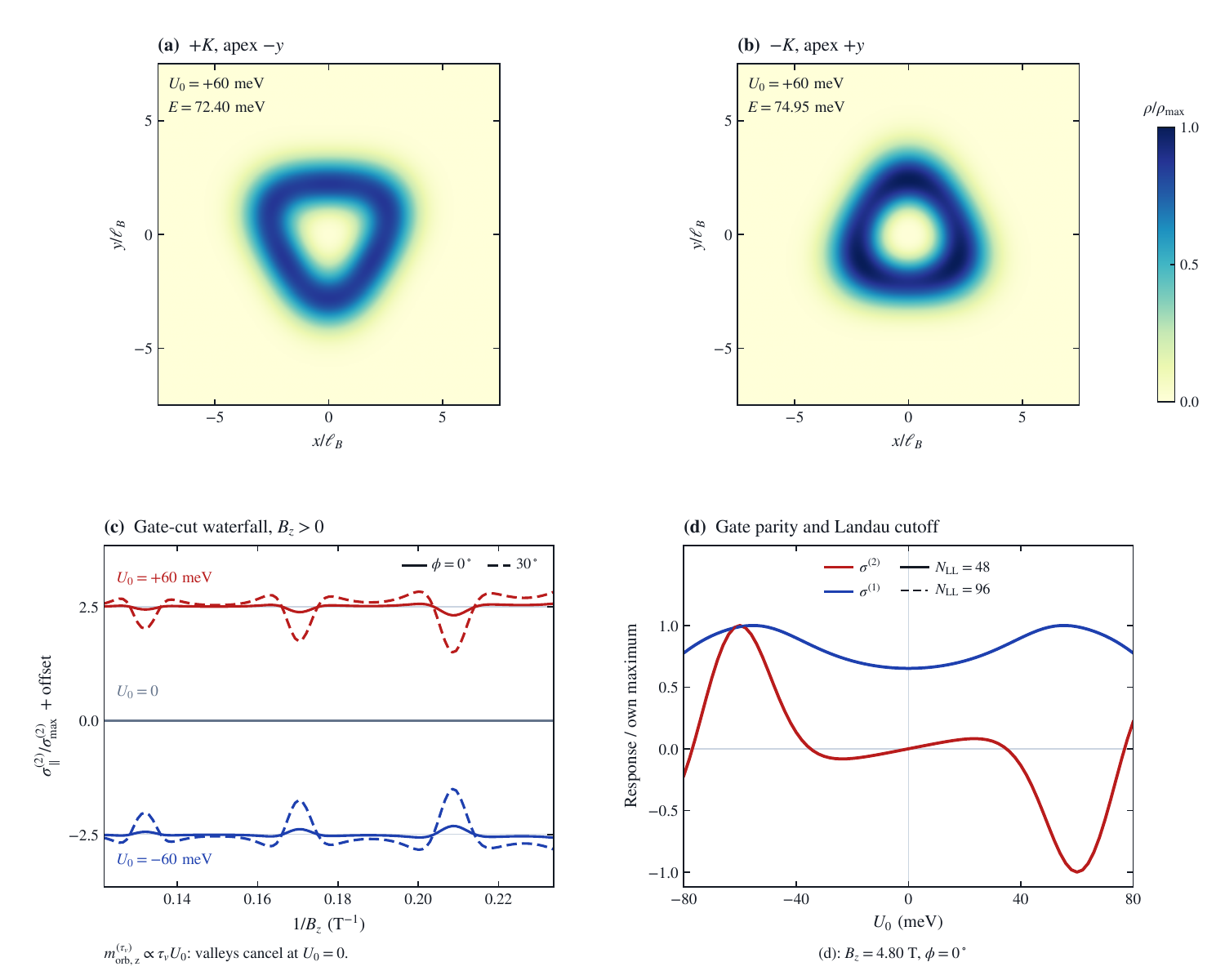}
\caption{Orbital mechanism and gate parity in realistic four-band continuum BLG [Eq.~\eqref{eq:sm_blg4}; Table~\ref{tab:sm_blg_params}, $B_z>0$]. (a),(b)~Resonant $m=0$ Landau densities near $\mu=75\;\mathrm{meV}$ at $U_0=+60\;\mathrm{meV},B_z\simeq 7.41\;\mathrm{T}$ showing opposite triangles for $+K$ ($\triangledown$) and $-K$ ($\triangle$). (c)~Valley-summed $\sigma_\parallel^{(2)}$ at $\phi=0^\circ$ (solid) and $30^\circ$ (dashed) for $U_0=\pm 60,0\;\mathrm{meV}$. (d)~Even $\sigma_{xx}^{(1)}(U_0)$ vs.\ odd $\sigma_\parallel^{(2)}(U_0)$ at $B_z=4.80\;\mathrm{T}$ ($N_{\mathrm{LL}}=48$ vs.\ $96$).}
\label{fig:sm_blg_valleys}
\end{figure}

Three-dimensional inversion $\mathcal P=\Sigma_x P_o$ ($\Sigma_x$ swapping $A_1\leftrightarrow B_2,A_2\leftrightarrow B_1$; $P_o:\bm r\mapsto -\bm r$) leaves $B_z$ fixed and maps $\mathcal P H_{\mathrm{BLG}}^{(\tau_v)}(\bm\Pi,U_0)\mathcal P^{-1}=H_{\mathrm{BLG}}^{(-\tau_v)}(\bm\Pi,-U_0)$ while flipping $\bm\eta\mapsto -\bm\eta$ and $v_\parallel\mapsto -v_\parallel$. Summing over valleys yields opposite gate parities [confirmed in realistic-parameter continuum calculations, Fig.~\ref{fig:sm_blg_valleys}]:
\begin{equation}
 \sigma_{ab}^{(1)}(B_z,-U_0)=+\sigma_{ab}^{(1)}(B_z,U_0),\qquad
 \chi_{o,e}(B_z,-U_0)=-\chi_{o,e}(B_z,U_0),\qquad
 \chi_{o,e}(B_z,0)=0.\label{eq:sm_blg_gate_parity}
\end{equation}
With static disorder $V_{\mathrm{imp}}(\bm r)$, $\mathcal P H(B_z,+U_0,V_{\mathrm{imp}})\mathcal P^{-1}=H(B_z,-U_0,V_{\mathrm{imp}}^{\mathcal P})$ [$V_{\mathrm{imp}}^{\mathcal P}(\bm r)=V_{\mathrm{imp}}(-\bm r)$], pairing realizations one-to-one so Eq.~\eqref{eq:sm_blg_gate_parity} survives ensemble averaging over any inversion-symmetric disorder distribution. Applying $U_0\neq 0$ splits $+K$ and $-K$ via opposite valley orbital moments $m_{\tau_v,z}(\bm k)=\tau_v e\hbar U_0\varepsilon_k/[2m_{\mathrm{BLG}}(\varepsilon_k^2+U_0^2/4)]$ [with $\varepsilon_k\equiv\hbar^2 k^2/(2m_{\mathrm{BLG}})$] and asymmetric four-band dimer coupling~\cite{McCann2006PRB,XiaoChangNiu2010}.

\begin{figure}[!htbp]
\centering
\SupplementFigure{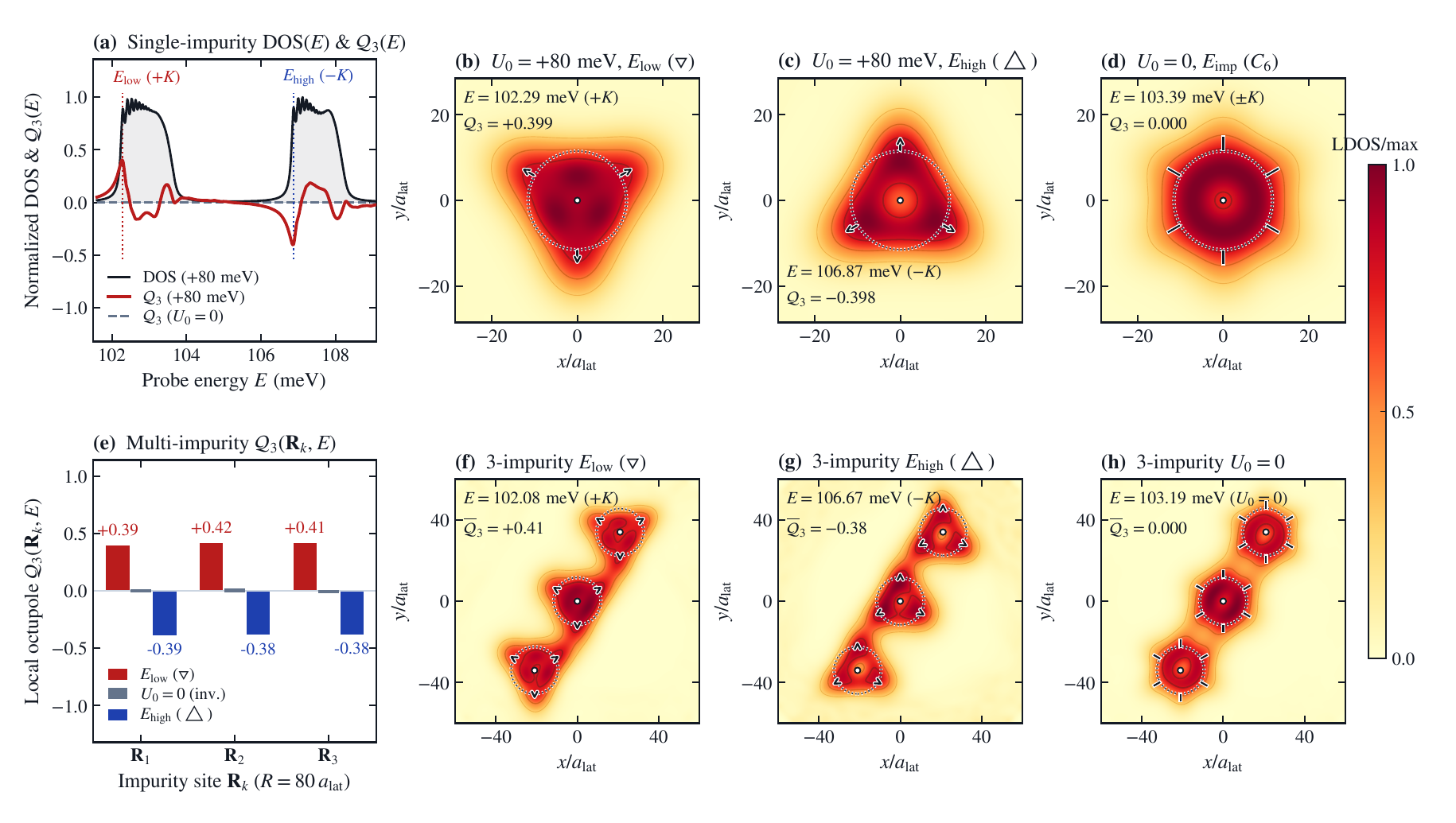}
\caption{Unfiltered real-space $\mathrm{LDOS}(\bm r,E)$ and local octupole $\mathcal Q_3(\bm R_k,E)$ on open Bernal BLG disks ($\Phi/\Phi_0=+1/180$, $\gamma_3/\gamma_0=0.30$; Table~\ref{tab:sm_blg_params}). (a)~Single-impurity $\mathrm{DOS}(E)$ and $\mathcal Q_3(E)$ across the complete untruncated $+K$ and $-K$ Landau bands at $U_0=+80\;\mathrm{meV}$ (black/red) vs.\ $U_0=0$ (dashed gray). (b)--(d)~2D LDOS maps at $E_{\mathrm{low}}=102.29\;\mathrm{meV}$ ($+K$, $\mathcal Q_3=+0.399$), $E_{\mathrm{high}}=106.87\;\mathrm{meV}$ ($-K$, $\mathcal Q_3=-0.398$), and $U_0=0$ ($C_6$, $\mathcal Q_3=0.000$). (e)--(h)~Three-impurity disk ($92{,}846$ sites) showing uniform crystal-locked $\triangledown$ ($\overline{\mathcal Q}_3=+0.41$), $\triangle$ ($\overline{\mathcal Q}_3=-0.38$), and compensated $U_0=0$ halos ($\overline{\mathcal Q}_3=0.000$).}
\label{fig:sm_blg_impurity}
\end{figure}

On open-boundary disks with smooth Gaussian impurities $V_{\mathrm{imp}}(\bm r)=-\sum_k V_0\exp[-|\bm r-\bm R_k|^2/(2\xi_{\mathrm{imp}}^2)]$, we evaluate the unfiltered $\mathrm{LDOS}(\bm r,E)$, conserved orbital bond current $I_{i\to j}^Q(E)=\frac{2e}{\hbar}\Im[H_{ij}\rho_{ji}(E)]$ [Fig.~3(b)], and normalized local octupole $\mathcal Q_3(\bm R_k,E)=\int_{|\delta\bm r_k|\le r_c}\dd^2\bm r\,\mathrm{LDOS}(3\delta x_k^2\delta y_k-\delta y_k^3)/[\bar r_k^3 W_k]$ ($r_c=21.5\,a_{\mathrm{lat}}$). Figure~\ref{fig:sm_blg_impurity} verifies crystal-locked $\triangledown$ ($+K$, $\mathcal Q_3=+0.399$) and $\triangle$ ($-K$, $\mathcal Q_3=-0.398$) impurity halos at $U_0=+80\;\mathrm{meV}$ collapsing to a $C_6$ hexagram ($\mathcal Q_3=0.000$) at $U_0=0$, persisting across three off-center impurities ($\overline{\mathcal Q}_3=+0.41,-0.38,0.000$).

For the finite-torus projected-channel illustration in Fig.~3(c,d), we place Eq.~\eqref{eq:sm_blg_tb} on a skewed magnetic torus ($\bm L_1=q\bm a_1^{\mathrm{TB}}$, $\bm L_2=n_2\bm a_2^{\mathrm{TB}}$, $n_2=6$, $\Phi/\Phi_0=\pm 1/q$, $q\in[62,156]$, $\mu=0.24\;\mathrm{eV}$, $U_0=80\;\mathrm{meV}$) and evaluate the velocity-vertex trace~\eqref{eq:sm_sigma2_velocity_reduced} via sparse LU factorization across $20$ independent disorder realizations per $(q,U)$ (Gaussian puddles $\xi_{\mathrm{imp}}=2.2\,a_{\mathrm{lat}}$, $\sigma_{\mathrm{pud}}=1.15 W_{\mathrm{nom}}$ plus Anderson disorder $\sigma_{\mathrm{And}}=0.45 W_{\mathrm{nom}}$, $W_{\mathrm{nom}}=3.20\;\mathrm{meV}$, $\Gamma_0=2.50\;\mathrm{meV}$; clean reference strip uses $n_{2,\mathrm{clean}}=12$, $\Gamma_{\mathrm{clean}}=4.80\;\mathrm{meV}$ with its $(+U_0,0,-U_0)$ curves assigned via exact inversion parity $(+c,0,-c)$, while all $20$ disordered realizations are evaluated independently at all three biases), triad-averaged over $C_{3z}$-equivalent angles to isolate the $3\phi$ channels. Because this scaled-hopping torus has transverse width $L_2/\ell_B\simeq 1.3\text{--}2.1$ without twist averaging or transverse-size convergence sweeps, it serves strictly as an illustrative demonstration of disorder survival and gate reversal, whereas exact bulk-limit gate parity and unprojected realistic-field ($5\text{--}7\;\mathrm{T}$) transport are established analytically by Eq.~\eqref{eq:sm_blg_gate_parity} and numerically in Fig.~\ref{fig:sm_blg_valleys}. Waveforms in Fig.~3(c,d) display $\tilde\chi_{o,e}(q,U)=\chi_{o,e}^{\mathrm{red}}(q,U)/\max_{q'}|\chi_{o,e;\mathrm{clean}}^{\mathrm{red}}(q',+U_0)|$ with $\chi_o^{\mathrm{red}}=(\chi_o/\Gamma_{\mathrm{diss}})(100/q)^2$ ($\Gamma_{\mathrm{diss}}=\Gamma_{\mathrm{clean}}$ or $\Gamma_0$) and $\chi_e^{\mathrm{red}}=\chi_e(100/q)^2$.

\begin{figure}[!htbp]
\centering
\SupplementFigure{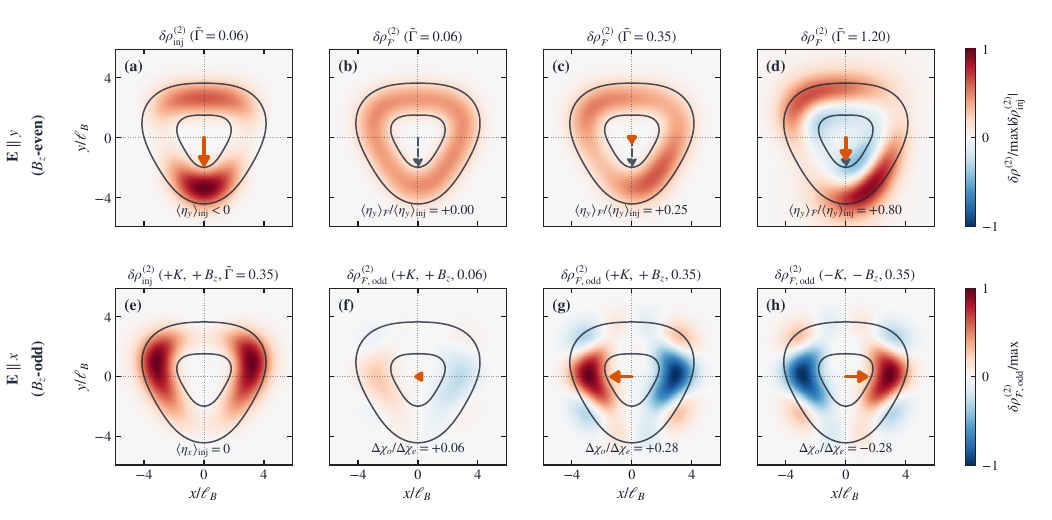}
\caption{Purely triangular real-space local-filling and steady-state density evolution in realistic four-band continuum Bernal BLG [Eq.~\eqref{eq:sm_blg4}; $v_F=10^6\;\mathrm{m/s}$, $\gamma_1=0.40\;\mathrm{eV}$, $v_3/v_F=0.12$, $U_0=+60\;\mathrm{meV}$, $|B_z|=7.41\;\mathrm{T}$, paired-quadruplet cutoff $N_{\mathrm{osc}}=64$, rising flank of the $E=88.81\;\mathrm{meV}$ conduction Landau level; solid contour marks $0.35\max|\Psi|^2$ with density-weighted ring ratio $|\sin 3\theta_{\bm r}/\cos 6\theta_{\bm r}|=11.0$ and $\langle\eta_y^3\rangle=-2.16\,\ell_B^3$]. (a)--(d)~$\bm E\parallel y$ ($B_z$-even, $+K,+B_z$): symmetrized local-filling density $\delta\rho_{\mathrm{inj,sym}}^{(2)}$ at $\tilde\Gamma=0.06$ (a) vs.\ steady-state kernel $\delta\rho_{\mathcal F}^{(2)}$ across $\tilde\Gamma=0.06,0.35,1.20$ (b)--(d), tracking $\langle\eta_y\rangle_{\mathcal F}/\langle\eta_y\rangle_{\mathrm{inj}}=+0.00,+0.25,+0.80$. (e)--(h)~$\bm E\parallel x$ ($B_z$-odd, $\Theta$-paired valleys): symmetric shoulder injection ($\langle\eta_x\rangle_{\mathrm{inj}}=0$) (e) and odd-$x$ steady-state shift $\delta\rho_{\mathcal F,\mathrm{odd}}^{(2)}$ (minus $v_3=0$ at fixed $E$) at $(+K,+B_z)$ (f),(g) reversing identically in its time-reversed partner $(-K,-B_z)$ (h) ($\Delta\chi_o/\Delta\chi_e=+0.28\to-0.28$).}
\label{fig:sm_blg_rho}
\end{figure}

Finally, to contrast directly with the hexagonally faceted Dirac cone in End Matter Fig.~4---where opposite spins carry $\pm\lambda_w\sigma_z$ and leave a visible $C_{2z}$-even $\cos 6\theta_{\bm r}$ contour alongside the $C_{2z}$-odd $\sin 3\theta_{\bm r}$ distortion---Fig.~\ref{fig:sm_blg_rho} evaluates the exact same local-filling and steady-state density kernels for the realistic four-band continuum BLG model Eq.~\eqref{eq:sm_blg4} ($v_F=10^6\;\mathrm{m/s}$, $\gamma_1=0.40\;\mathrm{eV}$, $v_3/v_F=0.12$, $U_0=+60\;\mathrm{meV}$, $|B_z|=7.41\;\mathrm{T}$, paired-quadruplet basis cutoff $N_{\mathrm{osc}}=64$, resonant conduction Landau level at $E=88.81\;\mathrm{meV}$). Because single-valley $\gamma_3$ warping breaks $C_6$ directly at first order without opposite-spin cancellation, the equilibrium ring is purely triangular ($|\sin 3\theta_{\bm r}/\cos 6\theta_{\bm r}|=11.0$ over the $\rho_{\mathrm{eq}}>0.25\max\rho_{\mathrm{eq}}$ ring, $\langle\eta_y^3\rangle=-2.16\,\ell_B^3$). Along $\bm E\parallel y$, clean cyclotron circulation ($\tilde\Gamma=0.06$) washes out the sharp $-y$ corner hot spot ($\langle\eta_y\rangle_{\mathcal F}/\langle\eta_y\rangle_{\mathrm{inj}}=+0.001$), whereas stronger damping recovers it to $+0.25$ ($\tilde\Gamma=0.35$) and $+0.80$ ($\tilde\Gamma=1.20$). Along $\bm E\parallel x$, comparing $(+K,+B_z)$ with its exact time-reversed partner $(-K,-B_z)$ (identical energy $E=88.81\;\mathrm{meV}$ and identical downward-triangle octupole $\langle\eta_y^3\rangle=-2.16\,\ell_B^3$) shows mirror-symmetric shoulder injection ($\langle\eta_x\rangle_{\mathrm{inj}}=0$) evolving into an opposite left--right chiral dipole ($\Delta\chi_o/\Delta\chi_e=+0.06$ at $\tilde\Gamma=0.06$, $+0.28\to-0.28$ at $\tilde\Gamma=0.35$, matching the NLLK ratio $\tilde\Gamma/(1+2\tilde\Gamma^2)=0.281$; the 2D odd-$x$ maps subtract the $v_3=0$ odd-$x$ reference evaluated at the same $E$, while $\langle\eta_x\rangle_{\mathcal F}$ and $\Delta\chi_o/\Delta\chi_e$ are strictly reference-independent).

\clearpage
\section{Numerical implementation and figure-reproduction scripts}
\label{sec:sm_reproducibility}

All simulation solvers, figure scripts, and precomputed archives (\texttt{.npz}) supporting the main text and this Supplemental Material are available from the corresponding authors upon reasonable request and will be deposited in a public repository upon publication. Table~\ref{tab:sm_solvers} summarizes every numerical solver and discretization setting.

\begin{table}[!ht]
\caption{\label{tab:sm_solvers}Numerical solvers, reproduction scripts, and discretization settings supporting main Figs.~1--4 and Figs.~\ref{fig:sm_boltzmann}--\ref{fig:sm_blg_rho}.}
\footnotesize
\begin{ruledtabular}
\begin{tabular}{p{0.15\textwidth}p{0.34\textwidth}p{0.16\textwidth}p{0.29\textwidth}}
Target figures & Script (\texttt{src-code/}) & Method / library & Key discretization \& convergence settings \\
\hline
Main Fig.~1 & \raggedright\texttt{plot\_fig1\_landau\_mechanism.py} & \raggedright Mechanism schematic & \raggedright Double-column 3-channel guiding-center scattering schematic (linear SdH, static octupole, $\pm B_z$ chiral circulation); zero-overlap verified\tabularnewline
Main Fig.~4 (End Matter) & \raggedright\texttt{plot\_fig4\_endmatter\_rho.py} & \raggedright Warped Dirac LL eigensolver & \raggedright Paired-doublet cutoff $N_{\mathrm{osc}}=22$, $n=4$ flank $A_4'>0$, $\tilde\Gamma=0.06,0.35,1.20$, $\pm B_z$\tabularnewline
\raggedright Main Fig.~2(a,b), Fig.~\ref{fig:sm_boltzmann} & \raggedright\texttt{compute\_fig2\_dirac\_crossover.py}, \texttt{plot\_figS1\_S2\_boltzmann\_bridge.py} & \raggedright Periodic BTE + Eq.~\eqref{eq:sm_sigma2_velocity_reduced} & \raggedright $96$ angular nodes; finite-basis $85\le N_{\mathrm{LL}}\le 1{,}200$; NLLK $p\le 19$\tabularnewline
\raggedright Fig.~\ref{fig:sm_dirac_orbit}, Fig.~\ref{fig:sm_nllk_zeeman} & \raggedright\texttt{solver\_dirac\_landau\_kubo.py}, \texttt{plot\_figS3\_nllk\_zeeman\_phase.py} & \raggedright Warped Dirac LL eigensolver & \raggedright Rectangular $2N_{\mathrm{LL}}$ ($N_{\mathrm{LL}}=90\text{--}100$); $48$-pt Gauss--Legendre; $p\le 24$\tabularnewline
\raggedright Main Fig.~2(b,c), Fig.~\ref{fig:sm_hall_scba}(a,c,d) & \raggedright\texttt{plot\_fig2\_dirac\_exp.py}, \texttt{plot\_figS4\_hall\_rotation\_scba.py} & \raggedright Joint NLLK fit + proj.\ SCBA & \raggedright $p_{\max}=3$ ($1{,}396$ pts); $|\Delta n|\le 40\text{--}45$, tol.\ $2\times 10^{-10}$, $384$-pt quad.; $144$ BS nodes\tabularnewline
\raggedright Main Fig.~2(d), Fig.~\ref{fig:sm_hall_scba}(b) & \raggedright\texttt{compute\_fig2\_alpha\_sn\_envelopes.py} & \raggedright Full-quantum decay pipeline & \raggedright $N_{\max}=36$, $d=4$, $w_{\mathrm{cyc}}=1.6$ across $4$ detrending protocols\tabularnewline
Fig.~\ref{fig:sm_blg_lattice} & \raggedright\texttt{plot\_figS5\_blg\_lattice\_hopping.py} & \raggedright Lattice \& orbit schematic & \raggedright 4-sublattice Bernal BLG geometry, Peierls flux, $\Delta n=\pm 3$ continuum orbits\tabularnewline
Fig.~\ref{fig:sm_blg_valleys}(a--d) & \raggedright\texttt{plot\_figS6\_blg\_valleys\_parity.py} & \raggedright 4-band continuum BLG & \raggedright $N_{\mathrm{LL}}=48$ (verified vs.\ $N_{\mathrm{LL}}=96$)\tabularnewline
\raggedright Main Fig.~3(b), Fig.~\ref{fig:sm_blg_impurity} & \raggedright\texttt{compute\_fig3\_blg\_disk\_ldos.py}, \texttt{plot\_figS7\_blg\_impurity\_ldos.py} & \raggedright Open-disk Peierls TB & \raggedright Shift-invert Arnoldi (\texttt{eigsh}); $61{,}268$ / $92{,}846$ sites\tabularnewline
Fig.~\ref{fig:sm_blg_rho} & \raggedright\texttt{plot\_figS8\_blg\_quantum\_rho.py} & \raggedright 4-band continuum BLG & \raggedright Paired-quadruplet cutoff $N_{\mathrm{osc}}=64$, $E=88.81\;\mathrm{meV}$ flank, $\tilde\Gamma=0.06,0.35,1.20$, $\Theta$-paired valleys\tabularnewline
Main Fig.~3(c,d) & \raggedright\texttt{compute\_fig3\_blg\_torus\_disorder.py} & \raggedright Periodic-torus Peierls TB & \raggedright Sparse LU (\texttt{splu}); $n_2=6$, $20$ disorder realizations/bias\tabularnewline
\end{tabular}
\end{ruledtabular}
\end{table}

\FloatBarrier

\end{document}